\documentclass[printer]{aa}
\usepackage{txfonts}
\usepackage{graphicx}
\usepackage{url}
\usepackage{hyperref}
\usepackage{psfrag}
\DeclareGraphicsExtensions{.pdf,.png,.jpg,.ps,.eps}
\usepackage{natbib}
\bibpunct{(}{)}{;}{a}{}{,} 

\newcommand{\msun}{\mbox{M$_{\odot}$}}

\begin{document}
\title{Mapping the H$_{2}$D$^{+}$ and N$_{2}$H$^{+}$ emission towards prestellar cores}
\subtitle{Testing dynamical models of the collapse using gas tracers}

\author{E.~Koumpia\inst{\ref{inst1},\ref{inst7}}, L.~Evans\inst{\ref{inst1},\ref{inst2},\ref{inst3}}, J.~Di Francesco\inst{\ref{inst4}}, F.~F.~S.~van der Tak\inst{\ref{inst5},\ref{inst6}}, R.~D.~Oudmaijer\inst{\ref{inst1}}} 

\institute{School of Physics \& Astronomy, University of Leeds, Woodhouse Lane, LS2 9JT Leeds, UK\label{inst1}
\and
IRAP, Université de Toulouse, CNRS, UPS, CNES, F-31400 Toulouse, France\label{inst2}
\and
INAF-Osservatorio Astrofisico di Arcetri, Largo E. Fermi 5,I-50125, Florence, Italy\label{inst3}
\and
Herzberg Astronomy and Astrophysics Research Centre, National Research Council of Canada, Victoria, BC, Canada\label{inst4}
\and
SRON Netherlands Institute for Space Research, Landleven 12, 9747 AD Groningen, The Netherlands\label{inst5}
\and
Kapteyn Astronomical Institute, University of Groningen, The Netherlands\label{inst6}
\and \email{ev.koumpia@gmail.com}\label{inst7}
}
\date{Received date/Accepted date}

\abstract
    {The study of prestellar cores is critical as they set the initial conditions in star formation and determine the final mass of the stellar object. To date, several hypotheses are describing their gravitational collapse.}{We perform detailed line analysis and modelling of H$_{2}$D$^{+}$ 1$_{10}$-1$_{11}$ and N$_{2}$H$^{+}$ 4-3 emission at 372 GHz, using 2\arcmin$\times$2\arcmin maps (JCMT). Our goal is to test the most prominent dynamical models by comparing the modelled gas kinematics and spatial distribution (H$_{2}$D$^{+}$ and N$_{2}$H$^{+}$) with observations towards four prestellar (L1544, L183, L694-2, L1517B) and one protostellar core (L1521f).}{We perform a detailed non-LTE radiative transfer modelling using RATRAN, where we compare the predicted spatial distribution and line profiles of H$_{2}$D$^{+}$ and N$_{2}$H$^{+}$ with observations towards all cores. To do so, we adopt the physical structure for each core predicted by three different dynamical models taken from literature: Quasi-Equilibrium Bonnor-Ebert Sphere (QE-BES), Singular Isothermal Sphere (SIS), and Larson-Penston (LP) flow.}{Our analysis provides an updated picture of the physical structure of prestellar cores. We find that the SIS model can be cleary excluded in explaining the gas emission towards the cores, but a larger sample is required to differentiate clearly between the LP flow, the QE-BES and the static models. All models of collapse underestimate the intensity of the gas emission by up to several factors towards the only protostellar core in our sample, indicating that different dynamics take place in different evolutionary core stages. If the LP model is confirmed towards a larger sample of prestellar cores, it would indicate that they may form by compression or accretion of gas from larger scales. If the QE-BES model is confirmed, it means that quasi hydrostatic cores can exist within turbulent ISM.}


\titlerunning{Mapping the H$_{2}$D$^{+}$ and N$_{2}$H$^{+}$ emission towards prestellar cores} 
\authorrunning{Koumpia et al.} 
 \maketitle


 \section{Introduction}
 \label{intro}

Star formation begins within molecular clouds, where magnetic fields and turbulence dominate the dynamics, leading to the formation of filamentary structures of gas \citep{Arzoumanian2011,Pudritz2013,Andre2017}. In particular, {\it{Herschel}} observations revealed that prestellar cores and protostellar objects are forming in thermally critical and supercritical filaments \citep{Andre2010,Tafalla2015}. Understanding the physical and chemical processes that take place in these very early stages of star formation is of high importance. Not only do such cores set the initial conditions of star formation and determine the final mass of the star \citep{Bergin2007}, but they also have a strong influence on the multiplicity \citep{Pineda2015}. Although low-mass star formation is better understood than high-mass star formation, the initial conditions of the collapse remain uncertain. Prestellar cores are known to contract under gravitational forces \citep{Krumholz2005}, while thermal and magnetic pressure and the presence of turbulence support the core from collapsing \citep{Goodman1998}.

The most broadly known dynamical models of a spherically symmetric collapsing core are: {\it{i)}} the Larson-Penston flow \citep[LP flow;][]{Larson1969,Penston1969}, where at t$\sim$0 the inflow velocity at large radii reaches supersonic values ($\sim$3.3$\times$ speed of sound) and therefore is far from equilibrium, and the density profile follows a power-law; {\it{ii)}} the singular isothermal sphere (SIS) evolving via an ``inside-out'' collapse \citep{Shu1977}; and {\it{iii)}} the hydrostatic Bonnor-Ebert sphere (BES) supported by turbulence, which when disturbed can lead to a contraction in quasi-equilibrium \citep[QE;][]{Broderick2010}. Constraining the dynamical structure of prestellar cores observationally is very challenging. Hence, detailed studies focusing on the gas emission, as well as the age determination of observed prestellar cores, are crucial to distinguish between the proposed models. 

\citet{Ward-Thompson1994} presented the first submillimeter continuum survey of a sample of cores without associated infrared emission shortward of 100~$\mu$m, therefore presumably starless cores, reporting their first detection using longer wavelengths ($>$ 450~$\mu$m). The lifetime of prestellar cores has been observationally estimated to be $\sim$ $10^4$-$10^6$ years \citep[e.g., L1544, L694-2, L183][]{Beichman1986,Caselli2008,Enoch2008}. The prestellar cores are characterized by very low temperatures (T $\approx 10$~K) that increase outwards, and high densities \citep[$>$ 10$^{5}$~cm$^{-3}$;][]{Keto2008} that show a flattened density profile in the center, resulting in the unique gas chemistry presented in the current paper. Various studies have shown a strong correlation between CO depletion and the degree of deuteration of hydrogen-based species in prestellar cores \citep{Caselli2003,Vastel2006}. Below the critical temperature of T $\approx 25$~K, gas-phase species, including almost all CNO species, are frozen out \citep[$\gtrsim 98\%$][]{Caselli2003} to dust grains and therefore are depleted \citep{Caselli2008}. The enrichment in H$_{2}$D$^{+}$ and N$_{2}$H$^{+}$ under those conditions can be understood if one takes a closer look at the production and destruction mechanisms of the relevant species (Equation \ref{eq.h2dp} to \ref{eq.co2}):

\begin{equation}
{\rm H}_3^+ + {\rm HD} \rightleftharpoons {\rm H}_2{\rm D}^+ + {\rm H}_2 + \Delta {\rm E}
\label{eq.h2dp}
\end{equation}
\begin{equation}
{\rm H}_3^+ + {\rm N}_2 \rightarrow {\rm N}_2{\rm H}^+ + {\rm H}_2
\label{eq.n2hp}
\end{equation}
\begin{equation}
{\rm H}_3^+ + {\rm CO} \rightarrow {\rm HCO}^+ + {\rm H}_2
\label{eq.co1}
\end{equation}
\begin{equation}
{\rm N}_2{\rm H}^+ + {\rm CO} \rightarrow {\rm HCO}^+ + {\rm N}_2
\label{eq.co2}
\end{equation}

As soon as CO returns to the gas-phase (T $>$ 20-30~K, i.e., after the formation of a protostar), the abundance of both H$_{2}$D$^{+}$ and N$_{2}$H$^{+}$ will subsequently decrease. Note, however, that N$_{2}$H$^{+}$ survives with an abundance independent of the distance from the center \citep{Pagani2012,Lique2015} and appears to remain longer \citep{Tafalla2006}.

Given the scarcity of available molecular tracers originating from prestellar cores, previous studies have naturally focused on H$_{2}$D$^{+}$ \citep{Caselli2002,vanderTak2005,Vastel2006,Harju2006,Caselli2008} and N$_{2}$H$^{+}$ \citep{Tafalla2004,Pagani2007,Lique2015}, using high resolution ground-based sub-mm observatories (14$\arcsec$ - 22$\arcsec$; JCMT, CSO). Those studies provided important constraints on the temperature structure of such cores and the column densities and abundances of molecular species at individual cores. Since then, the collisional data for both H$_{2}$D$^{+}$ \citep{Hugo2009} and N$_{2}$H$^{+}$ \citep{Lique2015} have been revised and therefore the reported properties of those clouds also need to be revised. 

In the past decade, there have been several attempts to trace substructure via the dust continuum and molecular line emission within the central 1000 au of prestellar cores using interferometers, but mostly without positive results. There have been no fruitful detections of dust continuum at mm wavelengths towards single objects or towards a sample of cores in Perseus or Ophiuchus in the past \citep[CARMA, IRAM, SMA;][]{Schnee2012,Schnee2010,Crapsi2007}, explained by the shallow density profiles of prestellar cores at $<$ few thousand au. Due to the spatial capabilities and high sensitivity of the Atacama Large Millimetre and sub-millimetre Array (ALMA), a few studies have been able to detect some compact emission but without resolving the substructure \citep{Friesen2014,Kirk2017,Friesen2018}. Only very recently have the inner regions of a single prestellar core been resolved for the first time \citep[L1544 with ALMA;][]{Caselli2019}. Given the very challenging nature of the continuum observations at those high angular resolutions (50-100 au), it is no surprise that H$_{2}$D$^{+}$ detection with ALMA is limited to a single low-mass prestellar core \citep[SM1N;][]{Friesen2014}. Therefore, the analysis and modelling of single-dish observations of dust and gas towards prestellar cores is still a powerful approach to probe these enigmatic stages of star formation. 

In this paper, we present the first dedicated study to test the proposed dynamical models using advanced non-LTE radiative transfer modelling to simulate the gas emission, making use of the optically thin H$_{2}$D$^{+}$ 1$_{10}$-1$_{11}$ and N$_{2}$H$^{+}$ 4-3 line emission mapped with the James Clerk Maxwell Telescope (JCMT) towards a sample of prestellar cores (4) and one protostellar core. In Sect.~\ref{observ}, we describe the target selection, observations and data reduction. In Sect.~\ref{obs_res}, we present the observed spatial distributions and line analyses of H$_{2}$D$^{+}$ and N$_{2}$H$^{+}$ towards all cores and we also investigate the thermal and non-thermal contributions to the observed line width. In Sect.~\ref{col_densi}, we present the average column densities of H$_{2}$D$^{+}$ and N$_{2}$H$^{+}$ as determined using a non-LTE radiative transfer code to fit the line emission at the central sub-mm peak position towards all cores. In Sect.~\ref{mod_ratran}, we use a detailed radiative transfer model to simulate the 2$\arcmin$ $\times$ 2$\arcmin$ maps of the H$_{2}$D$^{+}$ and N$_{2}$H$^{+}$ line emission towards the cores, adopting three different dynamical models: Quasi-Equilibrium Bonnor-Ebert Sphere (QE-BES), Singular Isothermal Sphere (SIS), and Larston-Penston (LP) flow, and a static sphere. Our main results and conclusions are presented in Sect.~\ref{conclusions}.

\section{Observations}

\label{observ}

\subsection{Targets}

In this study, we present the analysis of five cores. Their coordinates, distances and continuum brightnesses at 850~$\mu$m and the core masses (4-10~\msun) based on dust and gas mm observations are presented in Table~\ref{sources}. These cores have been previously observed by \citet{Crapsi2005} in lines of N$_{2}$H$^{+}$ and N$_{2}$D$^{+}$ to explore the use of deuterium enrichment to constrain the evolutionary status of starless cores. The five cores of the present sample were specifically selected from the 31 cores presented in \citet{Crapsi2005} to provide a sample whereby detection of H$_{2}$D$^{+}$ could be considered favorable, typically by a high ratio of N(N$_{2}$D$^{+}$/N$_{2}$H$^{+}$) \citep[see Table 7;][]{Crapsi2005}. The present sample consists of four cores that are considered to be more evolved based on a variety of chemical and kinematical probes (L1544, L1521f, L694-2, L183) and one less evolved core (L1517B).  

\begin{table}[h]
\caption{Source sample: positions, continuum brightness, and distance}
\begin{tabular}{@{}*{10}{l}}
\hline
Name &RA (J2000) & Dec. (J2000) & S$_{850}$ & d & Mass \\
& hms& dms & Jy & pc & \msun\\
\hline
L1521f & 04:28:39.3 & +26:51:33 & 0.639 & 140 & 6$^{a}$ \\
L694-2 & 19:41:04.5 & +10:57:02 & 0.397 & 250 & 7$^{b}$ \\
L183 & 15:54:08.6 & -02:52:45 & 0.480 & 110 & 10$^{c}$ \\
L1544 & 05:04:17.2 & +25:10:44 & 0.470 & 140 & 8$^{d}$\\
L1517B & 04:55:18.3 & +30:37:48 & 0.261 & 140 & 4$^{e}$\\
\hline

\end{tabular}

\tiny {\bf{Notes}}: The coordinates for all cores are the positions of the respective peak 850~$\mu$m emission found in the maps available in the SCUBA Legacy Catalogue data \citep[see][]{DiFrancesco2008}. \\
$^{a}$ \citet{Crapsi2004}, $^{b}$ \citet{DiFrancesco2008}, $^{c}$ \citet{Crutcher2004}, $^{d}$ \citet{Tafalla1998}, $^{e}$ \citet{Fu2011}
\label{sources}
\end{table}

Of the five cores in Table~\ref{sources}, four contain no known young stellar objects and, therefore, can be considered starless. Only towards one core, L1521f, a VEry Low Luminosity Object (VELLO) was identified by \citet{Bourke2006}, and therefore it is considered to be protostellar.

\subsection{Submillimetre single-dish observations}

The five cores in this study were observed from January 2007 to January 2008 as part of three semesters of queue-mode observing at the James Clerk Maxwell Telescope\footnote{At the time of the described observations, the James Clerk Maxwell Telescope was operated by the Joint Astronomy Centre on behalf of the Science and Technology Facilities Council of the United Kingdom, the National Research Council of Canada and the Netherlands Organisation for Scientific Research.} (JCMT) on Mauna Kea, Hawaii, under observing programs M06BC11, M07AC15, and M07BC06. All observations were made under very dry weather conditions, i.e., $\tau_{225}$ $<$ 0.05, to ensure maximum sensitivity to the ortho-H$_{2}$D$^{+}$ 1$_{10}$-1$_{11}$ line and the N$_{2}$H$^{+}$ 4--3 line \citep[at 372.42 GHz and 372.67 GHz respectively;][]{Pickett1998}, which are adjacent to a broad atmospheric H$_{2}$O feature. JCMT has a beamwidth of 14$\arcsec$ at the relevant frequency band (350 GHz).

The observations were made using the 16-element Heterodyne Array Receiver Programme (HARP) focal-plane array that operates over 325-375 GHz and the AutoCorrelation Spectrometer Imaging System (ACSIS) back-end \citep{Buckle2009}. HARP was tuned to 372.5 GHz, and ACSIS configured to provide nominally 500 MHz total bandwidth with 61 kHz wide channels or 0.048~km~s$^{-1}$ spacing. This setup allowed both the H$_{2}$D$^{+}$ 1$_{10}$-1$_{11}$ and N$_{2}$H$^{+}$ 4--3 lines to be observed simultaneously in the same spectral window. The velocity resolution of ACSIS data is a factor of $\sim$1.2 less than the configured channel spacing, or in this case, 0.058~km~s$^{-1}$. 

The observations were defined in minimum schedulable blocks (MSBs) and were preceded by standard focus calibration observations at 345 GHz on nearby bright objects like R Aql. Also, pointing calibrations at 345 GHz were conducted on objects such as CRL618, both before the program MSBs and every $\sim$ 60-90 minutes. Flux calibration was monitored by observing when possible spectral line standards, like W75N, at a variety of frequencies. The aperture efficiency, $\eta_{\alpha}$, of HARP at 372.5 GHz is $\sim$0.5.

HARP consists of 16 receptors arranged in a 4$\times$4 square pattern with an on-sky spacing between receptors of 30$\arcsec$. Program MSBs were executed by pointing one of the inner four HARP receptors at the target positions of the cores (Table~\ref{sources}). Data were obtained by pointing HARP at the target positions, and integrating using a ``stare-mode'' fashion, through chopping between the target positions and positions 180$\arcsec$ distant in azimuth at 7.8125 Hz. Each MSB consisted of five repeats of 300~s. Each target was observed for $\sim$4 hours in total. When H$_{2}$D$^{+}$ 1$_{10}$-1$_{11}$ was detected at the target position by at levels $\geq$ $\sim$ 5 $\sigma$, the telescope was shifted diagonally in position by $\sim$22.5$\arcsec$, i.e., the target position was centered in the array between the center four receptors, and further staring observations were to be executed for up to another $\sim$ 4 hours. This strategy allowed the acquisition of high sensitivity data of faint lines in a ``checkerboard'' pattern across each core at a spatial sampling better than given by a single HARP footprint. Observing each core with HARP in a jiggle pattern would have resulted in better spatial sampling but at the cost of sensitivity in each source. Since HARP is aligned on the JCMT in azimuth and elevation, the final data per pointing sometimes include samples of more than 16 positions on the sky due to field rotation. Also, the total integration duration varies from core to core, given the nature of queue-mode observing and the scarcity of very dry weather conditions.

\subsection{Data reduction}

The data were reduced using standard routines and procedures in the STARLINK reduction package\footnote{The Starlink software is currently supported by the East Asian Observatory.} \citep{Currie2014}. Each integration was visually checked for baseline ripples, absent detectors, or large spikes, and specific integrations where these occurred were removed from the data ensemble. In particular, receptor H03 was unavailable during most of the observations; at various times, no more than two other receptors were also unavailable. Less strong spikes were identified and removed using a standard methodology created by developers at the Joint Astronomy Centre. Spectral baselines were subtracted, frequency axes converted to velocities and spectra trimmed using STARLINK scripts. To increase the signal-to-noise ratio (SNR), we re-binned the velocity axes of all the data by applying a factor of 2, resulting in a spectral resolution of $\sim$0.12~km~s$^{-1}$. 

\section{Observational results}

\label{obs_res}

\subsection{Spatial distribution of H$_{2}$D$^{+}$ and N$_{2}$H$^{+}$}

\begin{figure*}[]
\begin{center}$
\begin{array}{cc} 
\includegraphics[scale=0.4]{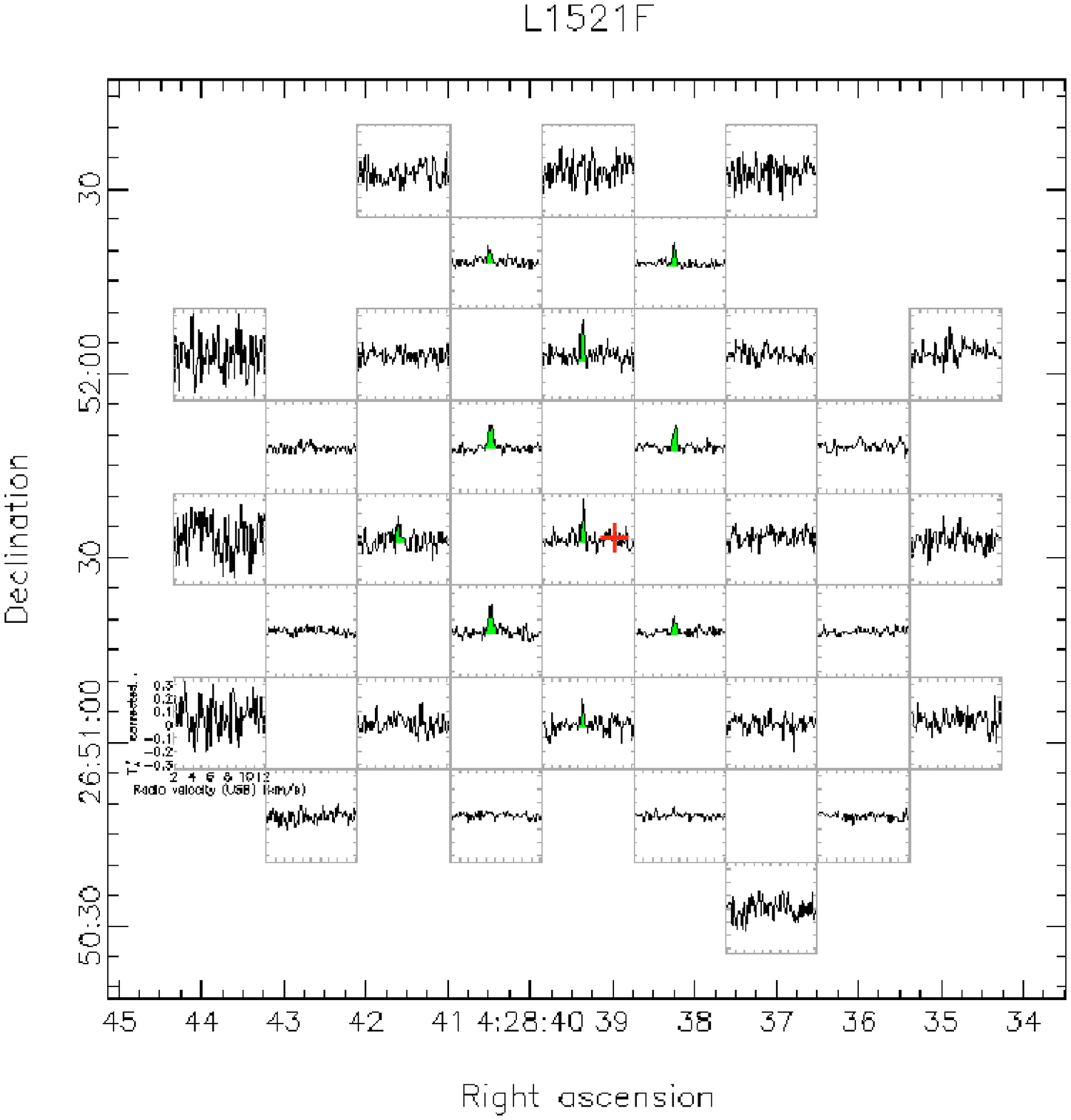} & \includegraphics[scale=0.4]{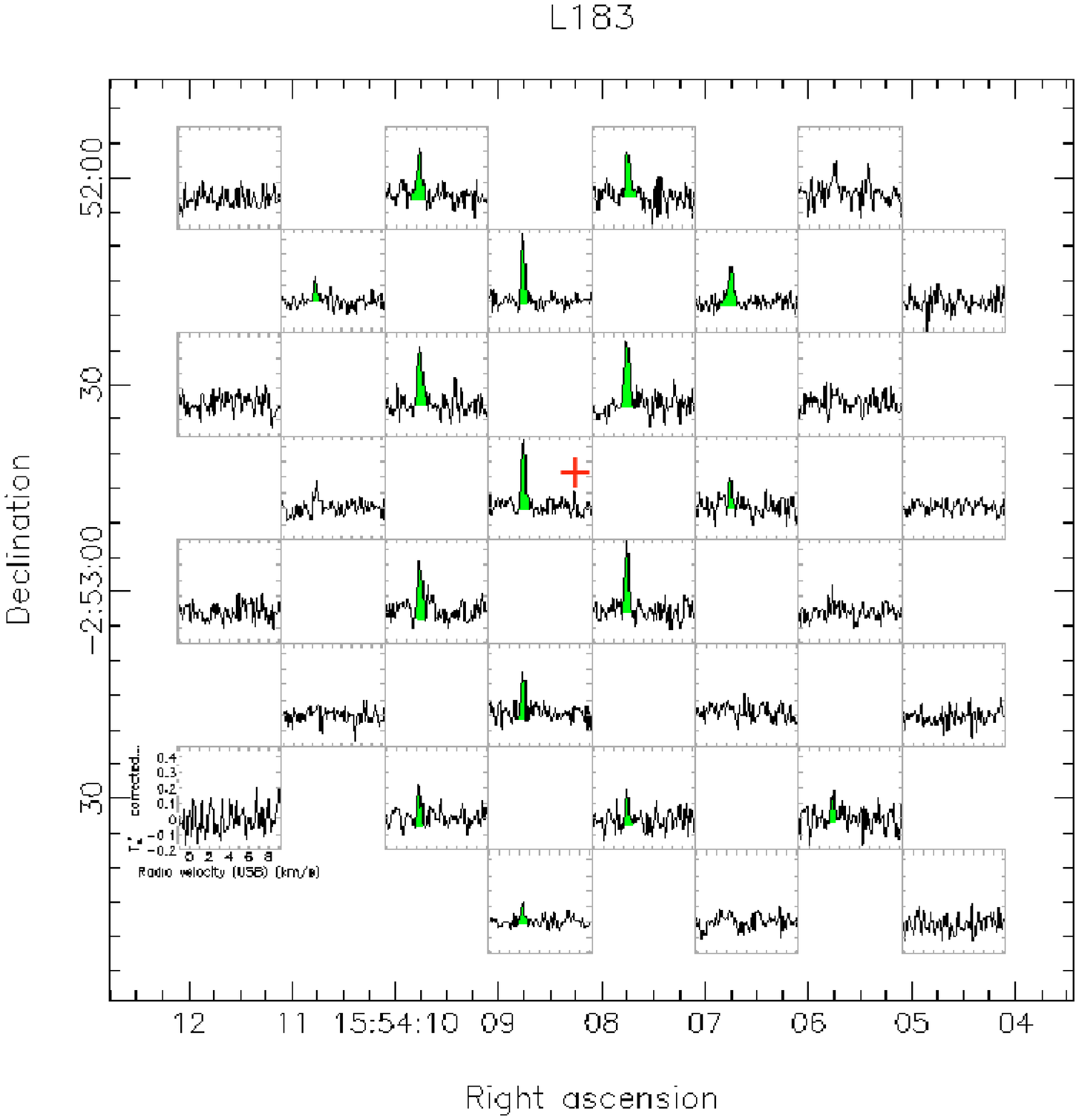} \\
\includegraphics[scale=0.4]{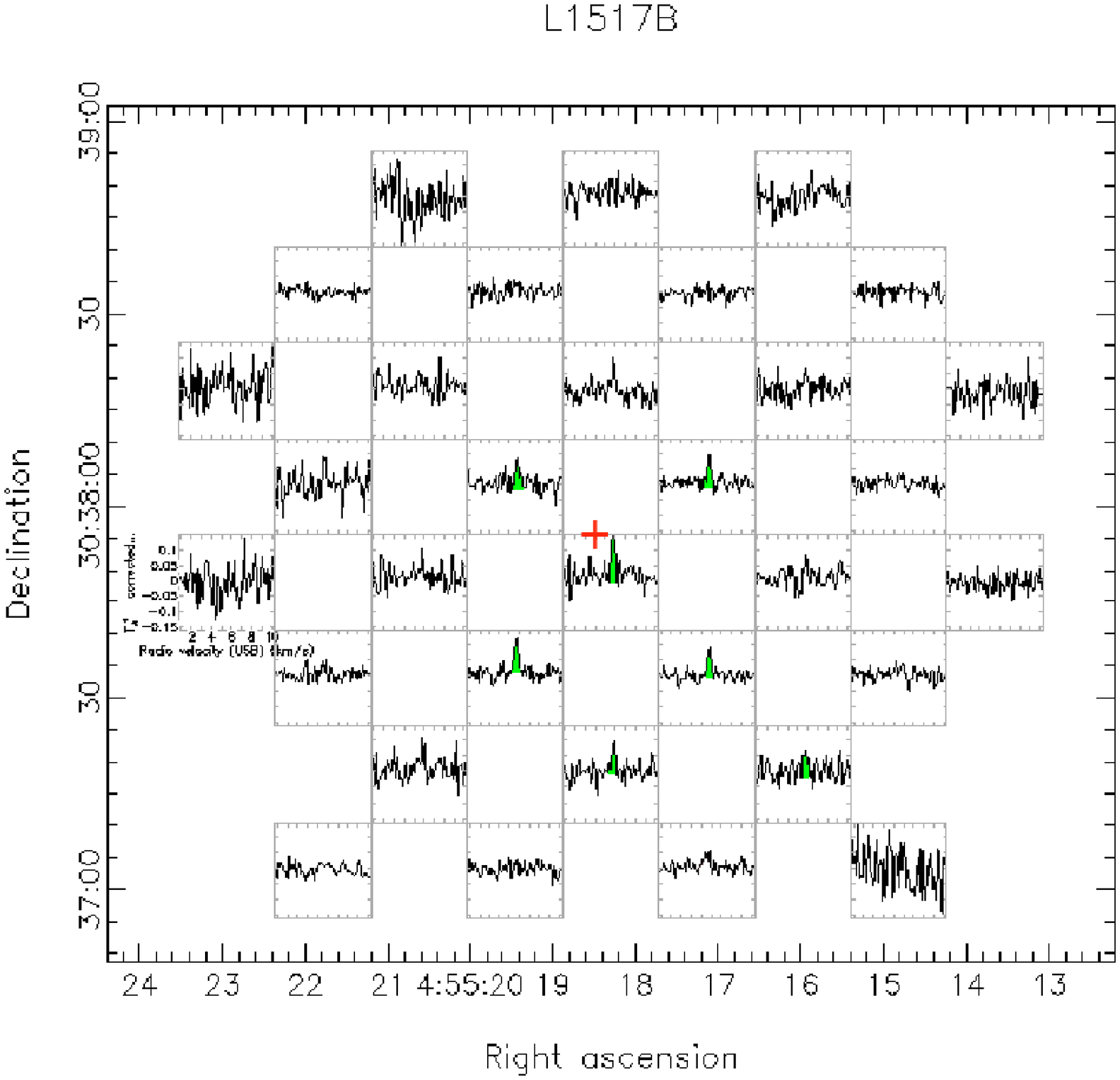} & \includegraphics[scale=0.4]{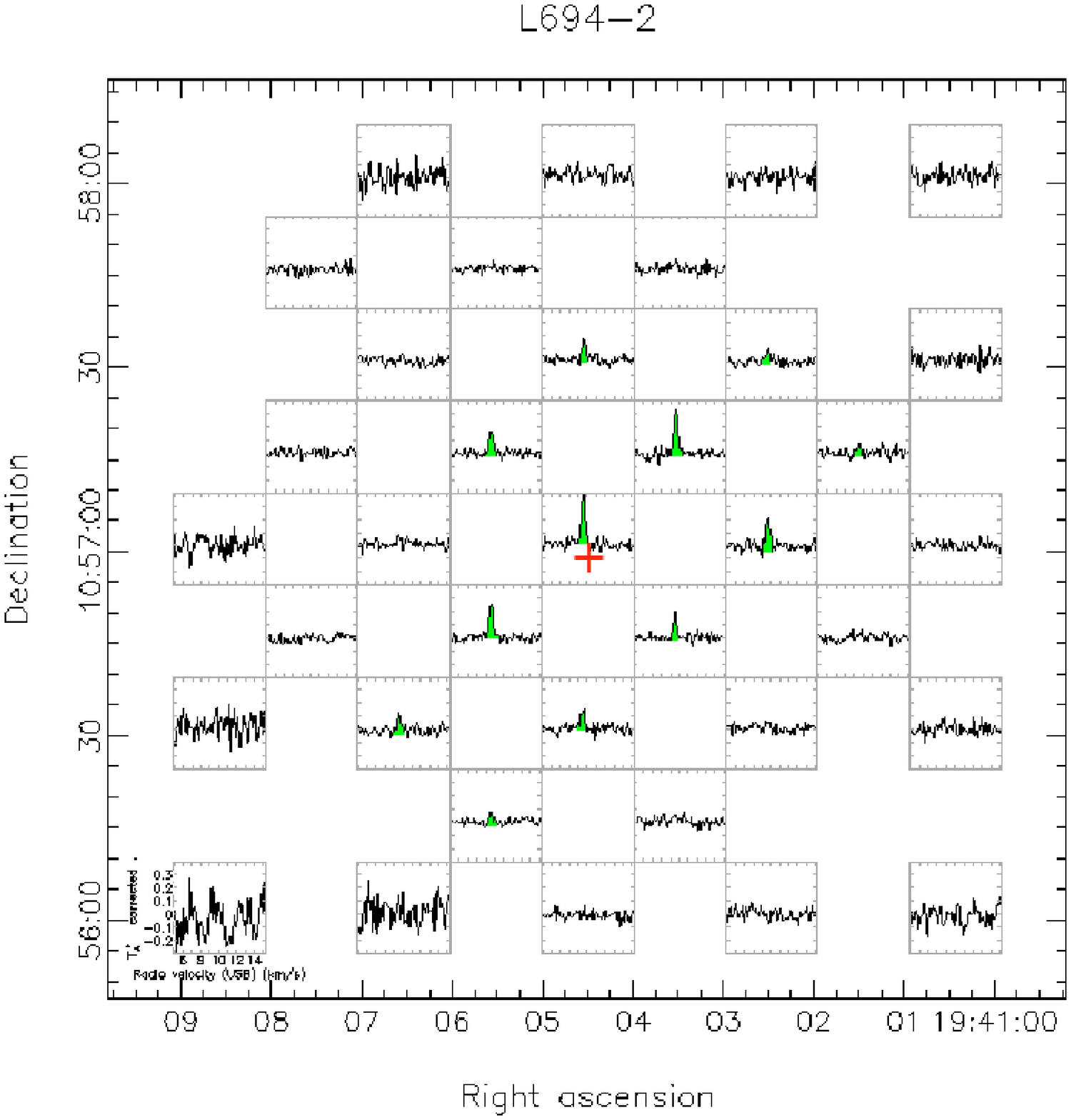} 
\end{array}$  
\end{center}
\begin{center}
\includegraphics[width=.55\textwidth]{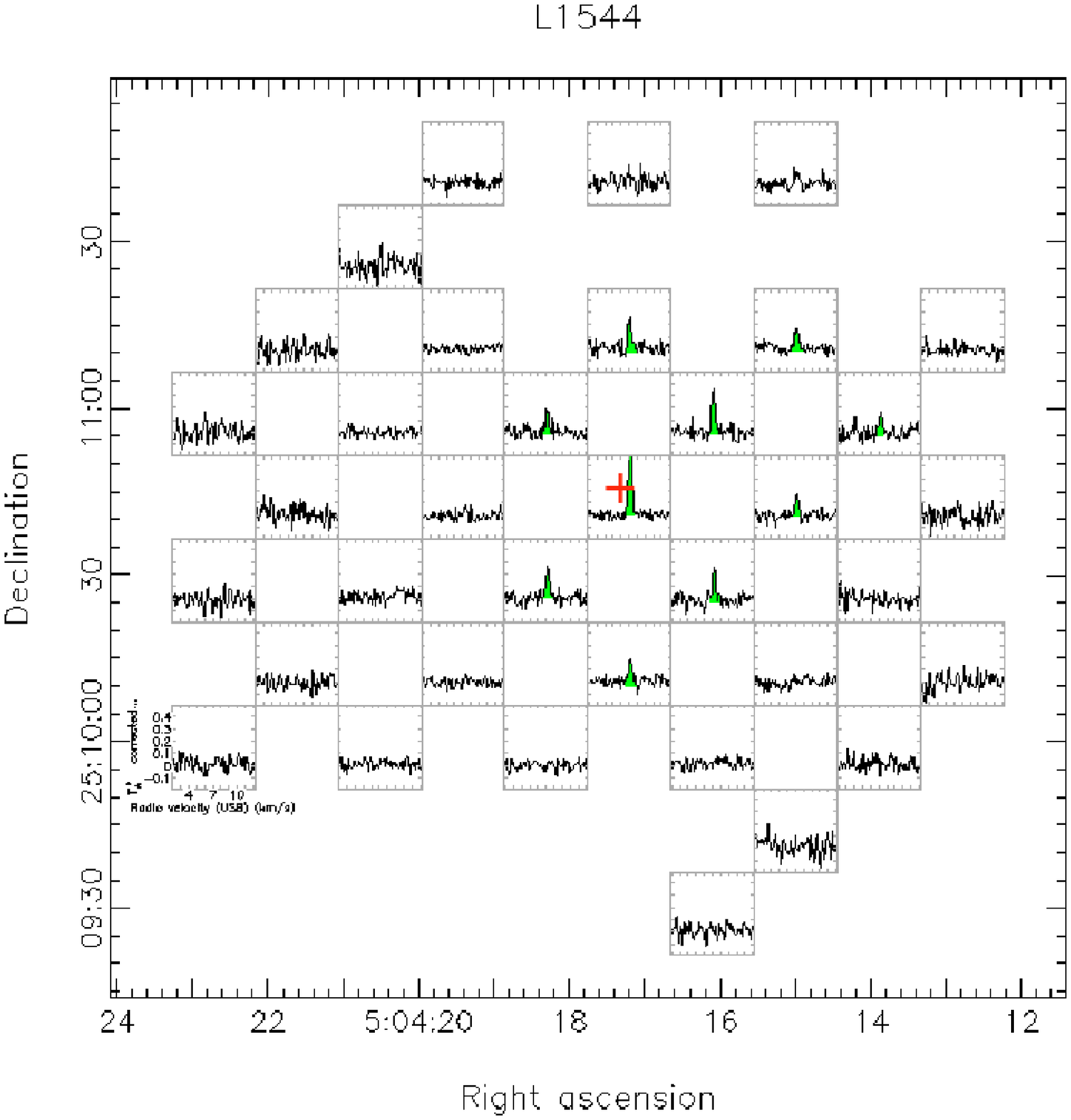} 
\end{center}  
\caption{Spatial distribution of H$_{2}$D$^{+}$ line emission as observed with JCMT towards the cores L1521f, L183, L1517B, L694-2, and L1544. The red cross indicates the peak of the 850~$\mu$m emission. The area filled with light green indicates the clear ($>$3~$\sigma$) detections and the possible detections ($\sim$2~$\sigma$) based on the line shape and peak velocity position.}
\label{fig.clinplot_h2}
\end{figure*}

\begin{figure*}[ht]
\begin{center}$
\begin{array}{ll} 
\includegraphics[scale=0.4]{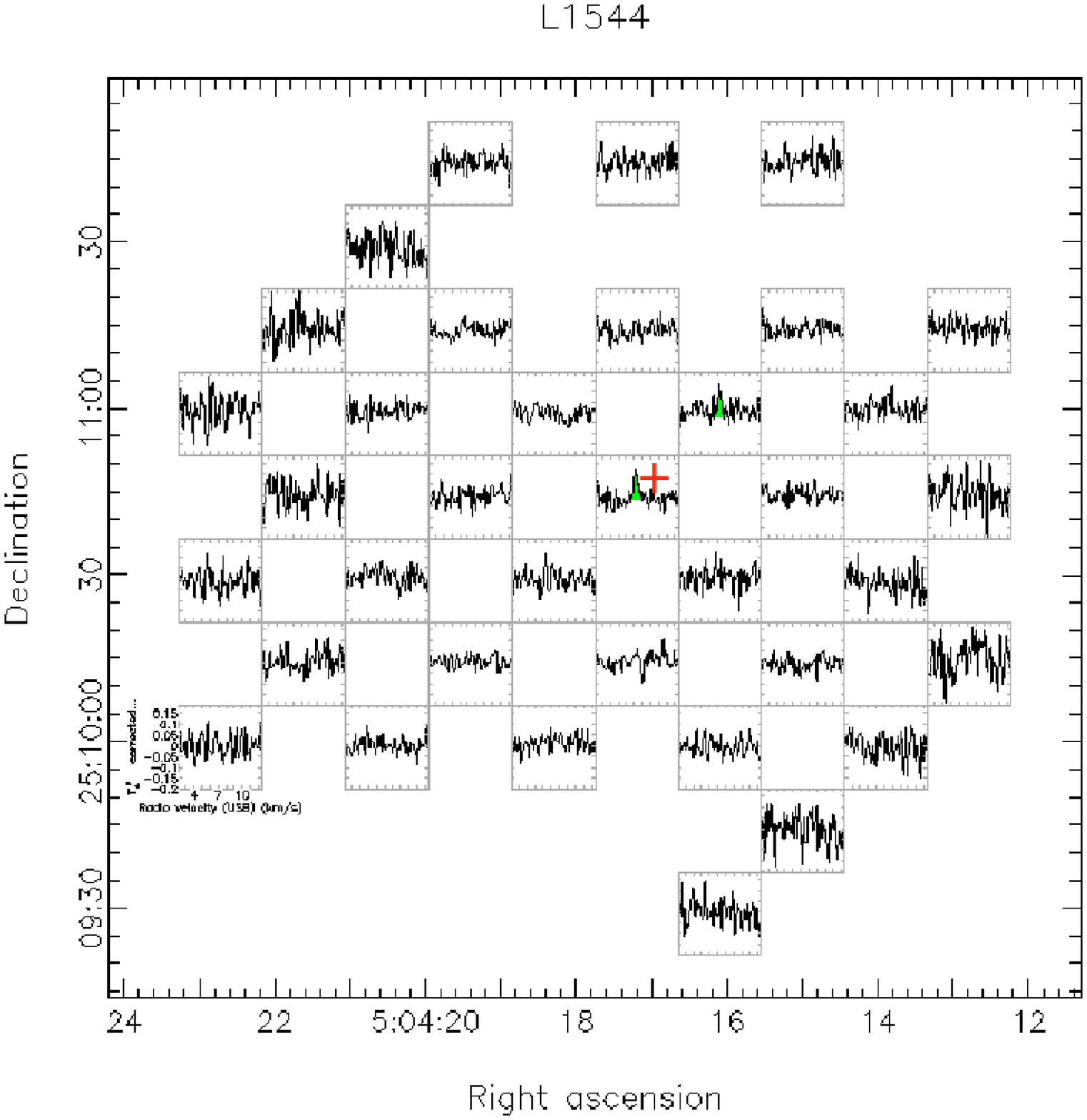} & \includegraphics[scale=0.4]{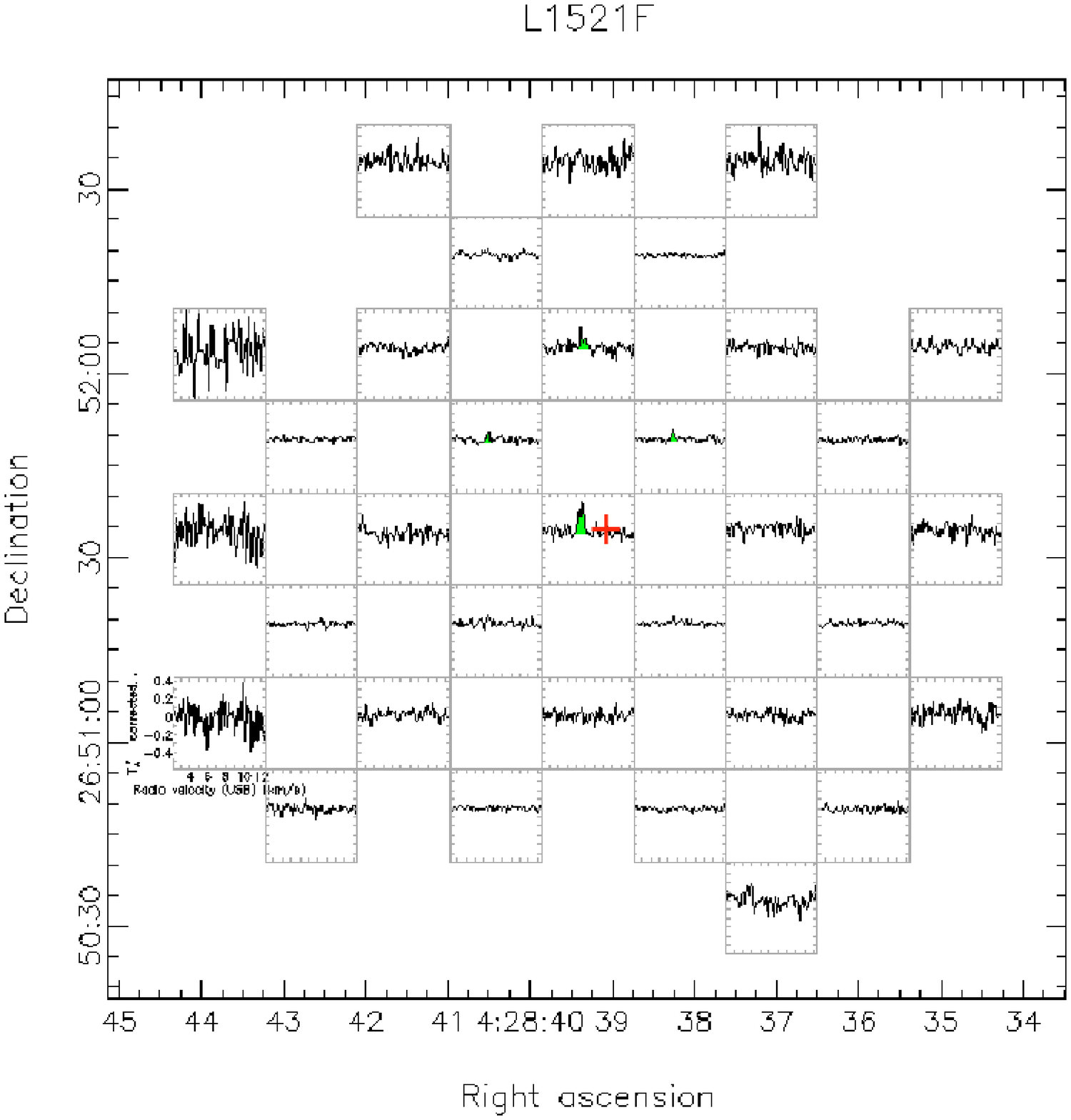} 
\end{array}$  
\end{center}
\caption{Spatial distribution of N$_{2}$H$^{+}$ 4-3 as observed with JCMT towards L1544 and L1521f. These two are the only cores with N$_{2}$H$^{+}$ 4-3 line emission detections. The area filled with light green indicates the clear ($>$3~$\sigma$) detections and the possible detections ($\sim$2~$\sigma$) based on the line shape and peak velocity position.}
\label{fig.clinplot_n2}
\end{figure*}

The observed spatial distributions of H$_{2}$D$^{+}$ 1$_{10}$-1$_{11}$ emission toward all cores are presented in Figure~\ref{fig.clinplot_h2}. H$_{2}$D$^{+}$ in emission is clearly seen towards each core of the sample. In all cases, the peak line brightness is found at the position of the 850~$\mu$m peak continuum emission (i.e.,, within a $\sim$ 14$\arcsec$ beam width). The line in emission is also clearly detected at several positions offset from the central position towards all cores. The observations of N$_{2}$H$^{+}$ 4-3 from the cores where it was clearly detected (L1521f and L1544) are presented in Figure~\ref{fig.clinplot_n2}. The detection of this emission is spatially limited to two peaks towards L1521f and possibly L1544, and in particular, the second peak is at a location $\sim$ 15$\arcsec$ offset from the local peak of the submillimeter continuum emission.

\subsection{Line profiles}

\label{line_anal}

We fit single Gaussians to the H$_{2}$D$^{+}$ 1$_{10}$-1$_{11}$ and N$_{2}$H$^{+}$ 4-3\footnote{There is no detectable hyperfine structure of N$_{2}$H$^{+}$ 4-3, therefore it can be fitted with a single Gaussian (see Splatalogue: https://www.cv.nrao.edu/php/splat/advanced.php).} lines observed for each core and at each position where the emission is detected (T$^{*}_{A}$ $\geq$ 3$\times$rms). Table~\ref{table.dataa} lists the peak brightness (T$^{*}_{A}$), central line velocity (V$_{LSR}$), and the line width (FWHM) with the associated positions expressed in offsets from the local peak of the submillimeter 850~$\mu$m continuum emission. 

\begin{figure}[h]
\begin{center}
\includegraphics[width=.4\textwidth]{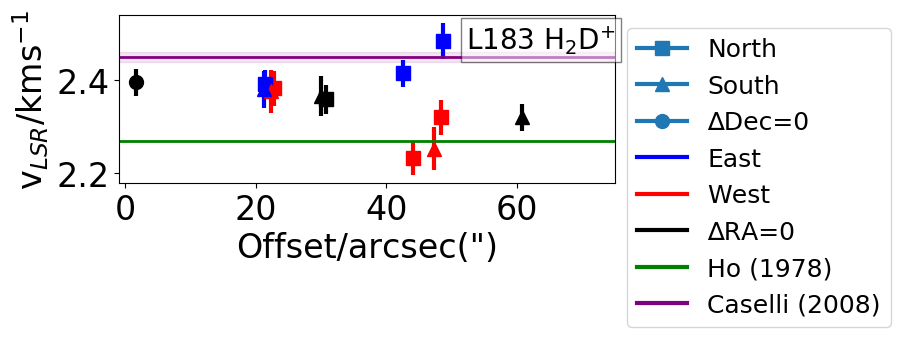}
\includegraphics[width=.4\textwidth]{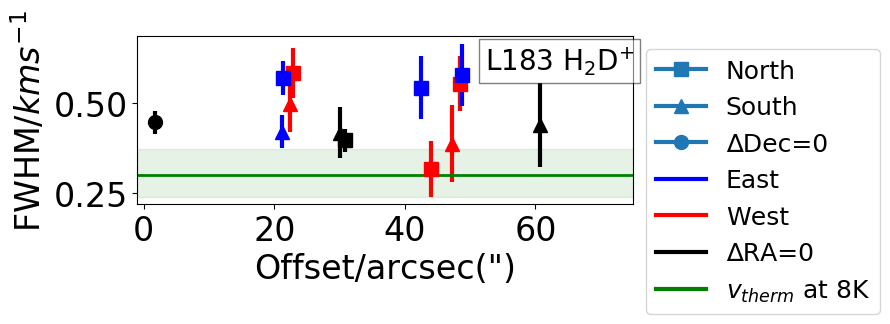}
\includegraphics[width=.4\textwidth]{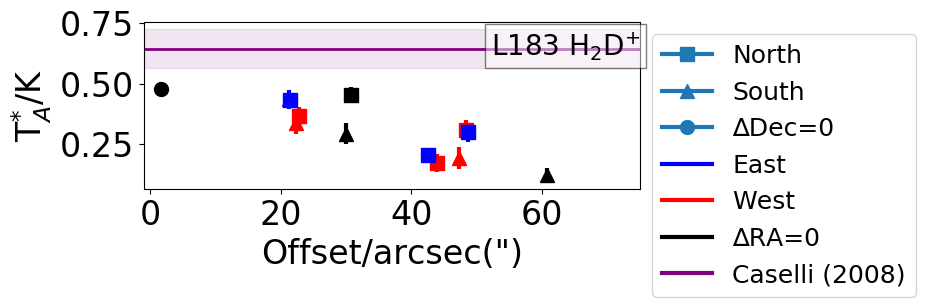}
\caption{$V_{LSR}$, FWHM, and $T_A^*$ measurements of H$_{2}$D$^{+}$ plotted over offset from the 850~$\mu$m dust peak of L183. The shaded region in the middle plot corresponds to the thermal broadening within the 5-12~K temperature range. The shaded regions in purple correspond to the errors reported in \citet{Caselli2008}. The square, triangle and circle refer to North, South, and 0" declination offsets respectively. The blue and red symbols refer to offsets towards the East and West respectively, while the black symbols correspond to 0" Right Ascension offsets. For example, a ~40" offset can be represented by a red square (40" North-West), a black triangle (40" South), a blue circle (40" East), and so on.}
\label{fig.plot_L183}
\end{center}
\end{figure}

\begin{figure}[h]
\begin{center}
\includegraphics[width=.4\textwidth]{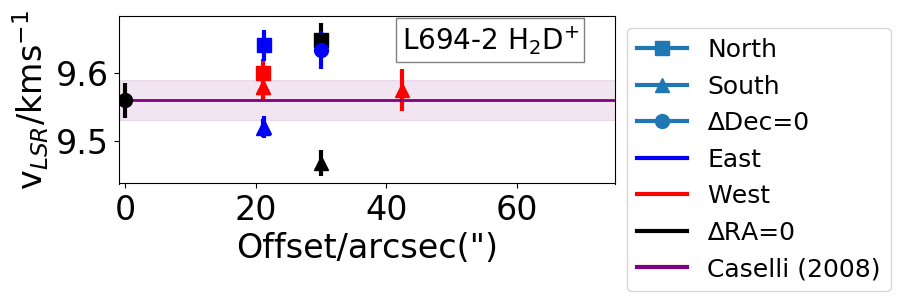}
\includegraphics[width=.4\textwidth]{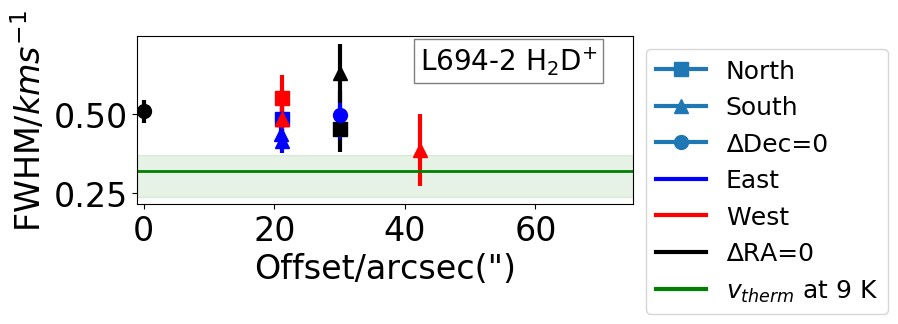}
\includegraphics[width=.4\textwidth]{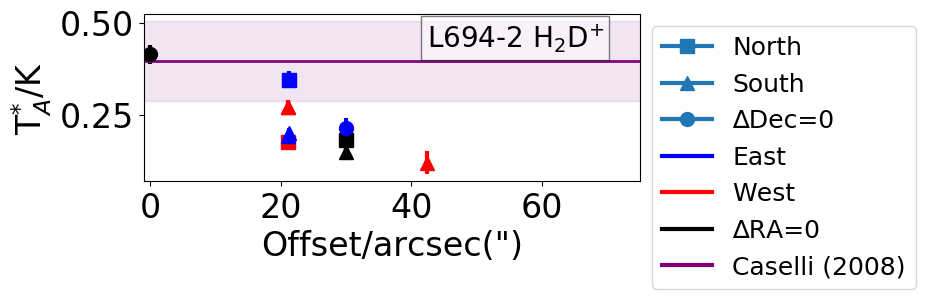}
\caption{As Fig.~\ref{fig.plot_L183}, but for L694-2.}
\label{fig.plot_L694-2}
\end{center}
\end{figure}

\begin{figure}[h]
\begin{center}
\includegraphics[width=.4\textwidth]{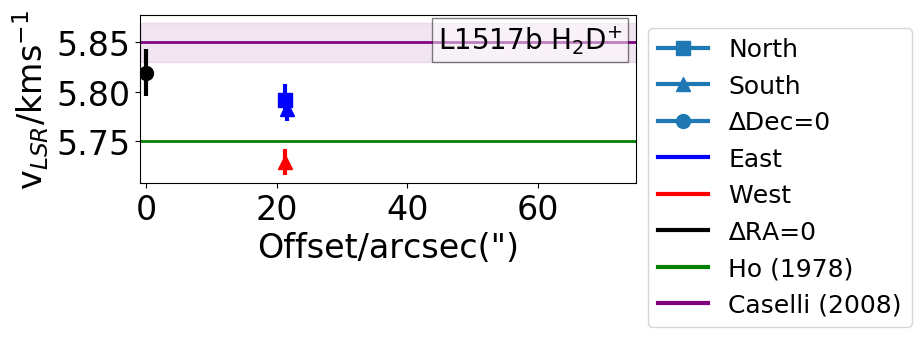}
\includegraphics[width=.4\textwidth]{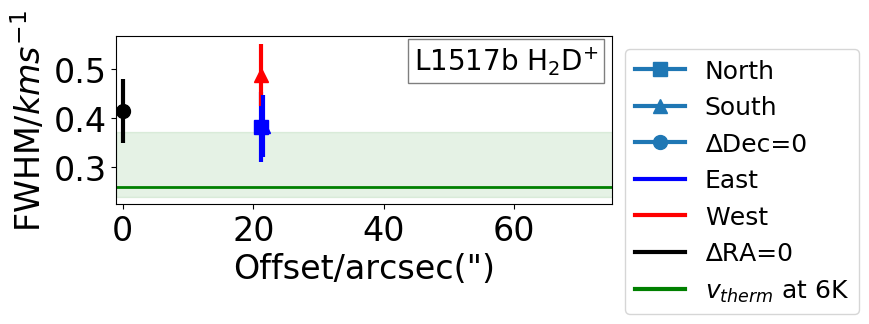}
\includegraphics[width=.4\textwidth]{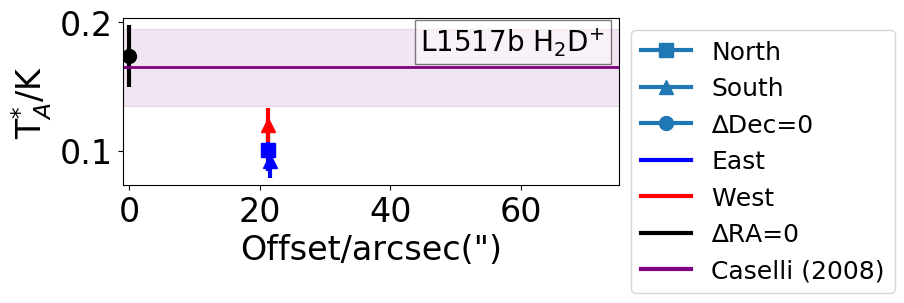}
\caption{As Fig.~\ref{fig.plot_L183}, but for L1517B.}
\label{fig.plot_L1517B}
\end{center}
\end{figure}

\begin{figure}[h]
\begin{center}
\includegraphics[width=.4\textwidth]{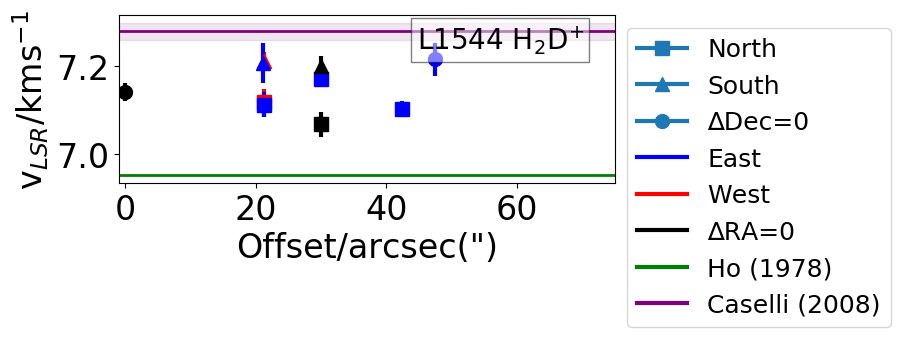}
\includegraphics[width=.4\textwidth]{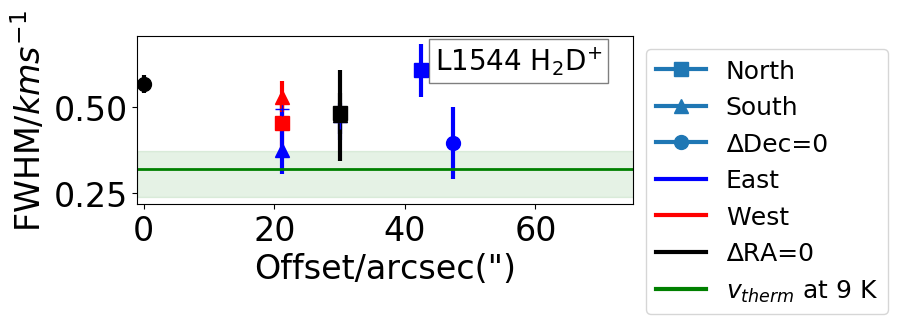}
\includegraphics[width=.4\textwidth]{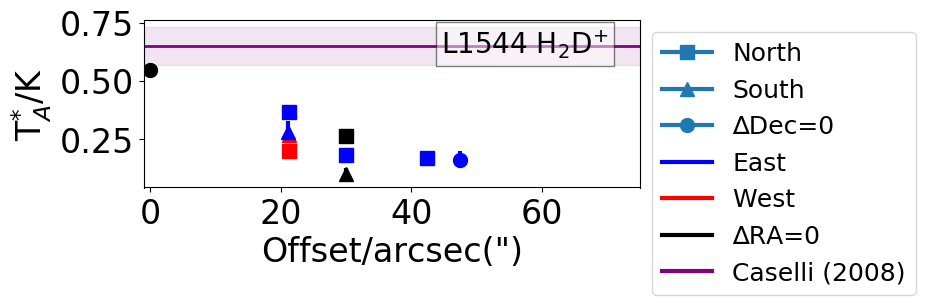}
\caption{As Fig.~\ref{fig.plot_L183}, but for L1544.}
\label{fig.plot_L1544}
\end{center}
\end{figure}

\begin{figure}[h]
\begin{center}
\includegraphics[width=.4\textwidth]{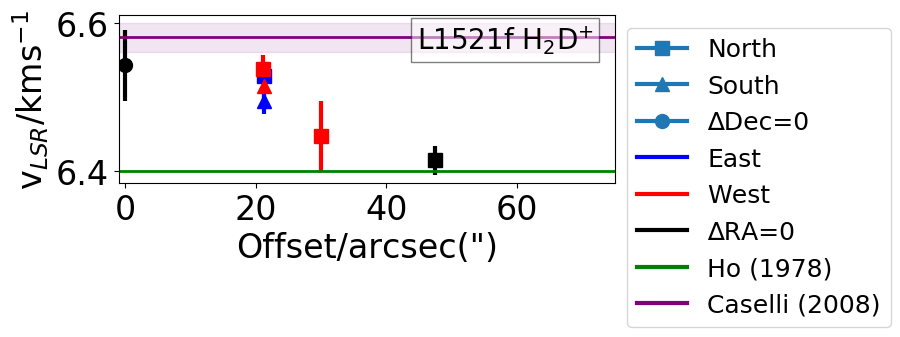}
\includegraphics[width=.4\textwidth]{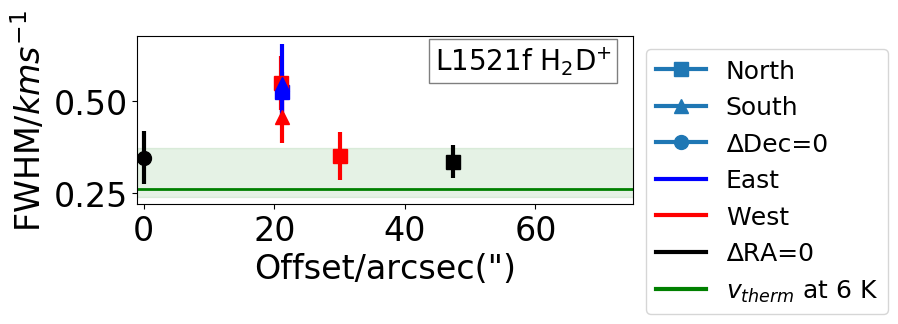}
\includegraphics[width=.4\textwidth]{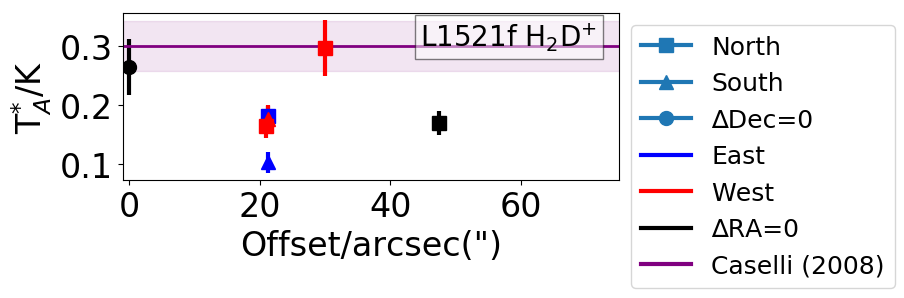}
\caption{As Fig.~\ref{fig.plot_L183}, but for L1521f.}
\label{fig.plot_L1521f}
\end{center}
\end{figure}

Figures~\ref{fig.plot_L183}--\ref{fig.plot_L1521f} show the derived line properties, V$_{LSR}$, FHWM, and T$_{A}^{*}$ of H$_{2}$D$^{+}$ and N$_{2}$H$^{+}$ emission with respect to their offsets from the 850~$\mu$m dust peak. Firstly, we plot the $V_{LSR}$ versus offsets and compare these to the known source velocities taken from the literature \citep[e.g.,][]{Caselli2008}. The upper plots in Figures \ref{fig.plot_L183}--\ref{fig.plot_L1521f} show that the measured $V_{LSR}$ at the central position is in very good agreement and mostly within the errors when compared to the values reported by \citet{Caselli2008} towards most of the cores. \citet{Caselli2008} observed H$_{2}$D$^{+}$ with CSO and got measurements of the emission at the central position of those cores at a similar spectral resolution to the current study but a lower angular resolution (22\arcsec compared to our 14\arcsec). In contrast, the source core velocity of L1544 reported by \citet{Ho1978} using NH$_{3}$ measurements is systematically lower than those measured in our study and in \citet{Caselli2008} for the central positions. Instead, the measured velocities reported in \citet{Ho1978} come to a better agreement with those we find at offsets further out in the cloud. We should note that the angular resolution in the 1978 study was larger by a factor of $>$ 4. Taking all the above into consideration, we conclude that our measurements are consistent with previous work.



\begin{center}
\begin{table*}[ht]
\scriptsize
\caption{Line measurements at all offsets with clear detections for each core - JCMT data}
\centering
\begin{tabular}{@{}*{11}{l}}
\hline
Core& R.A. Offset&Dec Offset &Species&T$_{A}^{*}$&Error&V$_{LSR}$&Error&FWHM&Error&rms\\
\hline
&(\arcsec)&(\arcsec)&&(K)&&(~km~s$^{-1}$)&&(~km~s$^{-1}$)&& (K)\\
\hline
L1544&0&0&H$_{2}$D$^{+}$&0.55&0.02&7.14&0.01&0.57&0.03&0.05\\ 
&0&30&&0.26&0.03&7.07&0.02&0.48&0.06&0.06\\
&-30&30&&0.17&0.02&7.10&0.03&0.61&0.08&0.05\\       
&15&15&&0.20&0.03&7.12&0.04&0.45&0.08&0.06\\ 
&-15&15&&0.37&0.03&7.11&0.02&0.49&0.05&0.06\\  
&-45&15&&0.16&0.04&7.21&0.04&0.4&0.1&0.07\\
&-30&0&&0.18&0.02&7.17&0.02&0.46&0.05&0.04\\ 
&15&-15&&0.27&0.02&7.21&0.02&0.53&0.05&0.05\\ 
&-15&-15&&0.28&0.05&7.21&0.03&0.38&0.07&0.05\\
&0&-30&&0.10&0.02&7.20&0.06&0.5&0.1&0.05\\ 
&0&0&N$_{2}$H$^{+}$&0.11&0.02&7.20&0.04&0.4&0.1&0.03\\ 
&-15&15&&0.12&0.03&7.26&0.05&0.4&0.1&0.07\\ 
\hline
L183&15&45&H$_{2}$D$^{+}$&0.31&0.04&2.32&0.03&0.56&0.08&0.09\\  
&-15&45&&0.30&0.04&2.48&0.04&0.58&0.09&0.09\\ 
&30&30&&0.17&0.04&2.23&0.03&0.32&0.08&0.06\\   
&0&30&&0.45&0.03&2.36&0.01&0.3&0.2&0.06\\                    
&-30&30&&0.21&0.03&2.41&0.04&0.1&0.1&0.07\\ 
&15&15&&0.37&0.04&2.38&0.03&0.58&0.07&0.09\\    
&-15&15&&0.43&0.03&2.39&0.02&0.57&0.05&0.07\\           
&0&0&&0.48&0.03&2.40&0.01&0.45&0.03&0.06\\  
&15&-15&&0.34&0.05&2.38&0.03&0.50&0.08&0.1\\
&-15&-15&&0.43&0.04&2.38&0.02&0.42&0.05&0.05\\ 
&0&-30&&0.29&0.04&2.36&0.03&0.42&0.07&0.09\\  
&15&-45&&0.19&0.05&2.25&0.05&0.4&0.1&0.09\\                 
&0&-60&&0.12&0.03&2.32&0.05&0.4&0.1&0.06\\
&-15&-45&&0.19&0.06&2.40&0.04&0.3&0.1&0.07\\
\hline                             
L694-2&0&30&H$_{2}$D$^{+}$&0.18&0.02&9.65&0.03&0.45&0.07&0.05\\             
&15&15&&0.18&0.02&9.60&0.03&0.55&0.07&0.05\\                   
&-15&15&&0.35&0.02&9.64&0.02&0.48&0.04&0.05\\                 
&0&0&&0.41&0.02&9.56&0.02&0.51&0.04&0.06\\                   
&-30&0&&0.21&0.03&9.63&0.03&0.50&0.08&0.06\\                       
&15&-15&&0.27&0.02&9.58&0.02&0.48&0.04&0.04\\                       
&-15&-15&&0.20&0.02&9.52&0.02&0.41&0.04&0.03\\                      
&30&-30&&0.12&0.03&9.57&0.05&0.4&0.1&0.06\\                       
&0&-30&&0.15&0.02&9.47&0.04&0.63&0.09&0.05\\                       
\hline                                  
L1517B&-15&15&H$_{2}$D$^{+}$&0.10&0.02&5.79&0.03&0.38&0.07&0.03\\   
&0&0&&0.17&0.02&5.82&0.03&0.07&0.08&0.05\\           
&15&-15&&0.12&0.01&5.73&0.03&0.49&0.06&0.03\\          
&-15&-15&&0.10&0.01&5.77&0.02&0.36&0.06&0.03\\
&0&-30&&0.10&0.03&5.88&0.04&0.4&0.1&0.04\\
\hline                                 
L1521f&-15&45&H$_{2}$D$^{+}$&0.17&0.02&6.41&0.02&0.34&0.05&0.07\\   
&0&30&&0.30&0.05&6.45&0.03&0.35&0.07&0.09\\
&0&-30&&0.21&0.06&6.43&0.03&0.24&0.07&0.06\\
&15&15&&0.18&0.02&6.53&0.02&0.53&0.05&0.04\\
&15&45&&0.10&0.02&6.35&0.06&0.6&0.1&0.03\\
&-15&15&&0.16&0.02&6.54&0.03&0.55&0.07&0.03\\           
&0&0&&0.27&0.05&6.54&0.03&0.35&0.07&0.09\\             
&15&-15&&0.18&0.02&6.51&0.03&0.46&0.07&0.05\\           
&-15&-15&&0.10&0.02&6.50&0.05&0.6&0.1&0.04\\
&-45&15&&0.07&0.02&6.62&0.08&0.6&0.2&0.03\\
&-15&15&N$_{2}$H$^{+}$&0.11&0.02&6.47&0.03&0.36&0.08&0.04\\       
&0&0&&0.30&0.04&6.56&0.04&0.6&0.1&0.06\\
&0&30&&0.22&0.06&6.53&0.04&0.29&0.09&0.07\\
\hline
\end{tabular}

\tiny {\bf{Notes}}: To convert to main beam temperature, T$_{mb}$, one should divide the antenna temperature, T$_{A}^{*}$, with the beam efficiency $\eta= 0.6$ ($T_{mb}=\frac{T_A^*}{\eta}$).
\label{table.dataa}
\end{table*}
\end{center}

\vspace{-0.9cm}

The FWHMs of the lines are of interest since they can provide information on whether the broadening of the emission is a result of only a thermal contribution or if non-thermal motions are also present at those very early stages of star formation. The line profiles of all five cores are characterized by narrow emission (0.2-0.6~km~s$^{-1}$). The thermal broadening ($\Delta V_{THERM}$) is estimated using: 

\begin{equation}
\Delta V_{THERM}=\sqrt{8\textrm{ln}2\frac{kT}{m}}
\label{eq.therm}
\end{equation}

\hspace{-0.7cm} where T is the gas kinetic temperature we determine for each individual core (6-9~K; see RADEX analysis in Sect.~\ref{col_densi}, Table~\ref{table.radex}), and $m$ is the mass of the ion. 

$\Delta V_{THERM}$ was thus found to be within 0.26~km~s$^{-1}$-0.32~km~s$^{-1}$ for H$_{2}$D$^{+}$ and 0.09~km~s$^{-1}$-0.11~km~s$^{-1}$ for N$_{2}$H$^{+}$. By determining the $\Delta V_{THERM}$ and measuring the observed line width of the species at each position in the maps, we can quantify how much of the observed line broadening is due to non-thermal contributions \citep[$\Delta V_{NT}=\sqrt{\Delta V_{OBS}^2 - \Delta V_{THERM}^2}$;][]{Myers1991} for each source (i.e., turbulence and magnetic fields). To address the significance of non-thermal contributions, we approach the temperature errors such that the temperature is within the range 5-12~K (typical values for dense cores). 

Here we present our findings regarding the entire sample, while more details on each individual core can be found in Appendix~\ref{anal_core}. The middle plots in Figures \ref{fig.plot_L183}--\ref{fig.plot_L1521f} show that thermal broadening alone is insufficient to explain the observed line widths towards most of the offset positions towards all cores. In particular, there is an indication that the non-thermal contributions become less significant for both H$_{2}$D$^{+}$ and N$_{2}$H$^{+}$ as one moves outwards from the central position for three out of five cores. This finding contradicts the theory and what was previously observed towards molecular clouds, where the velocity dispersion increases with scale as the density decreases \citep[e.g.,][]{Maloney1988}. 

For the two cores for which both H$_{2}$D$^{+}$ and N$_{2}$H$^{+}$ measurements are available, we compare the non-thermal contribution at the central position. In particular, we find non-thermal contributions for H$_{2}$D$^{+}$ to be $0.47^{+0.05}_{-0.06}$~km~s$^{-1}$ and $0.23^{+0.15}_{-0.10}$~km~s$^{-1}$ for L1544 and L1521f, respectively, while for N$_{2}$H$^{+}$ we find these to be $0.42^{+0.10}_{-0.08}$~km~s$^{-1}$ and 0.6$\pm$0.1~km~s$^{-1}$, respectively. For L1544, the regions traced by the two species are characterized by similar non-thermal contributions (within the uncertainties) and, therefore, originate from regions with similar internal motions. For the more evolved, L1521f core, the non-thermal contributions are higher by a factor of 3 for the N$_{2}$H$^{+}$ line compared to the H$_{2}$D$^{+}$, which is beyond the uncertainties. This latter difference indicates that even though both lines are observed at a central position of the protostellar core, the H$_{2}$D$^{+}$ emission originates from more quiescent gas than the N$_{2}$H$^{+}$ emission does. This difference can be explained by the more complex structure of protostellar cores (i.e., outflow, disk) compared with prestellar cores. 

Lastly, the $T_A^*$ at each offset can provide valuable information regarding the gas density and temperature properties towards the cores. The lower plots in Figures \ref{fig.plot_L183}--\ref{fig.plot_L1521f} suggest that $T_A^*$ generally decreases with increasing offset. This behavior can be explained by the decreasing density or decreasing excitation temperature, T$_{ex}$ (i.e., sub-thermal excitation), as one moves outwards from the center of the core, and it is suggestive of the low optical depth of the line (See also Sect.~\ref{abundance}). Note that the kinetic temperature in prestellar cores generally increases outwards (up to 5~K difference) and therefore it cannot be behind the observed decrease in $T_A^*$. The observed small change in temperature does not seem to be the one driving the brightness of the emitting line. Fluctuations in the $T_A^*$ values of the lines either at the same offsets (see, e.g., L1544) or at increasing offsets (see, e.g., L183) indicate the presence of a non-homogeneous medium (e.g., clumps, cavities).

\section{Column densities and gas temperatures}

\label{col_densi}

\subsection{Radiative transfer modelling}

The determination of the temperature and column densities of the targeted species is a first step to help us study the chemistry, and, therefore, evolutionary status of the cores. We estimate the column density of H$_{2}$D$^{+}$ and N$_{2}$H$^{+}$ and kinetic temperature towards all cores by modelling the velocity-integrated main beam temperature at the peak/central position using the non-LTE radiative transfer code RADEX \citep{vanderTak2007}. We assume an isothermal homogeneous sphere. RADEX predicts the line intensities of the molecular transitions of interest for a given set of physical parameters: kinetic temperature, H$_{2}$ density, molecular column density, Rayleigh-Jeans equivalent background temperature of a black body shaped radiation field (in this case, the cosmic microwave background (CMB) temperature of 2.73~K was adopted), and the observed line width. To convert to the main beam temperature, we adopt the main beam efficiency, $\eta_{mb}$, of 0.6 ($\eta_{\alpha}$/$\eta_{mb}$ $\sim$0.8) measured by \citet{Buckle2009}.

\subsubsection{H$_{2}$D$^{+}$}

The observed spatial distribution of the H$_{2}$D$^{+}$ emission is extended for all cores, and therefore the implicit assumption of unit beam filling factor is proper. The most recent available collisional properties of the H$_{3}^{+}$-H$_{2}$ system and all its isotopic variations were studied and presented in \citet{Hugo2009}. To determine the excitation of H$_{2}$D$^{+}$, we adopt the rates presented in \citet{Hugo2009} with two simplifications. First, our calculations only consider the lowest two energy levels of o-H$_{2}$D$^{+}$, the 1$_{11}$ ground state, and the 1$_{10}$ excited state which lies 17~K above the ground. This two-level approximation is valid at the low temperatures of prestellar cores (T$\sim$10~K), and therefore for the present sample, since the next highest level, 2$_{12}$, lies at 109~K above the ground and its excitation is therefore negligible. The second simplification is that our calculations ignore reactive collisions. Note that in general the collisions between H$_{3}^{+}$ and H$_{2}$ may lead to reaction (i.e., the fully elastic case), excitation (i.e., the fully inelastic case), or both. Our assumption is justified by the work of \citet{Hugo2009}, who show that at the low temperatures of interest here, the reactive collision rates are factors $>$ 100 lower than the inelastic ones.

Our calculations include inelastic collisions of o-H$_{2}$D$^{+}$ with o-H$_{2}$ and p-H$_{2}$. Since we are dealing with very low temperatures, the ortho-to-para ratio of H$_{2}$ is assumed to be equal to its thermal value ($\sim$10$^{-4}$ in our case) instead of the high-temperature limit of 3. In several previous studies \citep[][]{vanderTak2005,Vastel2006,Caselli2008}, the H$_{2}$D$^{+}$ abundances towards the cores of our sample (when studied) were calculated using scaled radiative rates adopted by \citet{Black1990}, and therefore the column densities and abundances derived in those studies should be updated with the most recent rates, as we do here.

\subsubsection{N$_{2}$H$^{+}$}

In this work, we make use of the newest available collisional data of N$_{2}$H$^{+}$ \citep{Lique2015}. As opposed to former collisional data, where the collisional partner was Helium (He), the new ones are calculated based on collisions with the most abundant partner in the interstellar medium, which is H$_{2}$, and are found to be significantly different \citep{Lique2008}.

\subsection{RADEX results}

\label{rad_results}
 
Table~\ref{table.radex} presents the adopted gas densities $n$(H$_{2}$) of the cores along with their derived column densities (N(H$_{2}$D$^{+}$) and N(N$_{2}$H$^{+}$)), and derived temperatures (T$_{KIN}$) at the central position of each core. The average volume density, at least at the core center where we are focusing in this part of the analysis, was adopted as the values presented in \citet{Caselli2008}. In particular for this sample, we report low kinetic temperatures between 6~K and 9~K, and column densities for both species lying within the range of 0.4-2$\times$10$^{13}$cm$^{-2}$. The derived kinetic temperatures are used to assess the line thermal broadening per core presented in Section~\ref{line_anal}.

Comparing our observables with those presented in \citet{Caselli2008}, we see that the reported $T_{mb}$ values are similar in both studies, which indicates that the difference in beam size between JCMT and CSO plays a minor role, supporting our conclusion that the emission is extended and the beam is filled. The line widths are also similar except for L1521f and L1544. The observed difference is likely due to the more limited spectral resolution of the CSO backends. The column densities among the studies are consistent (mostly within a factor of 2), given the differences in adopted kinetic temperatures (up to 3~K), observed line fluxes, and molecular data. Increasing the number of observed transitions per species, and at multiple offsets, would allow the accurate determination of the kinetic gas temperature, column, and volume density, and the production of the column and gas kinetic temperature maps of the cores.

\begin{table}[h]
\caption{Column density ($N_{mol}$) and gas kinetic temperature ($T_{KIN}$) estimates for the adopted volume density ($n_{H_2}$) at the central position of core using RADEX.}
\begin{tabular}{@{}*{10}{l}}
\hline
&Species&Core&T$_{KIN}$ $^{a}$&n$_{H_2}$ (10$^{6}$) &N(mol)$^{a}$ & $\Delta V_{th}$ $^{b}$ \\
&&&(K)&(cm$^{-3}$)&(10$^{13}$cm$^{-2}$) & (km~s$^{-1}$) \\
\hline
&H$_{2}$D$^{+}$&L1544&9/7&2.0&2/3.2 & 0.32/0.28\\
&&L183&8/7&2.0&1.4/2.5 & 0.30/0.28\\
&&L694-2&9/7&0.9&0.8/3.2 & 0.32/0.28\\
&&L1517B&6/9.5&0.2&0.4/0.4 & 0.26/0.33 \\
&&L1521f&6/9.3&1.1&0.5/0.7 & 0.26/0.33 \\
\hline
&N$_{2}$H$^{+}$&L1544&5&2.0&0.8 & 0.09 \\
&&L1521f&8&1.1&0.9 & 0.11 \\
\hline

\end{tabular}

\tiny {\bf{Notes}}: The volume density for each core is adopted by \citet{Caselli2008}. $^{a}$ The first values of T$_{KIN}$ and N(mol) refer to this work followed by the values reported in \citet{Caselli2008} for direct comparison. $^{b}$ The thermal broadening is computed based on T$_{KIN}$. Accounting for kinetic temperatures within a range of 5-12~K, the reported column densities can vary by up to $\sim$60-80\%.
\label{table.radex}
\end{table}

\section{Modelling the spatial distribution of H$_{2}$D$^{+}$ and N$_{2}$H$^{+}$}

\label{mod_ratran}

A primary goal of this paper is to distinguish between dynamical models that turn a prestellar core to a protostar that have been proposed in the literature. To do so, we model the spatial distribution and line profiles of H$_{2}$D$^{+}$  and N$_{2}$H$^{+}$ (detected towards only two cores) for each core, as predicted from each dynamical model using advanced radiative transfer modelling, and compare the predicted emission with our observed maps. 
\subsection{RATRAN - initializing the model}

To model the spatial distribution and line profiles of H$_{2}$D$^{+}$ and N$_{2}$H$^{+}$ (where present) emission from the cores, we ran the Monte Carlo radiative transfer code RATRAN \citep{Hogerheijde2000}. RATRAN calculates synthetic line emission for the species of interest, taking into account the temperature and density gradients of the source, the dust emission and absorption properties, and the small-scale (e.g., thermal motions, turbulence) and large-scale (e.g., infall motions) source kinematics. We fix the physical structure of the sources by adopting the corresponding predicted temperature, density, and velocity structure from proposed dynamical models in the literature. The dynamical models examined in this study are described in Sect.~\ref{dyn_models}. For our models, we assume that the dust temperature is equal to the gas temperature, which is valid for environments of high volume density ($>$10$^{4}$~cm$^{-3}$), and a gas-to-dust ratio of 100. 

For our calculations, and to simulate the 2\arcmin$\times$2\arcmin~observations, we defined a grid of 13 logarithmically spaced shells for each core, extending to an outer radius of 10000~au for all cores except L694-2, which was extended to 15000~au due to its larger distance (see Table~\ref{sources}). The density and velocity structures are shown in Figure~\ref{density_velocity_keto} and were adopted by \citet{Keto2015} for the different models which are described in more detail in Sect.~\ref{lp}--\ref{qe}.

The temperature profiles were adopted from the literature \citep[L183, L694-2, L1517B, L1544, L1521f][]{Pagani2007,Redaelli2018,Maret2013,ChaconTanarro2019,Crapsi2004}. These profiles were derived using radiative transfer models to fit the observed gas (e.g., N$_{2}$H$^{+}$) or dust emission (e.g., mm IRAM observations), and are specific for each core. The LP and SIS models are by definition isothermal, while the QE-BES model predicts a range of temperatures, and is more consistent with the observed temperature variations ($\sim$5~K). Our approach to adopting the same temperature profile in all dynamical models may affect the brightness of the line, but not the dynamics \citep{Keto2010}. In this study, we examine the line morphology, which is highly dependent on the dynamics, i.e., the distribution of the velocities of the gas, rather than the small variations in the gas kinetic temperature; therefore, our general conclusions are robust.         

The Doppler parameter, b, was calculated for each shell as equal to 0.6$\times$FWHM, where FWHM is the result of the Gaussian fitting process for each core and position presented in Table~\ref{table.dataa}. In Table~\ref{abundances_table} we present the initial abundance profiles adopted during the modelling process for H$_{2}$D$^{+}$ and N$_{2}$H$^{+}$ for each core with the associated references, and the relevant modifications per model to achieve a good fit. In particular, we proceed to a more accurate determination of the abundance profiles by fitting the emission at the central position and expand the solution to the entire map (see Sect.~\ref{abundance} below). Lastly, to compare the modelled emission with observations, the resulting modelled spatial distribution is convolved to the beam size of the JCMT (14$\arcsec$ at 372 GHz).

\subsection{Physical Structure- Dynamical Models}
\label{dyn_models}
Multiple hypotheses exist about the dynamical processes that take place for a prestellar core to collapse. To date, the determination of the most accurate model has been challenging. In this study, we examine three different dynamical models based on spherical symmetry that result in similar density profiles but very different velocity profiles across the cores. These models are the quasi-equilibrium contraction of a Bonnor-Ebert sphere (QE-BES), the Larson-Penston flow (LP), and the inside-out collapse of the singular isothermal sphere (SIS).

\begin{figure}[ht]
\begin{center}
\includegraphics[scale=0.3]{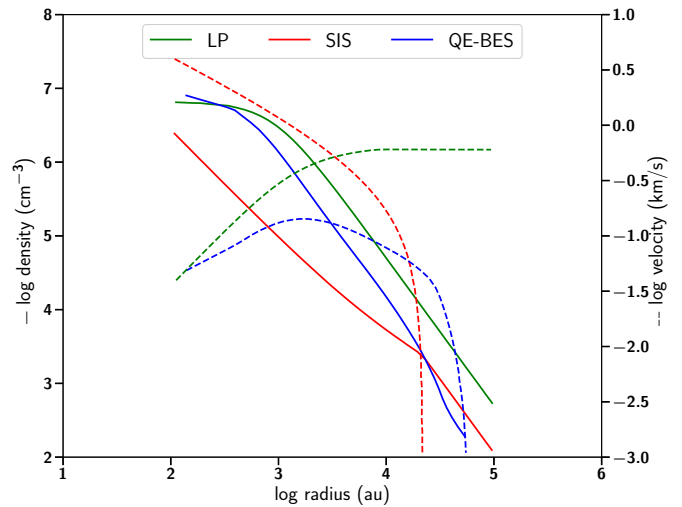}
\end{center}
\caption{The density (solid line) and velocity (dashed line) profiles of a prestellar core predicted by the three dynamical models: LP flow (green), SIS (red), and QE-BES (blue), as presented in \citet{Keto2015}. Each of the modelled profiles was adopted to model the H$_{2}$D$^{+}$ emission using RATRAN.}
\label{density_velocity_keto}
\end{figure}

We note that despite the similarity of the slopes in the density distribution, the different density profiles account for a total cloud mass of $\sim$10~\msun~for the QE-BES model, $\sim$ 27~\msun~for the LP model, and $\sim$4~\msun~for the SIS model, for a radius up to 30000 au. Based on mass estimations alone, the SIS and QE-BES models appear to be the most suitable in reproducing the estimated mass for the cores in our sample (based on dust continuum observations; see Table~\ref{sources}). Here we investigate the behavior of the gas. The quite distinct velocity profiles resulting from the different dynamical models are shown in Figure~\ref{density_velocity_keto}. Gas emission observations in the optically thin regime that can trace those kinematics are therefore crucial. The models of interest are described below.

\subsubsection{LP flow}
\label{lp}
\citet{Larson1969} attempted to solve numerically the dynamical equations concerning the collapse of a spherically symmetric isothermal core, known as Larson core, which is gravitationally unstable, yet all rotational, magnetic, and turbulent motions are neglected (free-fall collapse). The collapse, in this case, is non-homologous, with the central region collapsing to a high density and reaching hydrostatic equilibrium and the outer regions collapsing to a much lesser extent, creating a first hydrostatic core object. In the meantime, \citet{Penston1969} presented the dynamics of self-gravitating gaseous spheres, focusing on the isothermal case, which collapses in free-fall. The analytical solutions to the equations of free-fall collapse are used to produce the various symmetries (including spherical) at the instant when a dense hydrostatic object forms in the innermost dense regions of the core. For the hydrodynamic simulations, the diffusion approximation for radiative transfer was used. 

The density and velocity profiles of the LP model (Figure~\ref{density_velocity_keto}) were adopted by \citet{Keto2015}, and are products of integrating Equations 11 and 12 in \citet{Shu1977}. To produce a similar density profile (slope $\sim${\it{r}}$^{-2}$) for all models (LP, SIS, QE-BES), the sound speed was 0.2~km~s$^{-1}$ at $\sim$11~K, and the evolutionary time was 2.8$\times$10$^{11}$~s. The density follows a flat distribution with a central density {\it{n}} $=$ 6.3$\times$10$^{6}$~cm$^{-3}$ out to $\sim$ 850 au and follows an {\it{r}}$^{-2}$ slope outwards, with {\it{n}} = 3.5$\times$10$^{4}$~cm$^{-3}$ at $\sim$10000 au. The velocity gradually increases from 5$\times$10$^{-2}$~km~s$^{-1}$ at 100 au, reaching a maximum of 10$^{-1}$~km~s$^{-1}$ at $\sim$ 850 au, where it reaches and remains at a constant velocity through to the outermost regions of the core.

\subsubsection{SIS}
\label{sis}
\citet{Shu1977} proposed the Singular Isothermal Sphere (SIS) model to describe the gravitational collapse of isothermal gas clouds. In this case, the collapse occurs from the inside out, with the inner and outer envelopes modelled separately. The inner envelope is assumed to collapse under free-fall conditions while the outer envelope is modelled as essentially static. The proposed density distribution in this case follows the $r^{-3/2}$ for the inner envelope, and $r^{-2}$ for the outer envelope, adopting a sound speed, $\alpha$ $=$ 0.2~km~s$^{-1}$ at $\sim$11~K, and an evolutionary time t $=$ $\sim$2.5$\times$10$^{5}$~yr. In particular, the inner density is 2.5$\times$10$^{6}$~cm$^{-3}$ at 100 au, decreasing down to 1.4$\times$10$^{4}$~cm$^{-3}$ at $\sim$ 10000 au. The velocity profile in this dynamical model is characterized by a higher velocity at the center of the core (compared to the other dynamical models) of $\sim$ 4~km~s$^{-1}$ at $\sim$ 100 au and decreases down to 0.5~km~s$^{-1}$ at $\sim$ 10000 au, before starting to behave as a static sphere without infall in the outer parts of the core.

\begin{table*}
\caption{H$_{2}$D$^{+}$ and N$_{2}$H$^{+}$ abundances estimated with RATRAN for each core based on the JCMT observations. The table contains also the initial abundance profiles adopted from literature when available for direct comparison.} 
\centering
\small\addtolength{\tabcolsep}{-4.7pt}
\begin{tabular}{c c c c c c c c c c}
\hline\hline
& & \multicolumn{3}{c}{Literature} & & \multicolumn{3}{c}{This work} \\Core & Species & X$_{IN}$ & X$_{INTERMEDIATE}$ & X$_{OUT}$ & Authors & QE-BES & LP & SIS & Static\\
\hline\hline

{\bf{L183}} &  & (r $<$ 300 au) & (300 au $<$ r $<$ 5000 au) & (5000 au $<$ r $<$ 10000 au) & $\alpha$ & & Multiplication factor & \\
 & H$_{2}$D$^{+}$ & 1 $\times$ $10^{-10}$ & 2 $\times$ $10^{-10}$ & 4 $\times$ $10^{-11}$ & & 2 & 0.5 & 100 & 1 \\
\hline
{\bf{L1544}} & & (r $<$ 2500 au) & n/a & (2500 au $<$ r $<$ 10000 au) & $\beta$ & & Multiplication factor & \\
 & H$_{2}$D$^{+}$ & 1 $\times$ $10^{-9}$ & n/a & 5 $\times$ $10^{-10}$ & & 1 & 0.3 & 60 & 0.8 \\
 & & (r $<$ 2500 au) & n/a & (2500 au $<$ r $<$ 10000 au) & $\gamma$ & & Multiplication factor & \\
& N$_{2}$H$^{+}$& 0 & n/a & 2 $\times$ $10^{-10}$ & & 20 & 0.8 & 500 & 4 \\
\hline
{\bf{L694-2}} & & n/a & n/a & n/a & n/a & & Abundance (const.) & \\
 & H$_{2}$D$^{+}$& n/a & n/a & n/a & n/a & 4.4 $\times$ $10^{-10}$ & 5.2 $\times$ $10^{-11}$ & 2 $\times$ $10^{-8}$ & 2.5 $\times$ $10^{-10}$ \\
\hline
{\bf{L1517B}} & & n/a & n/a & n/a & n/a & & Abundance (const.) & \\
 & H$_{2}$D$^{+}$ & n/a & n/a & n/a & n/a &  5 $\times$ $10^{-11}$ & 1.2 $\times$ $10^{-11}$ & 5 $\times$ $10^{-9}$ & 1.5 $\times$ $10^{-10}$ \\
\hline
{\bf{L1521f}} & & (r $<$ 3500 au) & n/a & (3500 au $<$ r $<$ 10000 au) & $\delta$ & & Multiplication factor & \\
 & H$_{2}$D$^{+}$ & 3 $\times$ $10^{-10}$ & n/a & 6 $\times$ $10^{-11}$ & & 1.3 & 0.75 & 100 & 1 \\
& & (r $<$ 3500 au) & n/a & (3500 au $<$ r $<$ 10000 au) & $\delta$ & & Multiplication factor & \\
& N$_{2}$H$^{+}$ & 1 $\times$ $10^{-10}$ & n/a & 2 $\times$ $10^{-10}$ & & 1 & 0.3 & 15 & 1 \\
\hline
\end{tabular}

\tiny {\bf{Notes}}: The first two columns of the table report the core name and the observed species. The abundance profiles of the observed species per core as retrieved by the literature are presented in columns 3-5. In particular, we report the inner, intermediate and outer abundances (X$_{IN}$, X$_{INTERMEDIATE}$ and X$_{OUT}$) following their corresponding shell sizes. Lastly, the resulted abundance profiles of this work and per model (QE-BES, LP,SIS and Static) are reported in columns 7-10. The multiplication factor is the one applied to the abundance profiles taken from literature (when available) resulting in the best fit between the modelled and observed line intensity at the central position of the maps. For L1517B and L694-2 we report the resulted constant abundance. Authors: $\alpha$~\citet{Pagani2009}; For 300 au $<$ r $<$ 5000 au, the average value of the reported abundances was adopted, $\beta$~\citet{vanderTak2005}, $\gamma$~\citet{Caselli2002}, $\delta$~\citet{Crapsi2004}.
\label{abundances_table}
\end{table*}

\subsubsection{QE-BES}
\label{qe}
This model treats the prestellar core as a Bonnor-Ebert Sphere (BES) undergoing quasi-static contraction, in which the pressure in the system remains uniform and constant, due to a low rate of volume change. The pressure is not affected by small oscillations connected with small-scale processes within or near the cores. The notion of a BES was originally put forward in 1956 when it was shown that an isothermal sphere of gas in an external pressurized medium has a critical size above which it is gravitationally unstable \citep{Bonnor1956,Ebert1957}. This model predicts a density profile following $r^{-2}$ in the outer region of the core with an approximately constant density inwards \citep[e.g.,][]{Keto2010}. The central density in this case is {\it{n}} $=$ 10$^{7}$~cm$^{-3}$ \citep{Keto2015}. The velocity profile predicted by this contraction model shows a so-called $\Lambda$-shape, starting from 0~km~s$^{-1}$ in the center ($<$100 au) and increasing outwards, while reaching a maximum value ($\sim$1~km~s$^{-1}$) near the characteristic radius, r$_{f}$, ($\sim$1000 au in this case), where we observe the transition from a constant/flat density distribution to that of $r^{-2}$. After the velocity reaches its maximum value, it decreases again when moving towards the edge of the core.

\begin{table*}
\caption{Optical depth, $\tau$, of H$_{2}$D$^{+}$ and N$_{2}$H$^{+}$ estimated with RATRAN at the central position of each core.} 
\centering
\small\addtolength{\tabcolsep}{-4.7pt}
\begin{tabular}{c l c c c c}
\hline\hline
Core & Species & $\tau$ (QE-BES) & $\tau$ (LP) & $\tau$ (Static) & $\tau$ (SIS) \\
\hline\hline

{\bf{L183}}  & H$_{2}$D$^{+}$ 1$_{10}$-1$_{11}$ & 0.4 & 0.4 & 0.4 & 3.6 \\
{\bf{L694-2}} & H$_{2}$D$^{+}$ 1$_{10}$-1$_{11}$ & 0.7 & 0.2 & 0.3 & 0.4 \\
{\bf{L1517B}}  & H$_{2}$D$^{+}$ 1$_{10}$-1$_{11}$ & 0.1 & 0.05 & 0.08 & 0.8 \\
{\bf{L1544}}  & H$_{2}$D$^{+}$ 1$_{10}$-1$_{11}$ & 1.5 & 0.4 & 0.4 & 0.2 \\
& N$_{2}$H$^{+}$ 4-3 & 1.6 & 0.5 & 0.5 & 0.2 \\
{\bf{L1521f}}  & H$_{2}$D$^{+}$ 1$_{10}$-1$_{11}$ & 0.08 & 0.07 & 0.08 & 0.02 \\
& N$_{2}$H$^{+}$ 4-3 & 0.7 & 0.3 & 0.6 & 0.5 \\
\hline
\end{tabular}

\tiny {\bf{Notes}}: The optical depth, $\tau$, at the central position of each core is found to be $<$ 1 for both species for most models. 
\label{opacities_table}
\end{table*}

\subsection{Abundance derivation - dynamical approach}

\label{abundance}

\begin{figure*}[ht]
\begin{center}$
\begin{array}{cc} 
  \includegraphics[scale=0.35]{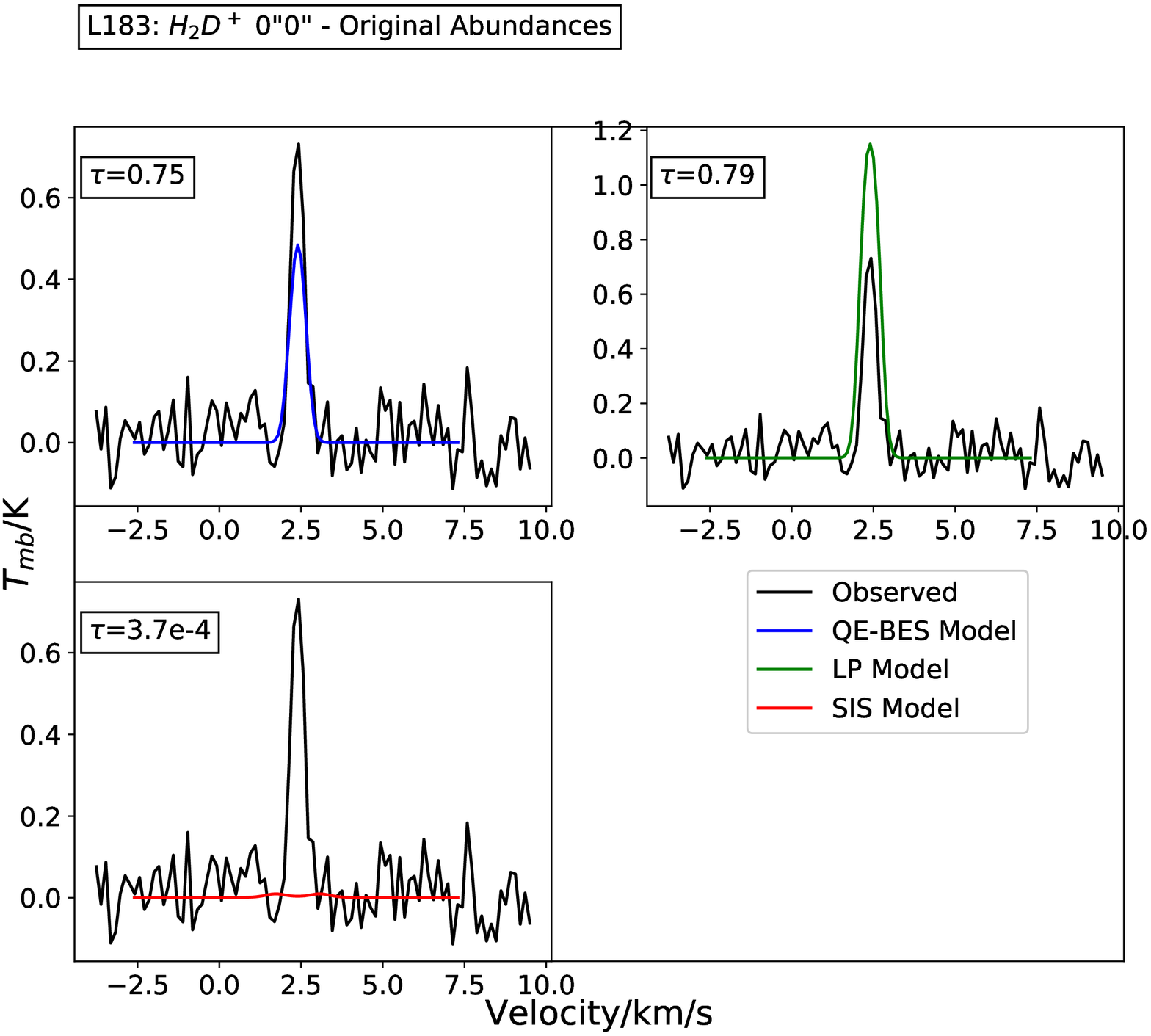} & \includegraphics[scale=0.35]{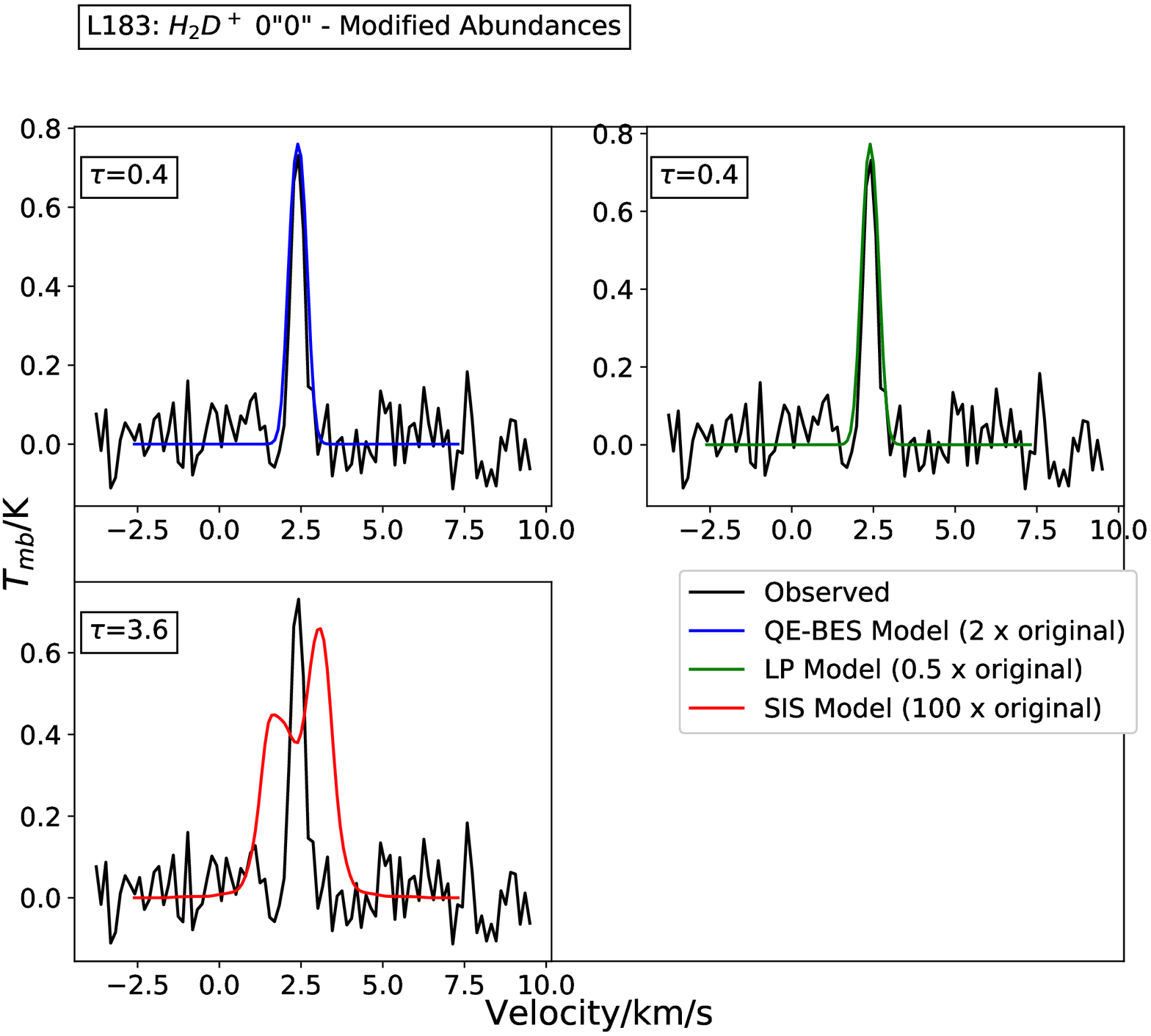}
\end{array}$  
\end{center}
\caption{An example of the abundance profile adjustment during the fitting process. The observed line profiles of H$_{2}$D$^{+}$ are overplotted with the modelled ones using RATRAN towards the central position of the L183 core. Three different dynamical models were adopted, QE-BES, SIS and LP flow. The original abundance profiles (left panel) were adopted by \citet{Pagani2009} and cannot reproduce the central peak of the observations. Adjusting the H$_{2}$D$^{+}$ abundance profile by applying a multiplication factor, resulted in a better reproduction of the observations for the LP and QE-BES models, while the SIS model failed to match the shape of the observed line profile.}
\label{fig.central_h_before_after}
\end{figure*}

Before determining the most representative dynamical model per source, we proceed with fine-tuning the abundance profiles of H$_{2}$D$^{+}$ and N$_{2}$H$^{+}$. We start the fitting process by adopting an abundance profile for the observed H$_{2}$D$^{+}$ per core from the literature and proceed to a more accurate determination of the abundance profiles by fitting the line emission at the central position for each core. To do so, we let the abundance vary by up to 2 orders of magnitude higher and lower from the initial abundance and using a step of a factor of 2. Table~\ref{abundances_table} lists the initial H$_{2}$D$^{+}$ abundance profiles adopted from the literature with their associated references and the multiplication factor we applied in this study to fit the peak line intensity of the central position. A constant abundance profile was also tested towards these cores but without resulting in a qualitative change on the results. Therefore, we adopt the more detailed abundance profiles when available throughout the paper. Note that for two cores, L694-2 and L1517B, initial abundance profiles could not be retrieved from the literature; therefore, we assumed a constant initial abundance of 1$\times$10$^{-10}$ which we varied as described above. Figure~\ref{fig.central_h_before_after} shows an example of the fitting process towards L183. 

Our observations concern only single transitions of both H$_{2}$D$^{+}$ and N$_{2}$H$^{+}$, while the physical structure is fixed. Therefore, although our fitting process results in an uncertainty of $\sim$14\%-18\% ($\sim$3$\times$$\chi^2_{min}$) for the derived abundances, this process somewhat underestimates the real uncertainties. Once we find the abundance profile that can reproduce the observed line profile in the central position of the core, we expand the solution to the entire map. In Table~\ref{opacities_table} we present the resulting optical depth for each line per model towards the central positions of each core. We find that the emission from both H$_{2}$D$^{+}$ 1$_{10}$-1$_{11}$ and N$_{2}$H$^{+}$ 4-3 is mostly in the optically thin regime ($\tau$ $<$ 1) justifying their good performance in tracing the dynamics of the cores. 


\subsection{Static sphere}

\label{stat_sph}

To provide an independent determination of the abundances, we proceed with applying the same radiative transfer technique, but this time we adopt the physical structure of each source based on their previous dust and gas observations. In this case we assumed a static envelope, without infall or expansion motions. The temperature and density profiles were adopted from the literature \citep[L183, L694-2, L1517B, L1544, L1521f][]{Pagani2007,Redaelli2018,Maret2013,ChaconTanarro2019,Crapsi2004}, and are based on continuum and gas observations, and they are, therefore, unique to each core. Again, the basic assumptions for all cores are that the gas and dust are thermally coupled, their temperatures are therefore set to be equal, and the gas to dust ratio is 100. As before, we define a grid of 13 logarithmically spaced shells for each core, while the density and temperature profiles per core are described as follows. 

\vspace{0.2cm}

{\bf{L183}}: For this core we adopt the density and temperature profiles presented in \citet{Pagani2007}, and are based on a detailed 1D radiative transfer modelling to fit both N$_{2}$H$^{+}$ and N$_{2}$D$^{+}$ emission (IRAM-30m cut observations), consistently with the dust emission \citep[MAMBO;][]{Pagani2004}. In particular, the density profile follows a broken power law with an inner density $\rho_0$ $=$ 2.34$\times$10$^{6}$~cm$^{-3}$, with the density proportional to r$^{-1}$ for a radius up to $\sim$4000 au, and to {\it{r}}$^{-2}$ for the rest of the core. The temperature profile has a constant value of 7 K up to 5600 au and increases outwards up to 12 K.  
\\

\vspace{0.2cm}
       {\bf{L694-2}}: For this core, we adopt the density and temperature profiles presented in \citet{Redaelli2018}. According to \citet{Arzoumanian2011}, the density profile is given by:

       \begin{equation}
\rho(r)=\left[\frac{\rho_0}{1+\left(\frac{r}{r_0}\right)^2}\right]^\frac{p}{2}
\label{densitylaw}
\end{equation}

\hspace{-0.55cm}which \citet{Redaelli2018} adjusted for L694-2 to fit its observed {\it{Herschel}}/SPIRE dust continuum at 250~$\mu$m, 350~$\mu$m, and 500~$\mu$m, using radiative transfer modelling. In particular, the inner density $\rho_0$ $=$ 1.8$\times$10$^{5}$~cm$^{-3}$, the half maximum density radius $r_0$ $=$ 2500 au, and $p$ $=$ 3.3. The temperature profile is 8~K in the center of the core and increases outwards to 12~K.
\\

\vspace{0.2cm}     

{\bf{L1517B}}: Here we adopt the density and temperature profiles presented in \citet{Maret2013} and \citet{Tafalla2004} respectively, based on radiative transfer modelling of the 1.2 mm continuum emission. In particular, the density profile of L1517B follows the form:

\begin{equation}
\rho(r)=\frac{\rho_0}{1+\left(\frac{r}{r_0}\right)^{\alpha}}
\label{densitylaw2}
\end{equation}

\hspace{-0.5cm}with a $\rho_0$ $=$ 2.2$\times$10$^{5}$~cm$^{-3}$, a half maximum density radius $r_0$ $=$ 4.9$\times$10$^{3}$~au, and $\alpha$=2.5. The temperature profile is also taken from those studies and is assumed to be isothermal and equal to 10~K throughout the core.
\\

\vspace{0.2cm}  

{\bf{L1544}}: For this core, we adopt the most recent density and temperature profiles presented in \citet{ChaconTanarro2019}, based on fitting the 1.1 mm and 3.3 mm continuum emission. The temperature profile follows the form:

\begin{equation}
T(r) = T_{out} - \frac{T_{out} - T_{in}}{1+\left(\frac{r}{r_0}\right)^{1.7}}
\label{templawl1544}
\end{equation}

{\hspace{-0.5cm}}with outer temperature T$_{out}$ = 12~K, inner temperature T$_{in}$ = 6.9~K, and critical radius {\it{r}}$_{0}$ = 28.07$''$. The density profile follows the same form as in Equation~\ref{densitylaw2}, but for $\rho_0$ $=$ 1.6$\times$10$^{6}$~cm$^{-3}$, a half maximum density radius $r_0$ $=$ 17.3$''$, and $\alpha$=2.6.
\\

\vspace{0.2cm}  

{\bf{L1521f}}: For L1521f, we adopt the density and temperature profiles presented in \citet{Crapsi2004}. In particular, the density profile follows the same form as in Equation~\ref{densitylaw2}, but for a central density $\rho_0$ $=$ 1$\times$10$^{6}$~cm$^{-3}$, a half maximum density radius $r_0$ $=$ 2800~au, and $\alpha$=2, while the temperature is assumed to be isothermal and equal to 10~K for the entire core.

\subsubsection{Abundance derivation - static approach}

We find that the initial abundance profiles reproduce the line strength of the central peak position, requiring no modification for L183 and L1521f, and a modification of a factor between 0.8 and 4 for the H$_{2}$D$^{+}$ and N$_{2}$H$^{+}$ emission respectively, towards L1544. The static model performs well in reproducing the central peak for all cores with minimum modification of the initial abundance profile. Therefore, we consider the H$_{2}$D$^{+}$ abundance determination to be more accurate compared to the one derived from the dynamical models towards the two cores for which we did not have prior knowledge of the H$_{2}$D$^{+}$ abundance. Our best fit results in [H$_{2}$D$^{+}$] of 2.5$\times$10$^{-10}$ for L694-2 and 1.5$\times$10$^{-10}$ for L1517B, and similar (within a factor of 2) to what was previously found for the rest of the cores. These results are summarized in Table~\ref{abundances_table}.

\subsection{RATRAN - results}

\label{rat_dyn}

In Figures~\ref{RATRAN_L183}--\ref{RATRAN_L1521f_n2hp}, we present the azimuthally averaged observed and modelled emission of H$_{2}$D$^{+}$ towards the cores for the different physical and kinematic structures predicted by the three dynamical models, LP, SIS, and QE-BES, and the static model. The corresponding plots using the original, before azimuthal averaging, maps of each core can be found in Appendix~\ref{ratran_ori}. The synthetic spectrum is compared to the observed one at each map position using a simple $\chi^2$ approximation given by:

\begin{equation}
  \chi^2=\displaystyle\sum_{i} \frac{\left(T^i_{MB,obs} -T^i_{MB,mod}\right)^{2}}{\sigma^2_{obs}} 
\label{chisquarelaw}
\end{equation}

\hspace{-0.7cm} where $T^i_{MB,obs}$ is the observed main beam temperature per velocity channel ({\it{i}}), $T^i_{MB,mod}$ is the modelled one, and $\sigma_{obs}$ is the observed rms.

\subsubsection{Individual cores}

\begin{figure*}[ht]
\begin{center}
\includegraphics[scale=0.5]{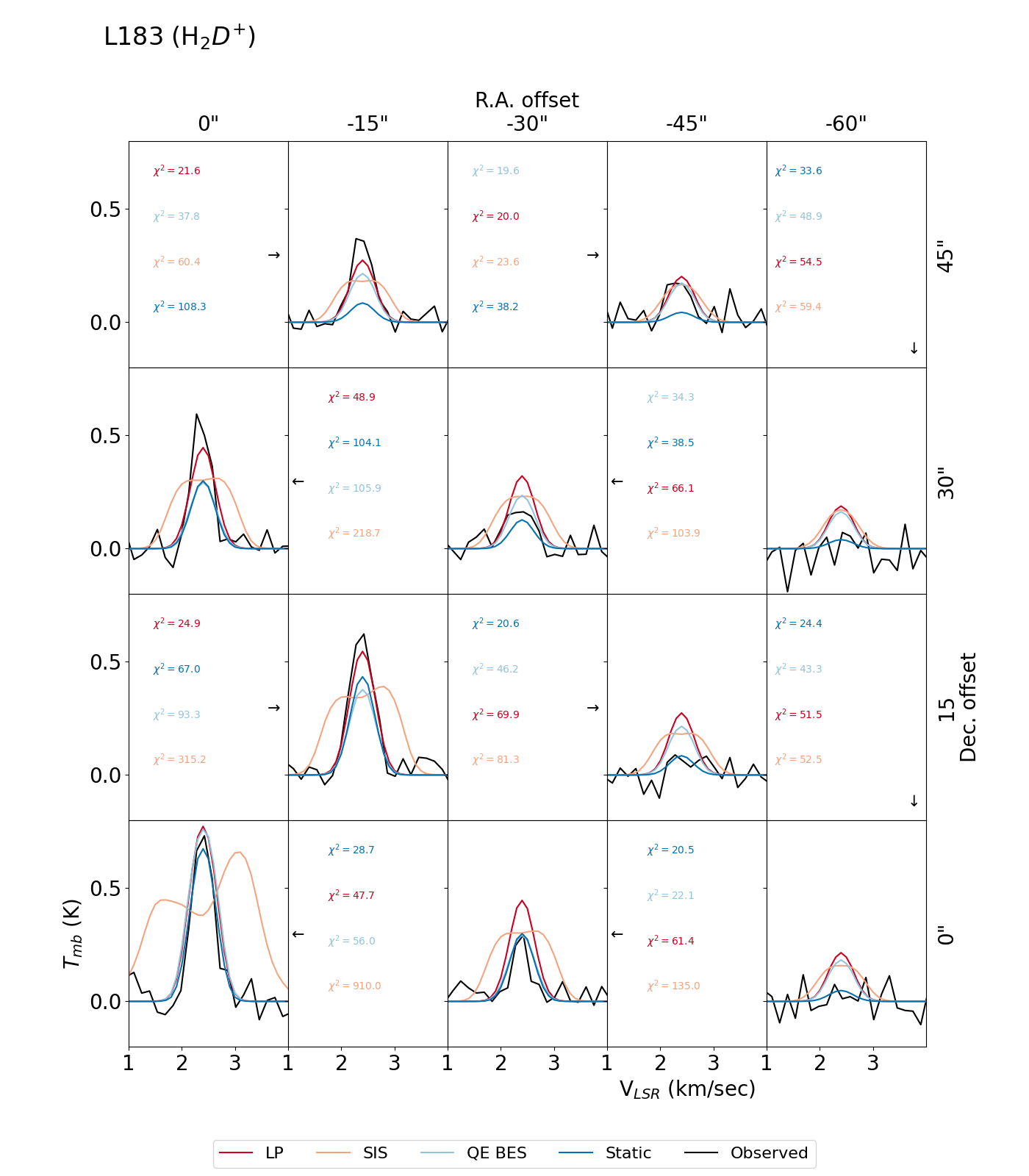}
\end{center}
\caption{Observed azimuthally averaged line profiles for each position of H$_{2}$D$^{+}$ maps towards L183, overplotted with the modelled ones for Static, LP, SIS and QE-BES models from RATRAN.}
\label{RATRAN_L183}
\end{figure*}

{\bf{L183:}} For L183, as seen in Figure~\ref{RATRAN_L183} and Figure~\ref{RATRAN_L183_ori}, the static, QE-BES and LP models all can reproduce the central (0$''$, 0$''$) observed peak intensity and line shape with $<$ 5\% deviation. This similarity is easily justified since the abundance profiles are adjusted accordingly in order to fit the central position. The SIS model, however, predicts $\sim$ 2 times larger line widths than observed, resulting in an order of magnitude larger $\chi^2$ ($\sim$ 900) compared to the rest of the models and demonstrating the poor quality of the fit produced by this model. 

Moving outwards, we see at 15$''$ offsets, that the LP model performs better in reproducing the observed peak intensities with absolute deviations of 10-30\%, followed by both the static and QE-BES models with absolute deviations of 30-50\%. The SIS model is again significantly worse ($>$ 3 times higher $\chi^2$ compared to the other three models) in reproducing the shape and intensity of the observed line profiles. At 30$''$ offsets, the LP model is still superior to the other models with typical deviations of $\sim$ 15\% from the observed emission, which is significantly closer to the observations compared to QE-BES and static models that show $>$ 40\% deviations. The exception to this rule is the 30$''$ North, where all models underestimate the emission by 50-65\%. Similarly, all models underestimate the observed emission by 50-90\% at 45$\arcsec$ offsets northwards (both east, west, and center). On the other hand, at 45$\arcsec$ offsets southwards, the core is characterized by weaker emission and the LP model can reproduce reasonably well the observations ($<$ 15\% deviations), which is indicative of a diversion from the spherical symmetry assumption (see Fig.~\ref{RATRAN_L183_ori}). Note that \citet{Pagani2007} did not report such diversions for other species (e.g., N$_{2}$D$^{+}$), but in that paper only the west-east cut is presented. Noticeably, the QE-BES model performs better than the static sphere at the larger offsets. 

We conclude that the LP model performs better in reproducing the observed line profiles and intensities globally for L183, followed by the QE-BES model, while the static model significantly underestimates the emission towards the outer parts of the core. The SIS model cannot reproduce the intensity and morphology of the line profiles globally, and therefore it is the least suitable among the models. 
\\
\vspace{0.2cm}

\begin{figure*}[ht]
\begin{center}
\includegraphics[scale=0.5]{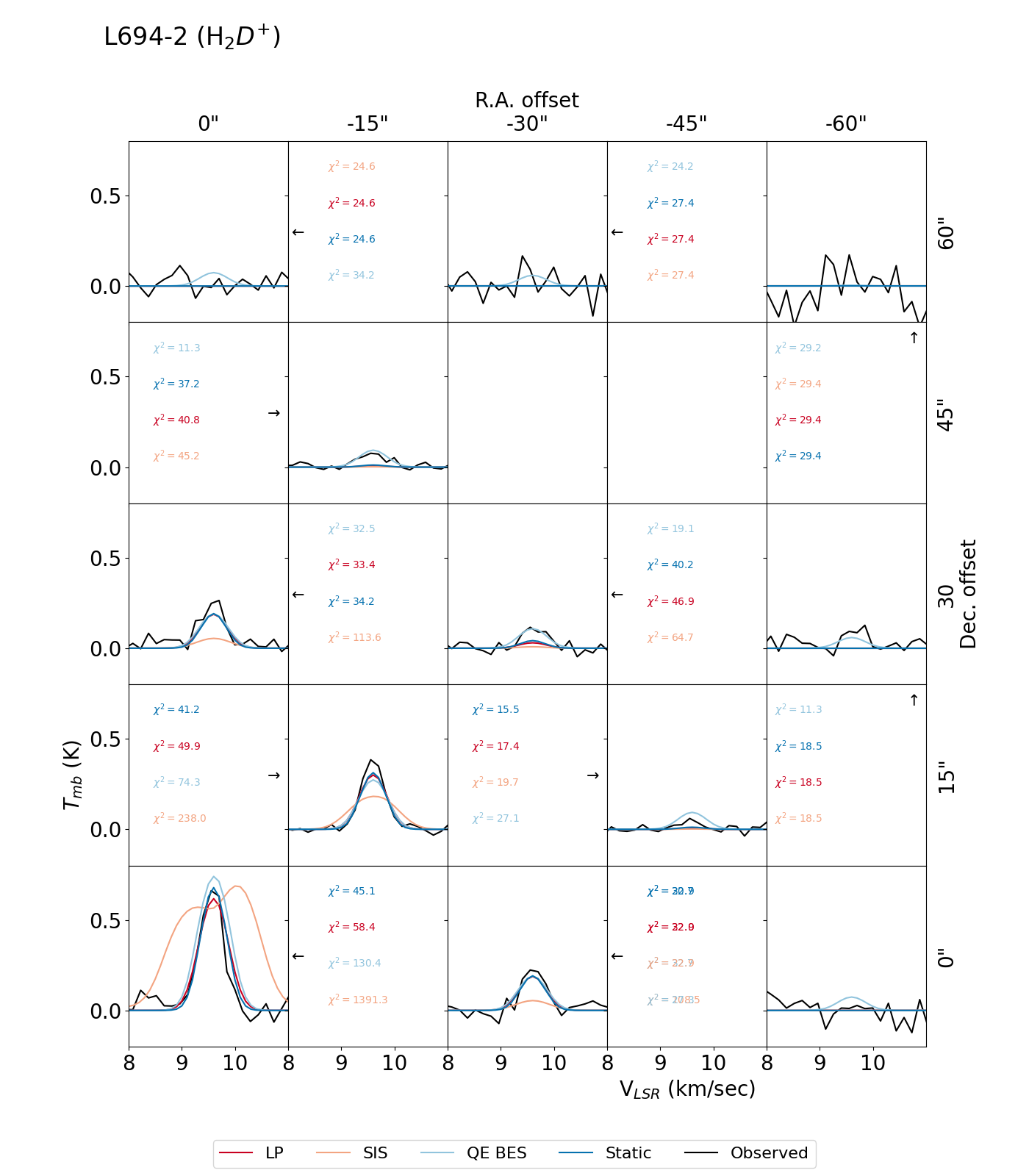}
\end{center}
\caption{As Fig.~\ref{RATRAN_L183}, but for L694-2.}
\label{RATRAN_L694-2}
\end{figure*}

{\bf{L694-2:}} For L694-2, as seen in Figure~\ref{RATRAN_L694-2} and Figure~\ref{RATRAN_L694-2_ori}, the static model reproduces the central (0$''$, 0$''$) observed peak intensity and line shape with only 1\% deviation, followed by the LP model with $<$ 10\% deviation. While the QE-BES model can also fairly reproduce the central emission, the SIS model fails to match the observations by producing a line profile with much stronger line wings than the observed, resulting in a $\chi^2$ $>$ 1000, which is two orders of magnitude higher than the LP and static models. 

Interestingly, when we examine the modelled versus the observed emission globally, we see that the LP, QE-BES and static models can all reproduce the observed peak intensities equally well with absolute deviations of 2-12\% at 15$''$ North-East and South-West from the central peak, while all three underestimate the observed line intensity at 15$''$ South-East and North-West by 30-50\%. The SIS model reproduces neither the strength nor the shape of the lines at those offsets. At offsets $>$ 30$''$, the QE-BES model predicts emission that is 30-60\% higher compared to the other models. This weak emission, however, is mostly within the rms limits, and therefore not possible to validate with the dataset in hand.

We conclude that for L694-2 we cannot clearly distinguish between LP, QE-BES, and the static models, as their predicted emission is very similar (mostly similar resulting $\chi^2$). Future deeper observations towards the outer parts of the core (e.g., $>$ 30$''$) will reveal if there is a weak emission similar to what the QE-BES model predicts, which will allow us to tell those models apart. The SIS model appears to be the least suitable among the models, failing to reproduce both the intensity and the shape of the line profiles globally. 
\\
\vspace{0.2cm}

\begin{figure*}[ht]
\begin{center}
\includegraphics[scale=0.5]{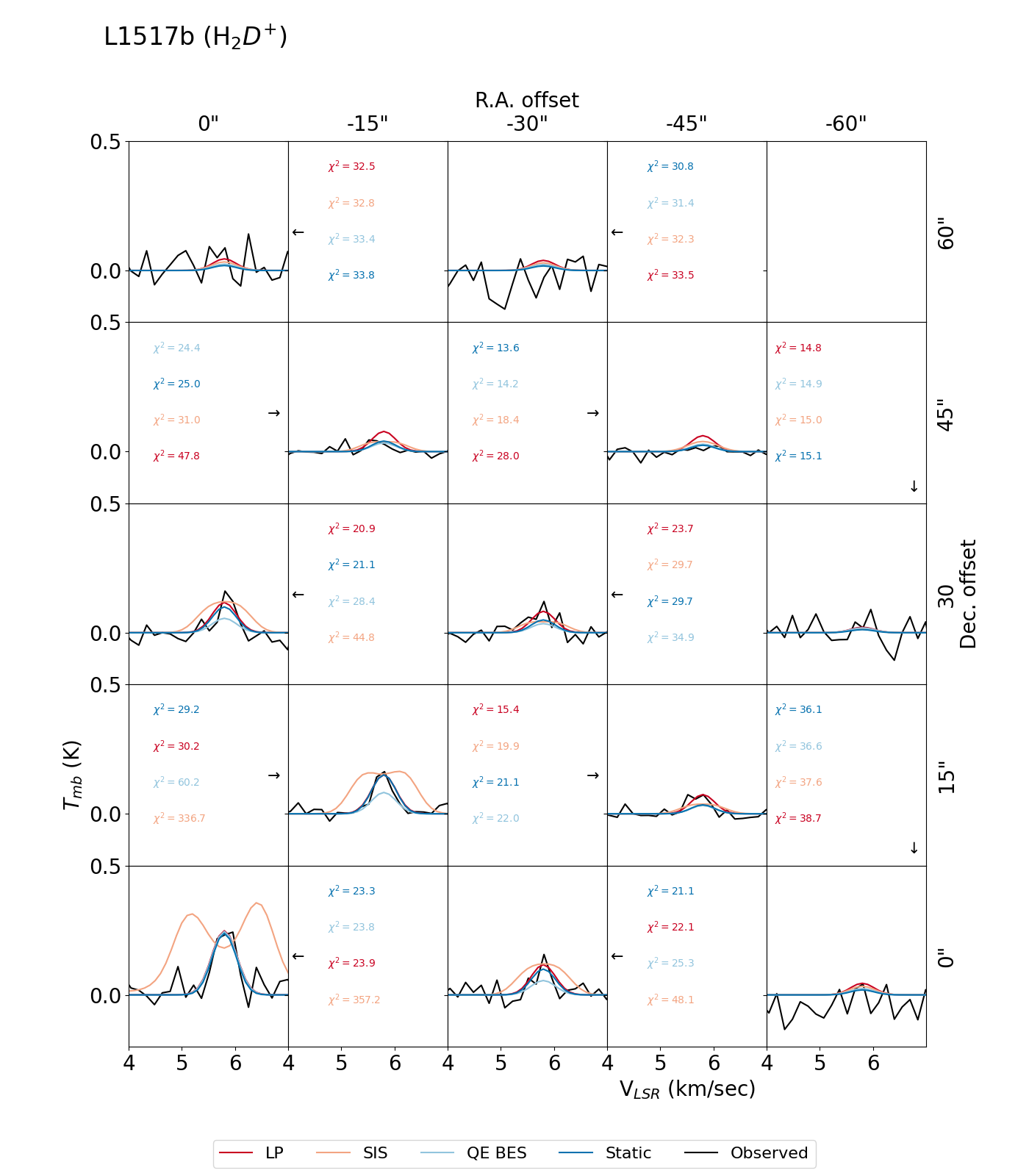}
\end{center}
\caption{As Fig.~\ref{RATRAN_L183}, but for L1517B.}
\label{RATRAN_L1517B}
\end{figure*}

{\bf{L1517b:}} For L1517b, as seen in Figure~\ref{RATRAN_L1517B} and Figure~\ref{RATRAN_L1517B_ori}, all three QE-BES, LP, and static models can reproduce the central (0$''$, 0$''$) observed peak intensity and line profile with $<$ 5\% deviation, while the SIS model predicts $\sim$15\% lower intensity at the source velocity.

When we examine the modelled versus the observed emission globally, we see that both the LP and the static model can reproduce the observed peak intensities with absolute deviations of 0.5-8\% at offsets within 15$''$, where the detections are more apparent, while the deviations of the models are $>$ 30\% at larger offsets where the observed emission gets weaker. For the same grid of offsets, the QE-BES model does not perform as well, showing larger deviations of 25-75\%, with the only exception towards the central position (0$''$, 0$''$). Lastly, the SIS model appears to reproduce the peak intensities with deviations of 5-30\% for within 15$''$, while it gets significantly worse at larger offsets (up to 80\% for $>$ 30$''$). When we look at the line profile of the modelled emission though, it becomes apparent that the SIS model fails to reproduce the observed emission globally. The unsuitability is reflected in high $\chi^2$ at most positions that are about an order of magnitude higher compared to those of the other three models at the same offsets (Fig.~\ref{RATRAN_L1517B}). At offsets with no detected emission (most offsets $>$ 30$''$), all four models predict some weak emission within the rms limits, which cannot be directly addressed in the current paper. Future observations of higher sensitivity will be able to reveal such emission.

We conclude that for L1517b, both the static and LP models are able to reproduce the observed peak intensities, line profiles, and spatial distribution of the emission equally well, and therefore we cannot clearly discriminate them. The QE-BES model underestimates the emission at most offsets for $>$ 50\%, while the SIS appears to be the least suitable among the models, and therefore can be safely excluded. 
\\
\vspace{0.2cm}

\begin{figure*}[ht]
\begin{center}
\includegraphics[scale=0.5]{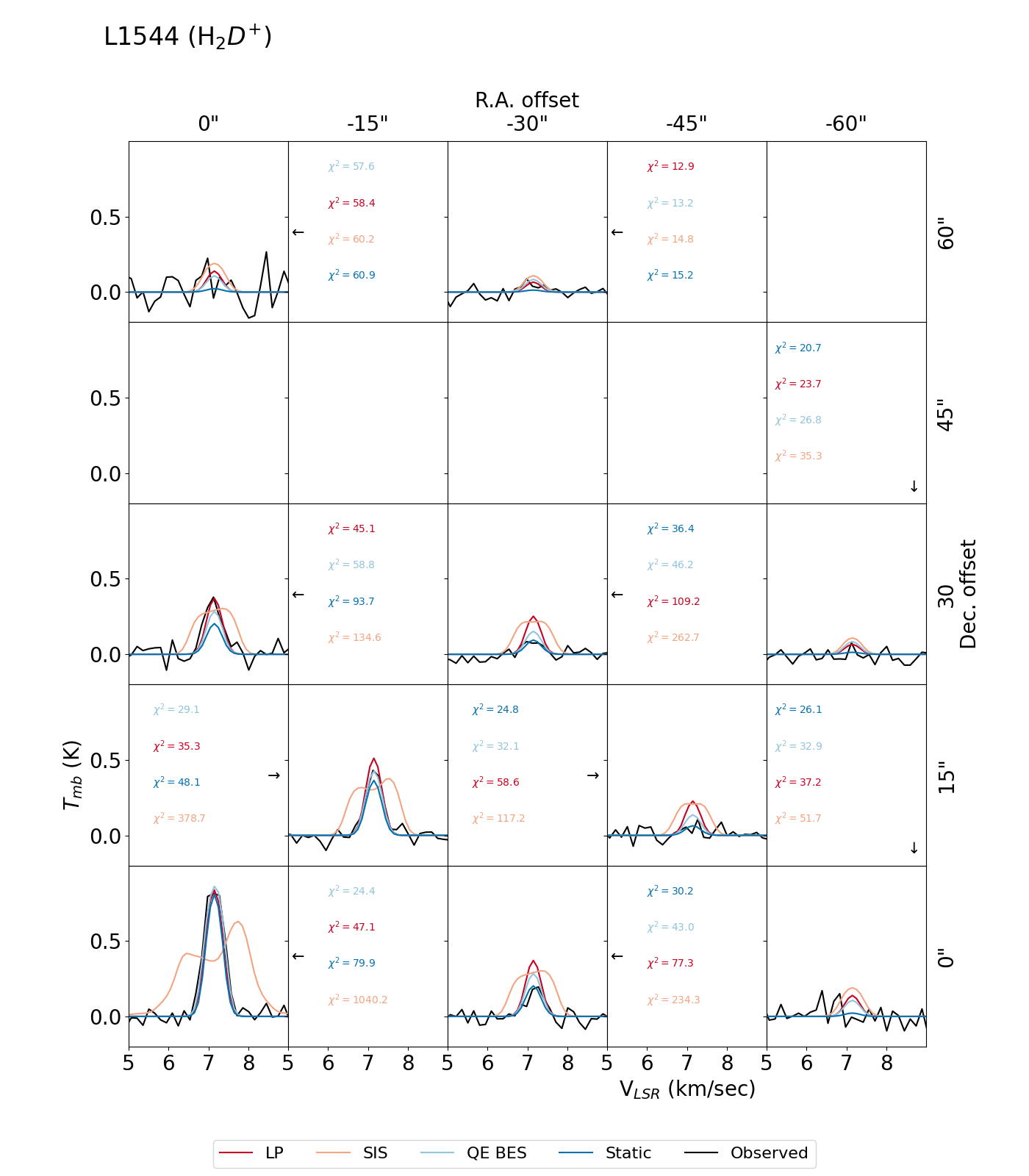}
\end{center}
\caption{As Fig.~\ref{RATRAN_L183}, but for L1544.}
\label{RATRAN_L1544}
\end{figure*}

{\bf{L1544:}} As seen in Figure~\ref{RATRAN_L1544} and Figure~\ref{RATRAN_L1544_ori} for L1544, both the QE-BES and LP models can reproduce the central (0$''$, 0$''$) observed peak intensity of H$_{2}$D$^{+}$ and line profile with $<$ 3\% deviation, followed by the static model ($\sim$ 8\%), while the SIS model predicts $\sim$55\% lower intensity at the source velocity.

When we examine the modelled versus the observed emission globally, we see that the LP model performs better in reproducing the observed peak intensities of detected emission with absolute deviations of 3-23\% at offsets within 30$''$. For the same grid of offsets, the QE-BES model is the next most suitable model showing deviations of 0.1-52\%, with the 0.1\% only towards the central position (0$''$, 0$''$), followed by the static model with deviations of 8-70\%. Lastly, the SIS model does a fairly good job in reproducing the peak intensities with deviations of 2-50\%. When we examine the line shape (profile and spatial distribution) of the modelled emission though, it becomes apparent that the SIS model fails to reproduce the observations globally. This failure is clearly depicted by the resulting $\chi^2$, which accounts for all velocity channels of the emission (not just the central peak), and is found to be systematically and significantly higher than the rest of the models at most positions where emission is present (e.g., up to 1000 at 0$''$ offset, compared to $<$ 100 for the rest of the models; Fig.~\ref{RATRAN_L1544}). 
The situation is different at offsets with no detected emission (i.e., most offsets $>$ 45$''$), where the static model is the only one to predict no emission, while all dynamical models predict weak emission but mostly within the rms limits. This discrepancy can be addressed in the future by obtaining higher sensitivity observations, which can reveal the predicted weak emission of the dynamical models. If the static model predictions are correct and there is indeed no emission towards the outer parts of the core, this result can be explained by the more realistic approach on density and temperature profiles, for this core, which are based on dust observations and not simulations.

\begin{figure*}[ht]
\begin{center}
\includegraphics[scale=0.5]{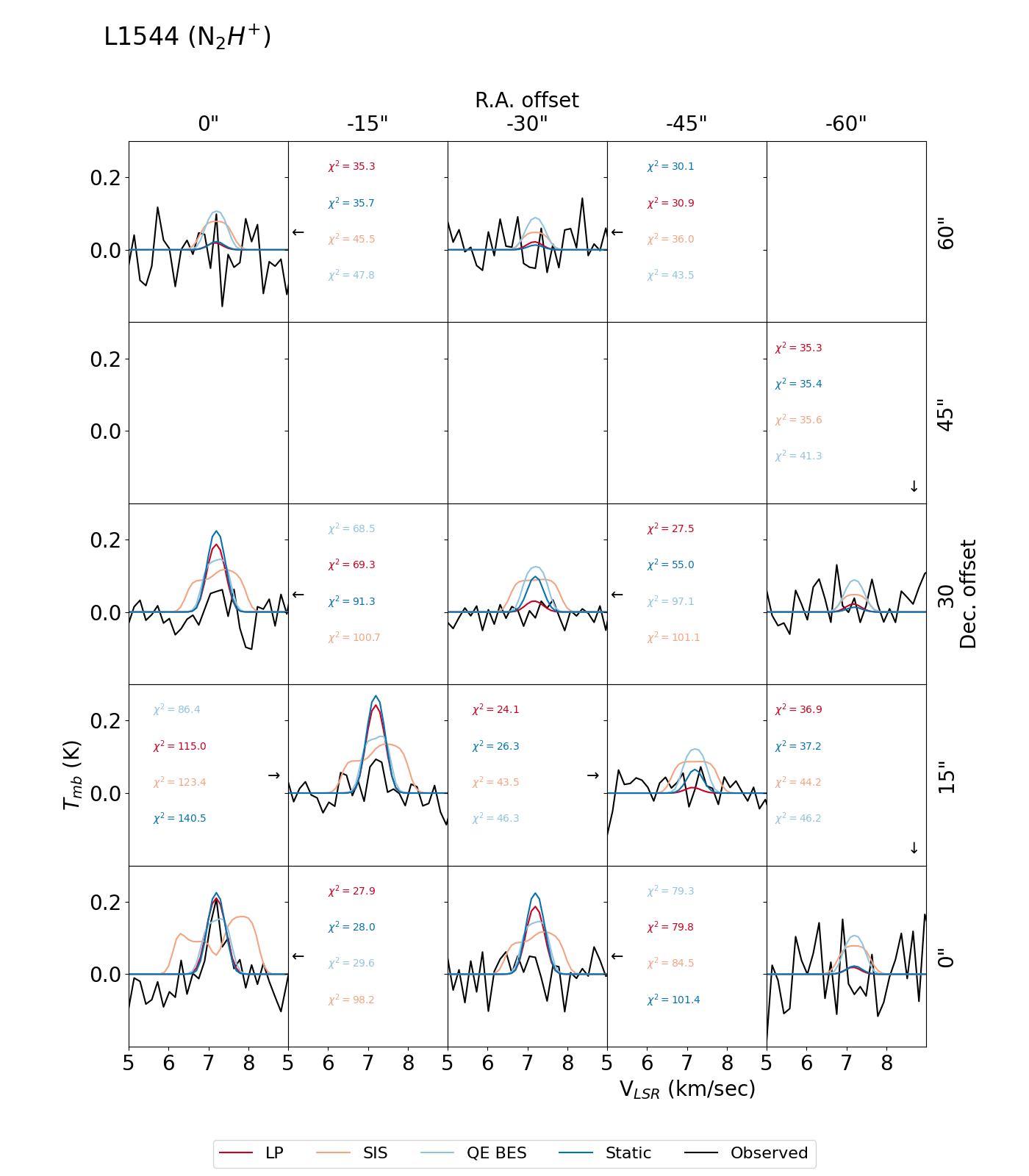}
\end{center}
\caption{As Fig.~\ref{RATRAN_L183}, but for N$_{2}$H$^{+}$ towards L1544.}
\label{RATRAN_L1544_n2hp}
\end{figure*}

The situation is less clear when simulating the N$_{2}$H$^{+}$ 4-3 emission. As we see in Figure~\ref{RATRAN_L1544_n2hp} and Figure~\ref{RATRAN_L1544_n2hp_ori}, all models apart from the SIS model can reproduce the observed peak intensity and line profile at the (0$''$, 0$''$) and (15$''$, 15$''$) offsets (where emission is present). The QE-BES model is also tested by \citet{Redaelli2019} towards L1544, and is found to reproduce the N$_{2}$H$^{+}$ 1-0 and 3-2 transitions at the central dust peak position reasonably well. All models in our study systematically and significantly overestimate the emission by an order of magnitude at all other inner positions and at offsets of 30$''$ regardless of the direction. Therefore, no model is able to explain the observed N$_{2}$H$^{+}$ 4-3 spatial distribution towards L1544 globally. 

We conclude that the LP model performs better in reproducing the observed line profiles and intensities of H$_{2}$D$^{+}$ globally for L1544, while the SIS model appears to be the least suitable among the models. 
\\

\vspace{0.2cm}

\begin{figure*}[ht]
\begin{center}
\includegraphics[scale=0.5]{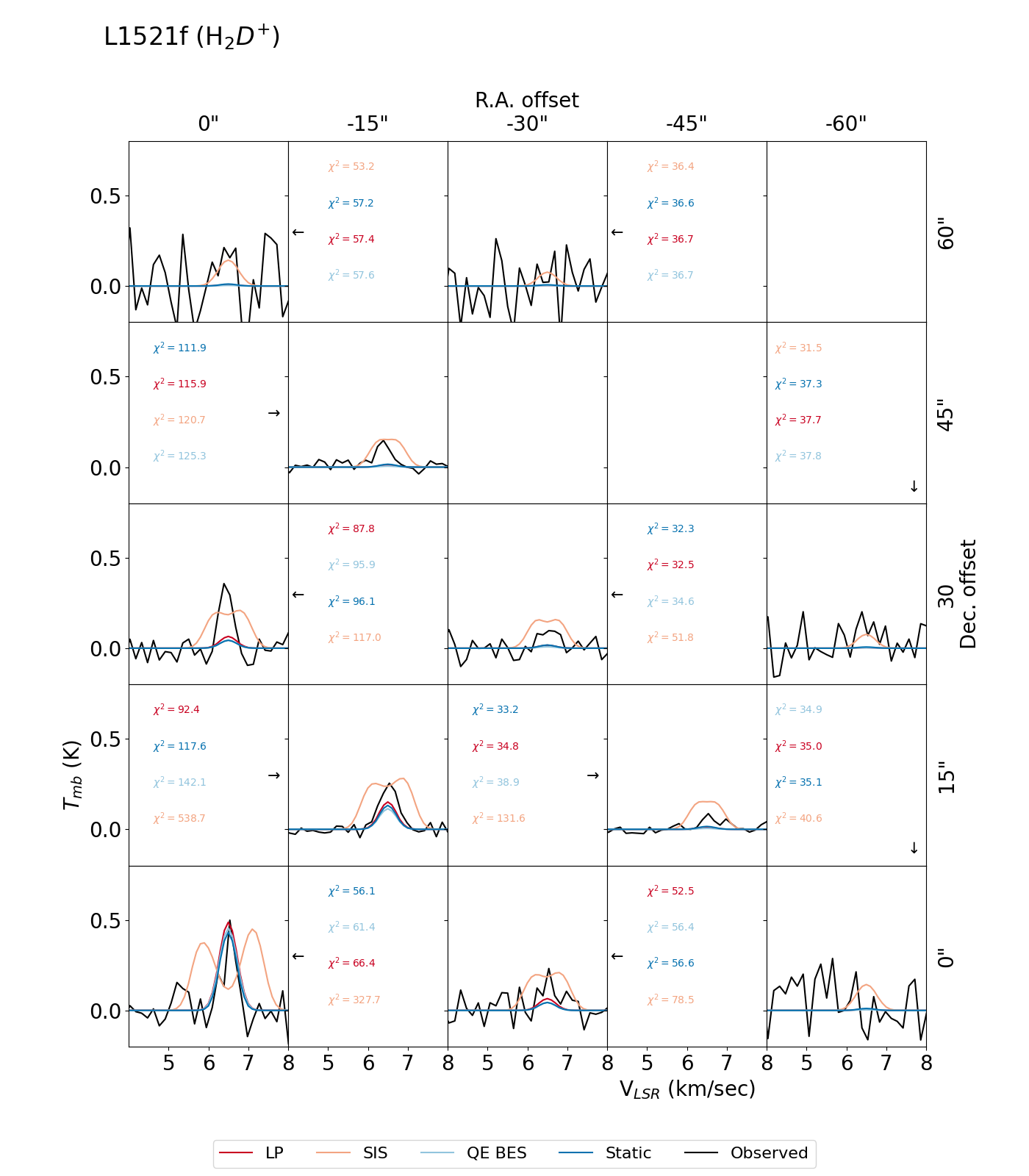}
\end{center}
\caption{As Fig.~\ref{RATRAN_L183}, but for the protostellar core L1521f.}
\label{RATRAN_L1521f}
\end{figure*}

{\bf{L1521f:}} For this core, as seen in Figure~\ref{RATRAN_L1521f} and Figure~\ref{RATRAN_L1521f_ori}, both QE-BES and LP models can reproduce the central (0$''$, 0$''$) observed peak intensity and line profile of H$_{2}$D$^{+}$ with $<$ 12\% deviation, followed by the static model ($\sim$ 17\%), while the SIS model predicts $\sim$75\% lower intensity at the source velocity.

When we examine the modelled versus the observed emission globally, we see that apart from two offsets at (0$''$, 0$''$) and (-15$''$, -15$''$), where the LP, QE-BES, and static models can reproduce the observed peak intensity similarly and with $<$ 35\% deviations, they systematically and significantly underestimate the peak intensity between 52\% and 96\% at the rest of the offsets with detections. The SIS is the only model providing a good representation of the observed peak intensity for the majority of offsets within 45$''$ from the center, with deviations of 8-40\%. Nevertheless, similarly to the other cores, the SIS model fails to reproduce the line profiles resulting in very high $\chi^2$ ($>$ 100), when all velocity channels are taken into account (Figure~\ref{RATRAN_L1521f}). The exceptions to this trend are the two observed offsets at 45$''$ north, where the SIS model appears to be the only one to predict the observed emission.

\begin{figure*}[ht]
\begin{center}
\includegraphics[scale=0.5]{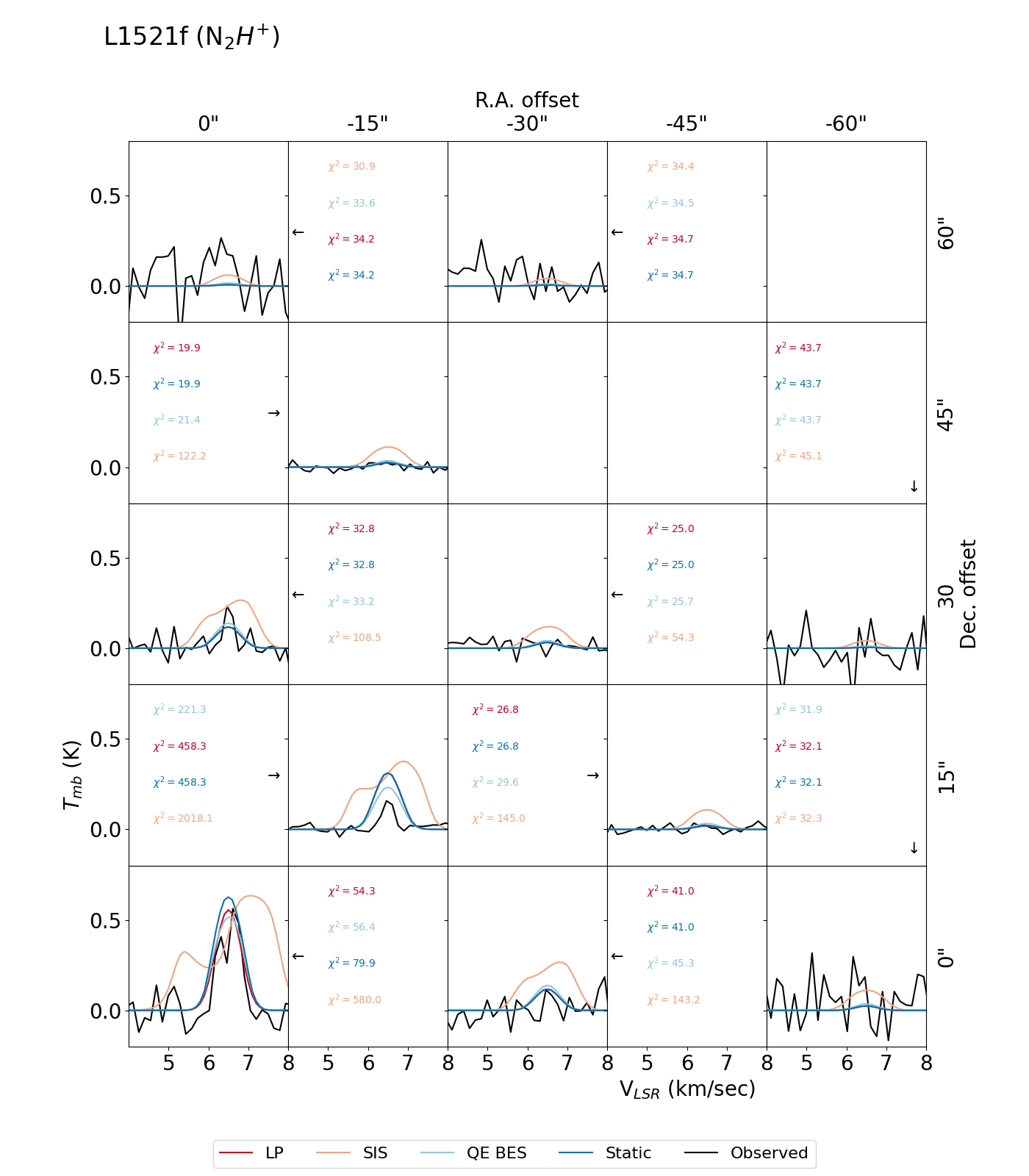}
\end{center}
\caption{As Fig.~\ref{RATRAN_L183}, but for N$_{2}$H$^{+}$ towards L1521f.}
\label{RATRAN_L1521f_n2hp}
\end{figure*}

When we examine the N$_{2}$H$^{+}$ emission in Figure~\ref{RATRAN_L1521f_n2hp} and Figure~\ref{RATRAN_L1521f_n2hp_ori} we see that all models apart from the SIS model can reproduce the central peak emission within 10\%, but overestimate the emission at 15$''$ offsets by $>$ 60\%, while they are all consistent with showing no emission at larger offsets ($>$ 30$''$). The amount of available information connected to N$_{2}$H$^{+}$ is not sufficient to draw clear conclusions.

We find that none of the models perform well in reproducing the observed emission globally, both line profiles, and the intensity and distribution of H$_{2}$D$^{+}$, towards L1521f. This outcome may be due to the fact that L1521f is a more evolved (protostellar) object compared to the rest of the sample of prestellar cores, and therefore it is characterized by more complex dynamics and physical structure. Protostellar cores, for example, are known to drive powerful outflows, and therefore the tested dynamical models that account only for infall motions may not be expected to reproduce the observations consistently with what we see. The situation is similar for the model of the static envelope. The good quality fit of the SIS model at the reported offsets may indicate that the dynamical evolution from a prestellar to a protostellar core is better represented by the SIS model rather than the LP model in opposition to what we see for prestellar cores. Given the poor ability of the model to reproduce the emission at a more global scale, however, we consider this alignment to be rather coincidental. The less problematic fit we get towards N$_{2}$H$^{+}$ indicates that H$_{2}$D$^{+}$ is dynamically decoupled from N$_{2}$H$^{+}$, with the former requiring a more complex structure.

\subsubsection{Summary of RATRAN results}

\begin{figure*}[ht]
\begin{center}
\includegraphics[scale=0.3]{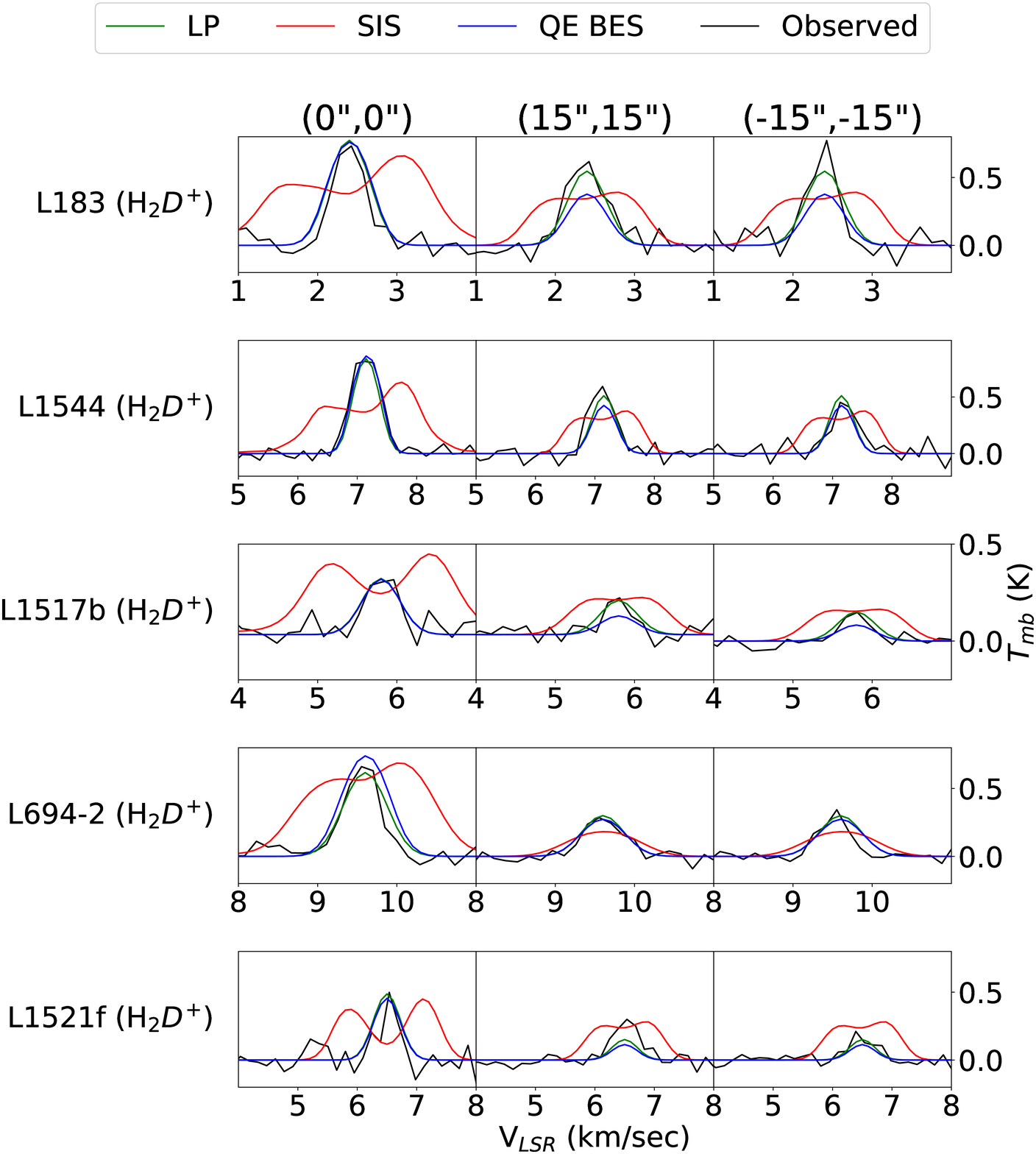}
\end{center}
\caption{Observed line profiles of H$_{2}$D$^{+}$, overplotted with the modelled ones for LP, SIS and QE-BES models from RATRAN at selected offsets: (0'',0''), (15'',15'') and (-15'',-15'').}
\label{RATRAN_all}
\end{figure*}

We find that the LP flow can reproduce the observed line profile and intensity of H$_{2}$D$^{+}$ fairly well on a global scale (most offsets), and appears to be superior to the static, SIS, and QE-BES models for two out of the four prestellar cores, L183 and L1544. The influence of the abundance profiles on the simulated line intensity makes clear differentiation among those models challenging. For L1517b, the situation is less clear, with the static sphere reproducing the observations equally as well as the LP model. For L694-2, we cannot clearly distinguish between the LP, QE-BES, and static sphere models. On the other hand, all models fail to reproduce the observed line profiles and distribution of H$_{2}$D$^{+}$ towards the protostellar core L1521f, indicating that the predicted model dynamics taking place during the prestellar phase are not suitable for a more evolved object, where outflows and a central heating source are present. These findings are summarized in Figure~\ref{RATRAN_all} at offsets along a cut in the South-West/North-East direction.

In addition, our modelling clearly demonstrates that the SIS dynamical model fails to reproduce the observed line profile for all cores, independent of the initial abundances. While the density profile of the SIS model shows a similar slope to the other two models (all models follow $\sim$ 1/{\it{r}}$^{2}$ profile), the exact density profile results in a smaller mass in the same volume of gas compared to the other two dynamical models. As a result, the overall density distribution of the SIS model is not sufficient to excite the H$_{2}$D$^{+}$ transition in the amount required to match the observations. This behaviour is clearly seen in the abundance calculations, where we find the H$_{2}$D$^{+}$ abundance needs to be 2-4 orders of magnitude higher for the SIS model compared to the other two models for all sources. The observed differences on the abundances can be depicted in the resulting column densities as well. For a given volume density an increase of a factor of 100 in the abundance corresponds to an increase in the molecular density by the same factor. For example, in the case of the H$_{2}$D$^{+}$ emission towards L1544 the LP model results in a column density of 1.6$\times$10$^{13}$cm$^{-2}$, while the SIS model results in a column density of 3.2$\times$10$^{15}$cm$^{-2}$. Referring back to the RADEX analysis (Sect.~\ref{rad_results}), we find N(H$_{2}$D$^{+}$) = 2.0$\times$10$^{13}$cm$^{-2}$ which is in very good agreement with the LP model but 2 orders of magnitude lower than the column density from the SIS model. This result demonstrates once more the poor performance of the SIS model compared to the LP and QE-BES models.

Moreover, the QE-BES model gives a core mass of $\sim$10~\msun, the LP model of $\sim$ 27~\msun~and the SIS model of 4~\msun, while the masses of the cores derived from dust emission are 4-10~\msun. We conclude that the differences in the modelled versus observed masses affect our abundance determinations, resulting in higher or lower H$_{2}$D$^{+}$ abundance depending on whether the modelled mass for a defined volume is higher or lower, respectively, than observed. To examine the effects of the assumed mass in the reproduction of the observations we notice that L1517b is as massive as 4~\msun (i.e., the same mass as the SIS model has), but the SIS model still fails to reproduce the observations; therefore, our conclusion on excluding the SIS model as a possible dynamical model to explain the observed H$_{2}$D$^{+}$ emission is independent from the mass assumptions. Similarly, the LP model has a mass $\sim$ 3 times higher than the core masses in our sample, yet it is able to reproduce the observations at all prestellar cores. We therefore conclude that although the mass differences introduce higher uncertainties in the abundance determinations, they do not affect the line profiles. The line profiles depend on the velocity structure of the models, making our analysis and conclusions in differentiating between the dynamical models robust. Referring to the results presented in Sect.~\ref{line_anal}, we found that there are significant non-thermal contributions towards all cores, which are traditionally attributed to turbulence and magnetic pressure. The ability of the LP flow to reproduce the observations without taking those phenomena into account, unlike the QE-BES model (i.e., turbulence), indicates that they are not significant in driving the core dynamics.

\section{Discussion}

In this section, we discuss the main findings of this paper. In particular, we attempt to explain the observed line broadening, the presence of N$_{2}$H$^{+}$ towards only two of the five cores in our sample, and, lastly, the core collapse dynamics, which is a debatable topic both in theory and observations.  

\subsection{Line broadening}

We first discuss our results from the line profile analysis presented in Sect.~\ref{line_anal}. We find that the line width of both N$_{2}$H$^{+}$ and H$_{2}$D$^{+}$ cannot be explained by thermal broadening alone, and the non-thermal contributions are variable with offsets and overall higher for N$_{2}$H$^{+}$ than H$_{2}$D$^{+}$ (Sect.~\ref{line_anal}). The observed variations with offsets of the non-thermal contributions are as high as $\sim$ 130\% and beyond the uncertainties. Traditionally, non-thermal motions are explained by presence of magnetic pressure or turbulence. In this study, we show that the LP flow can reproduce the line width of the H$_{2}$D$^{+}$ observations in multiple offset locations (e.g., see Figure~\ref{RATRAN_L183}), which by definition (free-fall) does not take into account turbulence or magnetic fields; Therefore, the non-thermal broadening can be merely due to infall/compression motions, characteristic of the dynamical structure of prestellar cores.

\subsection{N$_{2}$H$^{+}$ 4-3 emission as an evolution tracer}

Here we discuss the origin of the N$_{2}$H$^{+}$ 4-3 emission. We detect the N$_{2}$H$^{+}$ 4-3 transition towards two out of the five cores in our sample, in contrast to the ground ortho-H$_{2}$D$^{+}$ 1$_{10}$-1$_{11}$ emission which is present towards all cores. In addition, H$_{2}$D$^{+}$ is more spatially extended than N$_{2}$H$^{+}$, which is traced only within 15$\arcsec$ of the core centers.

To understand the spatial differences, and the presence or absence of these species towards these cores, we consider the critical densities of the observed transitions, calculated based on the available molecular data. In particular, the observed ortho-H$_{2}$D$^{+}$ 1$_{10}$-1$_{11}$ at 372.4~GHz is characterized by a critical density $n_{cr}$ $=$ 1.2$\times$10$^{5}$~cm$^{-3}$ at 10~K, while the observed N$_{2}$H$^{+}$ 4-3 transition at 372.67 GHz is characterized by a critical density n$_{cr}$ $=$ 7.7$\times$10$^{6}$~cm$^{-3}$ at 10~K, which is more than an order of magnitude higher. We also note that the N$_{2}$H$^{+}$ 1-0 transition at 93.17 GHz, which has a comparable critical density to the ground H$_{2}$D$^{+}$ transition (n$_{cr}$ $\sim$ 1.4$\times$10$^{5}$~cm$^{-3}$ at 10~K), has been previously detected towards all cores in our sample \citep{Crapsi2005}. We conclude that the observed spatial differences between the two species are due to the higher densities traced by N$_{2}$H$^{+}$ 4-3, which are unsurprisingly located towards the more central regions of the cores (see, e.g., density profiles in Sect.~\ref{stat_sph}). Indeed, this interpretation is supported by the observed lower J transitions 3-2 and 1-0 of N$_{2}$H$^{+}$ towards L1544 \citep{Redaelli2019}, which are found to be up to a factor of 4 more extended compared to the 4-3 transition.

The absence of N$_{2}$H$^{+}$ 4-3 emission for L694-2 and L1517B can be explained by their lower central densities of the order $\sim$10$^{5}$~cm$^{-3}$, i.e., an order of magnitude lower than the critical density of N$_{2}$H$^{+}$ 4-3. Similarly, the presence of this transition can be explained for L1521f and L1544, given their higher central densities of $\sim$10$^{6}$~cm$^{-3}$. Along this line of reasoning, the non-detection of N$_{2}$H$^{+}$ 4-3 towards L183 is more puzzling, given that its central density is found to be similar to those of L1521f and L1544, indicating that its central density may be overestimated. 

We conclude that N$_{2}$H$^{+}$ 4-3 could be used as a chemical clock, in the sense that it traces a fairly high density environment achievable only in more evolved prestellar/early protostellar core phases. Our conclusion is supported by the presence of this emission towards the only protostellar (more evolved) core in our sample (L1521f). This approach would mean that L694-2 and L1517B are less-evolved cores.

\subsection{Differentiating between the theories of collapse}

Here, we discuss our findings from the RATRAN analysis presented in Sect.~\ref{mod_ratran}, where we compare the performance of the different dynamical models of collapse. We find that the LP model can mostly reproduce the observed line profiles and spatial distribution of H$_{2}$D$^{+}$, followed in suitability by the QE-BES model. Meanwhile, the SIS model fails to reproduce the line width globally, predicting a very prominent double peak line profile towards the central position of all cores, in contrast to the observed single peak. Although the LP model seems to be superior to the QE-BES model for at least two of the prestellar cores in the sample, we cannot distinguish between the two models with high confidence. The widths of the line profiles can be mostly reproduced by both models, while in addition the emission resulting from adopting LP conditions is closer to the observed line intensity on a more global scale compared to the QE-BES model. All models fail to reproduce the emission from the one protostellar core in our sample, L1521f, which is indicative of the more complex structure of a protostellar object with respect to a prestellar core, i.e., presence of winds and outflows, disc, internal radiation, different grain sizes, and chemistry. 

The abundance profiles have a significant effect on the simulated line intensity. We consider the abundance derivation from the dynamical models to be less accurate compared to those of the static model, due to the differences in the total core mass that each model implies. As we explained in Sect.~\ref{rat_dyn} the exact density distribution over a specific volume (i.e., different mass) affects the degree of excitation of H$_{2}$D$^{+}$ and therefore its line intensity. For example, to match the H$_{2}$D$^{+}$ line intensity of L1544 ($\sim$8~\msun), the H$_{2}$D$^{+}$ abundance had to be lower in the case of the LP model ($>$ 10~\msun) compared to those of the QE-BES ($\sim$10~\msun) or SIS ($<$10~\msun) models. The static model, on the other hand, is based on continuum observations and therefore it accounts for the observed total mass of each individual core. Breaking the observed discrepancy between masses derived from continuum and gas observations will allow proper differentiation between the LP and QE-BES models. To achieve this goal, the dynamical models need to be ``tailored'' to match the observed mass per individual core. Such a procedure, however, requires prior knowledge of the exact evolutionary status of each core, and is beyond the scope of this work.

Unlike the line intensity, which in the optically thin regime ($\tau$ $<$ 1; see Table~\ref{opacities_table}) highly depends on the adopted abundances, the shape of the line profile for a given abundance profile depends mostly on the velocity structure, which is unique per dynamical model. Therefore, we can safely exclude the SIS model as a possible dynamic model to explain H$_{2}$D$^{+}$ emission for all cores. The situation is similar for N$_{2}$H$^{+}$, where present. Our qualitative results remain even when a constant abundance profile is adopted for each core.

To understand the differences in the shape of the line profile between the dynamical models, we explore the physical location of the H$_{2}$D$^{+}$ emission and we compare it to other tracers from previous work. The observed 1$_{10}$-1$_{11}$ ground-state transition of ortho-H$_{2}$D$^{+}$ at 372.4~GHz is characterized by a critical density $n_{cr}$ $=$ 1.2$\times$10$^{5}$~cm$^{-3}$ at 10~K, serving as a rather dense gas tracer. Looking at Figure~\ref{density_velocity_keto}, we see that in the case of the SIS model, the observed H$_{2}$D$^{+}$ transition would better probe the innermost regions ($<$ 1000 au). In the case of the QE-BES and LP models, however, it traces a greater volume of the gas ($<$ 10,000 au), but not the outer layers up to 100,000 au. The obvious mismatch between the predicted and observed H$_{2}$D$^{+}$ line profiles produced assuming an SIS collapse suggests that high-velocity inflow in the inner parts of the cores fails to explain the H$_{2}$D$^{+}$ emission. The line profiles produced by the LP and QE-BES models can mostly reproduce the observed line profiles, suggesting that more moderate contraction motions in the inner regions are more applicable (e.g., by three orders of magnitude difference in velocity). \citet{Keto2015} performed a similar analysis for H$_{2}$O and C$^{18}$O towards L1544 and found that the QE-BES model reproduces the shape and intensity of the line profiles for both species, while the LP model produced a more complex profile for the C$^{18}$O similar to what we see for H$_{2}$D$^{+}$ in the SIS model in this study. To understand better the observed differences in the line profiles produced from different models, we need to consider the origin of the emission. In particular, the C$^{18}$O transition presented in \citet{Keto2015} is characterized by a critical density about two orders of magnitude lower than that of the H$_{2}$D$^{+}$ line examined here. Therefore, it is better at tracing less dense gas compared to the ground H$_{2}$D$^{+}$ transition. Looking at Figure~\ref{density_velocity_keto}, we see that the outer regions of the core in the LP hypothesis are characterized by much higher velocities, explaining the broad wings modelled in C$^{18}$O. In conclusion, the QE-BES model can explain the observed line profiles of C$^{18}$O, H$_{2}$D$^{+}$ and H$_{2}$O, both in shape and intensity, for the central position of L1544, while the LP and SIS models cannot simultaneously explain the observed emission. 

In addition, we examine the performance of the static sphere. We find that the static model, which is based on the continuum observations of each core, can also reproduce the observations fairly well and similarly to the LP model for at least one core (L1517b), demonstrating the importance of the adopted abundance profiles. This model, however, does not take into account infall motions. With only a single available transition of H$_{2}$D$^{+}$, a more accurate determination of the abundance profile and its proper coupling to the specific core dynamics is very challenging.

We conclude that coupling gas core dynamics and chemistry with dust properties is crucial to differentiating between the QE-BES and LP models, and therefore, should be considered in future studies. In particular, chemical modelling is essential to constrain the chemical age of each core, which will subsequently lead to the coupling of the dynamical models to the observed mass of each individual core.

To be able to draw stronger conclusions it is necessary to expand this analysis to a larger sample of known prestellar and protostellar cores. If the LP model is confirmed towards a larger sample of cores, its specific initial conditions indicate that the prestellar cores may form by compression or accretion of gas from larger scales ($>$ 10000 au). On the other hand, if the QE-BES model is confirmed, it means that quasi hydrostatic cores can exist within turbulent ISM, which would be against predictions from numerical simulations \citep{Tilley2004}.

\section{Summary-Conclusions}

\label{conclusions}

This paper presents an analysis of H$_{2}$D$^{+}$ 1$_{10}$-1$_{11}$ and N$_{2}$H$^{+}$ 4-3 towards one protostellar and four prestellar cores using 2\arcmin$\times$2\arcmin~maps from JCMT. We present a basic line analysis for all positions in the maps where emission is detected. We estimated the average physical conditions ($T_{KIN}$ and $N_{COL}$) towards the center of the cores using the non-LTE radiative transfer code, RADEX, incorporating the latest collisional data available for both species. The derived kinetic temperatures per core are used to assess the thermal contributions on the observed line widths. Finally, we tested three different dynamical scenarios, using the advanced radiative transfer code RATRAN to model the spatial distribution of the gas traced by the optically thin H$_{2}$D$^{+}$ 1$_{10}$-1$_{11}$ and N$_{2}$H$^{+}$ 4-3 line emission and reproduce the sizes of the observed maps. The latter allowed us to not only test which of the available dynamical scenarios is the most favorable for the prestellar core phase but also to derive abundance profiles of the observed species. Our main conclusions are summarized as follows. 

\begin{itemize}
\item Thermal broadening alone is not sufficient to explain the line widths of H$_{2}$D$^{+}$ and N$_{2}$H$^{+}$ measured towards the majority of the positions in the maps with clear detections. This finding suggests that non-thermal effects play a significant role at the very early stages of star formation. The non-thermal contributions can be merely attributed to the infall/compression motions predicted by the LP flow. Therefore, in that scenario, turbulent and magnetic pressure contributions are not required to explain the observed line width.
\item The non-thermal contributions are generally found to be stronger in the innermost parts of L694-2 and L1544 ($<15''$; with the exceptions of few offsets which are similar) and are found to be higher for N$_{2}$H$^{+}$ than H$_{2}$D$^{+}$. We report variations to non-thermal contributions up to a factor of 2.5 at all cores for different offsets and beyond the uncertainties. The lack of sufficient N$_{2}$H$^{+}$ detections at offset positions does not allow a meaningful investigation on non-thermal variations with radius for that emission.
\item N$_{2}$H$^{+}$ 4-3 was only detected towards two of the cores, L1521f and L1544. Given that L1521f is known to be a VeLLO, and therefore more evolved than a prestellar core, and given that both cores are located at the same distance, these detections indicate that L1544 may also be more evolved compared to the rest of the sample. In that case, N$_{2}$H$^{+}$ 4-3 can be used as a chemical clock, tracing the higher densities achieved during core evolution.
\item The RADEX analysis results in similar mean column densities of H$_{2}$D$^{+}$ and N$_{2}$H$^{+}$ ($\sim$10$^{13}$~cm$^{-2}$) for L1521f and L1544. 
\item The RATRAN analysis reveals that the LP flow is superior in reproducing the H$_{2}$D$^{+}$ spatial distribution and line profiles towards two of the prestellar cores, while for the other two it is followed by the QE-BES and static scenarios, without being able to differentiate clearly between those. 
\item The SIS model fails to reproduce the shape and strength of the line profiles for the majority of the positions towards all cores and, therefore, does not appropriately describe the dynamical nature of these cores.
\item The RATRAN analysis reveals that none of the scenarios can reproduce the H$_{2}$D$^{+}$ emission towards the protostellar core, L1521f, indicating that more complex dynamical and chemical structures need to be considered, including outflow activity and radiation or wind pressure from the central heating source.
\end{itemize}

We explore the most prominent dynamical scenarios based on spherical contraction to explain the H$_{2}$D$^{+}$ and N$_{2}$H$^{+}$ (where present) emission towards four prestellar and one protostellar core. With the observations and sample in hand we cannot clearly differentiate between the LP flow and QE-BES models, but can clearly exclude the SIS model. Our conclusions remain consistent after adopting a jump-drop or a constant abundance profile. To differentiate between the other collapse scenarios is necessary to construct dynamical models adapted to the individual observed core mass. In addition, testing the models will benefit from transitions of more molecular species in both the optically thin and optically thick regimes \citep[e.g.,][]{vanderTak2005} and a larger sample of cores. Such data will inform us if the gas collapses as a free-fall (LP) or if turbulence is indeed crucial for the gas dynamics of cores.

In future, non-spherical models, such as oblate ones resulting from ambipolar diffusion, should also be considered. Applying chemical modelling `tailored' to the physical structure of each source to fit the H$_{2}$D$^{+}$ and N$_{2}$H$^{+}$ observed abundances, will provide us with the age \citep[e.g.,][]{Brunken2014} and therefore the evolutionary status of each core, allowing us to link the chemical to the dynamical evolution of prestellar cores overall. Lastly, high angular resolutions to reach the emission in the inner 100 au (e.g., ALMA) of the prestellar cores should prove to be a powerful tool to explore the gas dynamics, but the observed chemical inner holes towards those objects and the heavy freeze-out, make such observations very challenging.    









\begin{acknowledgements}

We would like to thank the anonymous referee for providing helpful comments and suggestions that improved the paper. EK is funded by the STFC (ST/P00041X/1). We thank Tim van Kempen and Werner Salomons for useful contributions during data reduction and analysis. 
 
\end{acknowledgements}

\bibliographystyle{aa} 
\bibliography{ref}

\begin{thebibliography}{66}
\expandafter\ifx\csname natexlab\endcsname\relax\def\natexlab#1{#1}\fi

\bibitem[{{Andr{\'e}}(2017)}]{Andre2017}
{Andr{\'e}}, P. 2017, Comptes Rendus Geoscience, 349, 187

\bibitem[{{Andr{\'e}} {et~al.}(2010){Andr{\'e}}, {Men'shchikov}, {Bontemps},
  {K{\"o}nyves}, {Motte}, {Schneider}, {Didelon}, {Minier}, {Saraceno},
  {Ward-Thompson}, {di Francesco}, {White}, {Molinari}, {Testi}, {Abergel},
  {Griffin}, {Henning}, {Royer}, {Mer{\'\i}n}, {Vavrek}, {Attard},
  {Arzoumanian}, {Wilson}, {Ade}, {Aussel}, {Baluteau}, {Benedettini},
  {Bernard}, {Blommaert}, {Cambr{\'e}sy}, {Cox}, {di Giorgio}, {Hargrave},
  {Hennemann}, {Huang}, {Kirk}, {Krause}, {Launhardt}, {Leeks}, {Le Pennec},
  {Li}, {Martin}, {Maury}, {Olofsson}, {Omont}, {Peretto}, {Pezzuto}, {Prusti},
  {Roussel}, {Russeil}, {Sauvage}, {Sibthorpe}, {Sicilia-Aguilar}, {Spinoglio},
  {Waelkens}, {Woodcraft}, \& {Zavagno}}]{Andre2010}
{Andr{\'e}}, P., {Men'shchikov}, A., {Bontemps}, S., {et~al.} 2010, \aap, 518,
  L102

\bibitem[{{Arzoumanian} {et~al.}(2011){Arzoumanian}, {Andr{\'e}}, {Didelon},
  {K{\"o}nyves}, {Schneider}, {Men'shchikov}, {Sousbie}, {Zavagno}, {Bontemps},
  {di Francesco}, {Griffin}, {Hennemann}, {Hill}, {Kirk}, {Martin}, {Minier},
  {Molinari}, {Motte}, {Peretto}, {Pezzuto}, {Spinoglio}, {Ward-Thompson},
  {White}, \& {Wilson}}]{Arzoumanian2011}
{Arzoumanian}, D., {Andr{\'e}}, P., {Didelon}, P., {et~al.} 2011, \aap, 529, L6

\bibitem[{{Beichman} {et~al.}(1986){Beichman}, {Myers}, {Emerson}, {Harris},
  {Mathieu}, {Benson}, \& {Jennings}}]{Beichman1986}
{Beichman}, C.~A., {Myers}, P.~C., {Emerson}, J.~P., {et~al.} 1986, \apj, 307,
  337

\bibitem[{{Bergin} \& {Tafalla}(2007)}]{Bergin2007}
{Bergin}, E.~A. \& {Tafalla}, M. 2007, \araa, 45, 339

\bibitem[{{Black} {et~al.}(1990){Black}, {van Dishoeck}, {Willner}, \&
  {Woods}}]{Black1990}
{Black}, J.~H., {van Dishoeck}, E.~F., {Willner}, S.~P., \& {Woods}, R.~C.
  1990, \apj, 358, 459

\bibitem[{{Bonnor}(1956)}]{Bonnor1956}
{Bonnor}, W.~B. 1956, \mnras, 116, 351

\bibitem[{{Bourke} {et~al.}(2006){Bourke}, {Myers}, {Evans}, {Dunham},
  {Kauffmann}, {Shirley}, {Crapsi}, {Young}, {Huard}, {Brooke}, {Chapman},
  {Cieza}, {Lee}, {Teuben}, \& {Wahhaj}}]{Bourke2006}
{Bourke}, T.~L., {Myers}, P.~C., {Evans}, Neal~J., I., {et~al.} 2006, \apjl,
  649, L37

\bibitem[{{Broderick} \& {Keto}(2010)}]{Broderick2010}
{Broderick}, A.~E. \& {Keto}, E. 2010, \apj, 721, 493

\bibitem[{{Br{\"u}nken} {et~al.}(2014){Br{\"u}nken}, {Sipil{\"a}}, {Chambers},
  {Harju}, {Caselli}, {Asvany}, {Honingh}, {Kami{\'n}ski}, {Menten}, {Stutzki},
  \& {Schlemmer}}]{Brunken2014}
{Br{\"u}nken}, S., {Sipil{\"a}}, O., {Chambers}, E.~T., {et~al.} 2014, \nat,
  516, 219

\bibitem[{{Buckle} {et~al.}(2009){Buckle}, {Hills}, {Smith}, {Dent}, {Bell},
  {Curtis}, {Dace}, {Gibson}, {Graves}, {Leech}, {Richer}, {Williamson},
  {Withington}, {Yassin}, {Bennett}, {Hastings}, {Laidlaw}, {Lightfoot},
  {Burgess}, {Dewdney}, {Hovey}, {Willis}, {Redman}, {Wooff}, {Berry},
  {Cavanagh}, {Davis}, {Dempsey}, {Friberg}, {Jenness}, {Kackley}, {Rees},
  {Tilanus}, {Walther}, {Zwart}, {Klapwijk}, {Kroug}, \&
  {Zijlstra}}]{Buckle2009}
{Buckle}, J.~V., {Hills}, R.~E., {Smith}, H., {et~al.} 2009, \mnras, 399, 1026

\bibitem[{{Caselli} {et~al.}(2002){Caselli}, {Benson}, {Myers}, \&
  {Tafalla}}]{Caselli2002}
{Caselli}, P., {Benson}, P.~J., {Myers}, P.~C., \& {Tafalla}, M. 2002, \apj,
  572, 238

\bibitem[{{Caselli} {et~al.}(2019){Caselli}, {Pineda}, {Zhao}, {Walmsley},
  {Keto}, {Tafalla}, {Chac{\'o}n-Tanarro}, {Bourke}, {Friesen}, {Galli}, \&
  {Padovani}}]{Caselli2019}
{Caselli}, P., {Pineda}, J.~E., {Zhao}, B., {et~al.} 2019, \apj, 874, 89

\bibitem[{{Caselli} {et~al.}(2003){Caselli}, {van der Tak}, {Ceccarelli}, \&
  {Bacmann}}]{Caselli2003}
{Caselli}, P., {van der Tak}, F.~F.~S., {Ceccarelli}, C., \& {Bacmann}, A.
  2003, \aap, 403, L37

\bibitem[{{Caselli} {et~al.}(2008){Caselli}, {Vastel}, {Ceccarelli}, {van der
  Tak}, {Crapsi}, \& {Bacmann}}]{Caselli2008}
{Caselli}, P., {Vastel}, C., {Ceccarelli}, C., {et~al.} 2008, \aap, 492, 703

\bibitem[{{Chac{\'o}n-Tanarro} {et~al.}(2019){Chac{\'o}n-Tanarro}, {Pineda},
  {Caselli}, {Bizzocchi}, {Gutermuth}, {Mason}, {G{\'o}mez-Ruiz}, {Harju},
  {Devlin}, {Dicker}, {Mroczkowski}, {Romero}, {Sievers}, {Stanchfield},
  {Offner}, \& {S{\'a}nchez-Arg{\"u}elles}}]{ChaconTanarro2019}
{Chac{\'o}n-Tanarro}, A., {Pineda}, J.~E., {Caselli}, P., {et~al.} 2019, \aap,
  623, A118

\bibitem[{{Crapsi} {et~al.}(2005){Crapsi}, {Caselli}, {Walmsley}, {Myers},
  {Tafalla}, {Lee}, \& {Bourke}}]{Crapsi2005}
{Crapsi}, A., {Caselli}, P., {Walmsley}, C.~M., {et~al.} 2005, \apj, 619, 379

\bibitem[{{Crapsi} {et~al.}(2004){Crapsi}, {Caselli}, {Walmsley}, {Tafalla},
  {Lee}, {Bourke}, \& {Myers}}]{Crapsi2004}
{Crapsi}, A., {Caselli}, P., {Walmsley}, C.~M., {et~al.} 2004, \aap, 420, 957

\bibitem[{{Crapsi} {et~al.}(2007){Crapsi}, {Caselli}, {Walmsley}, \&
  {Tafalla}}]{Crapsi2007}
{Crapsi}, A., {Caselli}, P., {Walmsley}, M.~C., \& {Tafalla}, M. 2007, \aap,
  470, 221

\bibitem[{{Crutcher} {et~al.}(2004){Crutcher}, {Nutter}, {Ward-Thompson}, \&
  {Kirk}}]{Crutcher2004}
{Crutcher}, R.~M., {Nutter}, D.~J., {Ward-Thompson}, D., \& {Kirk}, J.~M. 2004,
  \apj, 600, 279

\bibitem[{{Currie} {et~al.}(2014){Currie}, {Berry}, {Jenness}, {Gibb}, {Bell},
  \& {Draper}}]{Currie2014}
{Currie}, M.~J., {Berry}, D.~S., {Jenness}, T., {et~al.} 2014, Astronomical
  Society of the Pacific Conference Series, Vol. 485, {Starlink Software in
  2013}, ed. N.~{Manset} \& P.~{Forshay}, 391

\bibitem[{{Di Francesco} {et~al.}(2008){Di Francesco}, {Johnstone}, {Kirk},
  {MacKenzie}, \& {Ledwosinska}}]{DiFrancesco2008}
{Di Francesco}, J., {Johnstone}, D., {Kirk}, H., {MacKenzie}, T., \&
  {Ledwosinska}, E. 2008, \apjs, 175, 277

\bibitem[{{Ebert}(1957)}]{Ebert1957}
{Ebert}, R. 1957, \zap, 42, 263

\bibitem[{{Enoch} {et~al.}(2008){Enoch}, {Evans}, {Sargent}, {Glenn},
  {Rosolowsky}, \& {Myers}}]{Enoch2008}
{Enoch}, M.~L., {Evans}, Neal~J., I., {Sargent}, A.~I., {et~al.} 2008, \apj,
  684, 1240

\bibitem[{{Friesen} {et~al.}(2014){Friesen}, {Di Francesco}, {Bourke},
  {Caselli}, {J{\o}rgensen}, {Pineda}, \& {Wong}}]{Friesen2014}
{Friesen}, R.~K., {Di Francesco}, J., {Bourke}, T.~L., {et~al.} 2014, \apj,
  797, 27

\bibitem[{{Friesen} {et~al.}(2018){Friesen}, {Pon}, {Bourke}, {Caselli}, {Di
  Francesco}, {J{\o}rgensen}, \& {Pineda}}]{Friesen2018}
{Friesen}, R.~K., {Pon}, A., {Bourke}, T.~L., {et~al.} 2018, \apj, 869, 158

\bibitem[{{Fu} {et~al.}(2011){Fu}, {Gao}, \& {Lou}}]{Fu2011}
{Fu}, T.-M., {Gao}, Y., \& {Lou}, Y.-Q. 2011, \apj, 741, 113

\bibitem[{{Goodman} {et~al.}(1998){Goodman}, {Barranco}, {Wilner}, \&
  {Heyer}}]{Goodman1998}
{Goodman}, A.~A., {Barranco}, J.~A., {Wilner}, D.~J., \& {Heyer}, M.~H. 1998,
  \apj, 504, 223

\bibitem[{{Harju} {et~al.}(2006){Harju}, {Haikala}, {Lehtinen}, {Juvela},
  {Mattila}, {Miettinen}, {Dumke}, {G{\"u}sten}, \& {Nyman}}]{Harju2006}
{Harju}, J., {Haikala}, L.~K., {Lehtinen}, K., {et~al.} 2006, \aap, 454, L55

\bibitem[{{Ho} {et~al.}(1978){Ho}, {Martin}, \& {Barrett}}]{Ho1978}
{Ho}, P.~T.~P., {Martin}, R.~N., \& {Barrett}, A.~H. 1978, \apjl, 221, L117

\bibitem[{{Hogerheijde} \& {van der Tak}(2000)}]{Hogerheijde2000}
{Hogerheijde}, M.~R. \& {van der Tak}, F.~F.~S. 2000, \aap, 362, 697

\bibitem[{Hugo {et~al.}(2009)Hugo, Asvany, \& Schlemmer}]{Hugo2009}
Hugo, E., Asvany, O., \& Schlemmer, S. 2009, The Journal of Chemical Physics,
  130, 164302

\bibitem[{{Keto} \& {Caselli}(2008)}]{Keto2008}
{Keto}, E. \& {Caselli}, P. 2008, \apj, 683, 238

\bibitem[{{Keto} \& {Caselli}(2010)}]{Keto2010}
{Keto}, E. \& {Caselli}, P. 2010, \mnras, 402, 1625

\bibitem[{{Keto} {et~al.}(2015){Keto}, {Caselli}, \& {Rawlings}}]{Keto2015}
{Keto}, E., {Caselli}, P., \& {Rawlings}, J. 2015, \mnras, 446, 3731

\bibitem[{{Kirk} {et~al.}(2017){Kirk}, {Dunham}, {Di Francesco}, {Johnstone},
  {Offner}, {Sadavoy}, {Tobin}, {Arce}, {Bourke}, {Mairs}, {Myers}, {Pineda},
  {Schnee}, \& {Shirley}}]{Kirk2017}
{Kirk}, H., {Dunham}, M.~M., {Di Francesco}, J., {et~al.} 2017, \apj, 838, 114

\bibitem[{{Krumholz} {et~al.}(2005){Krumholz}, {McKee}, \&
  {Klein}}]{Krumholz2005}
{Krumholz}, M.~R., {McKee}, C.~F., \& {Klein}, R.~I. 2005, \nat, 438, 332

\bibitem[{{Larson}(1969)}]{Larson1969}
{Larson}, R.~B. 1969, \mnras, 145, 271

\bibitem[{{Lique} {et~al.}(2015){Lique}, {Daniel}, {Pagani}, \&
  {Feautrier}}]{Lique2015}
{Lique}, F., {Daniel}, F., {Pagani}, L., \& {Feautrier}, N. 2015, \mnras, 446,
  1245

\bibitem[{{Lique} {et~al.}(2008){Lique}, {Tobo{\l}a}, {K{\l}os}, {Feautrier},
  {Spielfiedel}, {Vincent}, {Cha{\l}asi{\'n}ski}, \& {Alexander}}]{Lique2008}
{Lique}, F., {Tobo{\l}a}, R., {K{\l}os}, J., {et~al.} 2008, \aap, 478, 567

\bibitem[{{Maloney}(1988)}]{Maloney1988}
{Maloney}, P. 1988, \apj, 334, 761

\bibitem[{{Maret} {et~al.}(2013){Maret}, {Bergin}, \& {Tafalla}}]{Maret2013}
{Maret}, S., {Bergin}, E.~A., \& {Tafalla}, M. 2013, \aap, 559, A53

\bibitem[{{Myers} {et~al.}(1991){Myers}, {Ladd}, \& {Fuller}}]{Myers1991}
{Myers}, P.~C., {Ladd}, E.~F., \& {Fuller}, G.~A. 1991, \apjl, 372, L95

\bibitem[{{Pagani} {et~al.}(2007){Pagani}, {Bacmann}, {Cabrit}, \&
  {Vastel}}]{Pagani2007}
{Pagani}, L., {Bacmann}, A., {Cabrit}, S., \& {Vastel}, C. 2007, \aap, 467, 179

\bibitem[{{Pagani} {et~al.}(2004){Pagani}, {Bacmann}, {Motte}, {Cambr{\'e}sy},
  {Fich}, {Lagache}, {Miville-Desch{\^e}nes}, {Pardo}, \&
  {Apponi}}]{Pagani2004}
{Pagani}, L., {Bacmann}, A., {Motte}, F., {et~al.} 2004, \aap, 417, 605

\bibitem[{{Pagani} {et~al.}(2012){Pagani}, {Bourgoin}, \& {Lique}}]{Pagani2012}
{Pagani}, L., {Bourgoin}, A., \& {Lique}, F. 2012, \aap, 548, L4

\bibitem[{{Pagani} {et~al.}(2009){Pagani}, {Vastel}, {Hugo}, {Kokoouline},
  {Greene}, {Bacmann}, {Bayet}, {Ceccarelli}, {Peng}, \&
  {Schlemmer}}]{Pagani2009}
{Pagani}, L., {Vastel}, C., {Hugo}, E., {et~al.} 2009, \aap, 494, 623

\bibitem[{{Penston}(1969)}]{Penston1969}
{Penston}, M.~V. 1969, \mnras, 144, 425

\bibitem[{{Pickett} {et~al.}(1998){Pickett}, {Poynter}, {Cohen}, {Delitsky},
  {Pearson}, \& {M{\"u}ller}}]{Pickett1998}
{Pickett}, H.~M., {Poynter}, R.~L., {Cohen}, E.~A., {et~al.} 1998, \jqsrt, 60,
  883

\bibitem[{{Pineda} {et~al.}(2015){Pineda}, {Offner}, {Parker}, {Arce},
  {Goodman}, {Caselli}, {Fuller}, {Bourke}, \& {Corder}}]{Pineda2015}
{Pineda}, J.~E., {Offner}, S. S.~R., {Parker}, R.~J., {et~al.} 2015, \nat, 518,
  213

\bibitem[{{Pudritz} \& {Kevlahan}(2013)}]{Pudritz2013}
{Pudritz}, R.~E. \& {Kevlahan}, N.~K.~R. 2013, Philosophical Transactions of
  the Royal Society of London Series A, 371, 20120248

\bibitem[{{Redaelli} {et~al.}(2018){Redaelli}, {Bizzocchi}, {Caselli}, {Harju},
  {Chac{\'o}n-Tanarro}, {Leonardo}, \& {Dore}}]{Redaelli2018}
{Redaelli}, E., {Bizzocchi}, L., {Caselli}, P., {et~al.} 2018, \aap, 617, A7

\bibitem[{{Redaelli} {et~al.}(2019){Redaelli}, {Bizzocchi}, {Caselli},
  {Sipil{\"a}}, {Lattanzi}, {Giuliano}, \& {Spezzano}}]{Redaelli2019}
{Redaelli}, E., {Bizzocchi}, L., {Caselli}, P., {et~al.} 2019, \aap, 629, A15

\bibitem[{{Schnee} {et~al.}(2010){Schnee}, {Enoch}, {Johnstone}, {Culverhouse},
  {Leitch}, {Marrone}, \& {Sargent}}]{Schnee2010}
{Schnee}, S., {Enoch}, M., {Johnstone}, D., {et~al.} 2010, \apj, 718, 306

\bibitem[{{Schnee} {et~al.}(2012){Schnee}, {Sadavoy}, {Di Francesco},
  {Johnstone}, \& {Wei}}]{Schnee2012}
{Schnee}, S., {Sadavoy}, S., {Di Francesco}, J., {Johnstone}, D., \& {Wei}, L.
  2012, \apj, 755, 178

\bibitem[{{Shu}(1977)}]{Shu1977}
{Shu}, F.~H. 1977, \apj, 214, 488

\bibitem[{{Tafalla} \& {Hacar}(2015)}]{Tafalla2015}
{Tafalla}, M. \& {Hacar}, A. 2015, \aap, 574, A104

\bibitem[{{Tafalla} {et~al.}(1998){Tafalla}, {Mardones}, {Myers}, {Caselli},
  {Bachiller}, \& {Benson}}]{Tafalla1998}
{Tafalla}, M., {Mardones}, D., {Myers}, P.~C., {et~al.} 1998, \apj, 504, 900

\bibitem[{{Tafalla} {et~al.}(2004){Tafalla}, {Myers}, {Caselli}, \&
  {Walmsley}}]{Tafalla2004}
{Tafalla}, M., {Myers}, P.~C., {Caselli}, P., \& {Walmsley}, C.~M. 2004, \aap,
  416, 191

\bibitem[{{Tafalla} {et~al.}(2006){Tafalla}, {Santiago-Garc{\'\i}a}, {Myers},
  {Caselli}, {Walmsley}, \& {Crapsi}}]{Tafalla2006}
{Tafalla}, M., {Santiago-Garc{\'\i}a}, J., {Myers}, P.~C., {et~al.} 2006, \aap,
  455, 577

\bibitem[{{Takahashi} {et~al.}(2013){Takahashi}, {Ohashi}, \&
  {Bourke}}]{Takahashi2013}
{Takahashi}, S., {Ohashi}, N., \& {Bourke}, T.~L. 2013, \apj, 774, 20

\bibitem[{{Tilley} \& {Pudritz}(2004)}]{Tilley2004}
{Tilley}, D.~A. \& {Pudritz}, R.~E. 2004, \mnras, 353, 769

\bibitem[{{van der Tak} {et~al.}(2007){van der Tak}, {Black}, {Sch{\"o}ier},
  {Jansen}, \& {van Dishoeck}}]{vanderTak2007}
{van der Tak}, F.~F.~S., {Black}, J.~H., {Sch{\"o}ier}, F.~L., {Jansen}, D.~J.,
  \& {van Dishoeck}, E.~F. 2007, \aap, 468, 627

\bibitem[{{van der Tak} {et~al.}(2005){van der Tak}, {Caselli}, \&
  {Ceccarelli}}]{vanderTak2005}
{van der Tak}, F.~F.~S., {Caselli}, P., \& {Ceccarelli}, C. 2005, \aap, 439,
  195

\bibitem[{{Vastel} {et~al.}(2006){Vastel}, {Caselli}, {Ceccarelli}, {Phillips},
  {Wiedner}, {Peng}, {Houde}, \& {Dominik}}]{Vastel2006}
{Vastel}, C., {Caselli}, P., {Ceccarelli}, C., {et~al.} 2006, \apj, 645, 1198

\bibitem[{{Ward-Thompson} {et~al.}(1994){Ward-Thompson}, {Scott}, {Hills}, \&
  {Andre}}]{Ward-Thompson1994}
{Ward-Thompson}, D., {Scott}, P.~F., {Hills}, R.~E., \& {Andre}, P. 1994,
  \mnras, 268, 276

\end{thebibliography}

\clearpage

\begin{appendix}

  \section{Line analysis per core}
  \label{anal_core}

Our line profile findings per source are described as follows.
\vspace{0.2cm}

{\bf{L183:}} For L183 (Fig.~\ref{fig.plot_L183}), the $V_{LSR}$ ranges between 2.25$\pm$0.05~km~s$^{-1}$ and 2.48$\pm$0.04~km~s$^{-1}$, and within the upper and lower velocities reported by \citet{Ho1978} and \citet{Caselli2008}. The fact that the measured $V_{LSR}$ is not constant but varies with offset indicates local changes in the global kinematics of the core. The FWHM ranges from 0.32$\pm$0.08~km~s$^{-1}$ to 0.58$\pm$0.07~km~s$^{-1}$. With the exception of only few offsets ($>$ 40 $\arcsec$ South-West and North-West), the observed FWHM cannot be explained by thermal broadening alone (beyond the errors). The FWHM is overall up to $\sim$90\% larger than that expected from thermal broadening alone. Lastly, the $T_A^*$ varies between 0.12$\pm$0.03~K and 0.48$\pm$0.03~K, with the highest measurements found towards the center and a gradual decrease outwards. The observed variations at similar offsets, as well as the fact that there is an increase in $T_A^*$ at the most outward position, can be explained by inhomogeneities present in the medium.

{\bf{L694-2:}} The $V_{LSR}$ for L694-2 (see Fig.~\ref{fig.plot_L694-2}) ranges from 9.47$\pm$0.04~km~s$^{-1}$ to 9.65$\pm$0.03~km~s$^{-1}$. There is not a specific trend on the observed $V_{LSR}$ variations, but the difference of up to $\sim$0.2~km~s$^{-1}$ indicates local kinematic differences throughout the core. In this case, the FWHM ranges from 0.39$\pm$0.1~km~s$^{-1}$ to 0.63$\pm$0.09~km~s$^{-1}$. With the exception of a single offset at $\sim$ 42$\arcsec$, where thermal broadening alone can explain the observed FWHM, the non-thermal contributions are significant at all other offsets inwards ($<$~40$\arcsec$). The line is generally up to $\sim$96\% broader than expected from thermal broadening alone. The decreased FWHM at a larger offset, may indicate a decrease in non-thermal contributions at the outermost parts of the core. Finally, the $T_A^*$ values gradually decrease with offset, with local variations in $T_A^*$ pointing towards medium inhomogeneities.

{\bf{L1517B:}} If one compares Figure~\ref{fig.plot_L1517B} to Figures \ref{fig.plot_L183}--\ref{fig.plot_L694-2}, \ref{fig.plot_L1544}--\ref{fig.plot_L1521f}, it is noticeable that the H$_{2}$D$^{+}$ emission of L1517B is less spatially extended than seen towards the rest of the cores; Indeed, the line emission in this core is rather concentrated around the central position of the 850~$\mu$m dust peak ($<$~30$\arcsec$). The $V_{LSR}$ of L1517B ranges from 5.73$\pm$0.03~km~s$^{-1}$ to 5.82$\pm$0.03~km~s$^{-1}$, and decreases with increasing offset. The FWHM ranges from 0.36$\pm$0.06~km~s$^{-1}$ to 0.49$\pm$0.06~km~s$^{-1}$. Thermal broadening alone can explain the observed FWHM at most offsets for core temperatures $>$10~K. This temperature is higher by a factor of 2 compared to the one we determined in Sect.~\ref{rad_results} for this source, but still within the acceptable temperature range of prestellar cores. The line is $\sim$85\% broader than expected from thermal broadening alone. Lastly, the $T_A^*$ is once again the highest in the central position and decreases with increasing offset without signs of inhomogeneities.

{\bf{L1544:}} The H$_{2}$D$^{+}$ analysis for this source (Fig.~\ref{fig.plot_L1544}) shows that the $V_{LSR}$ ranges between 7.07$\pm$0.03~km~s$^{-1}$ and 7.21$\pm$0.02~km~s$^{-1}$, with a mean $V_{LSR}$ value of 7.15$\pm$0.03~km~s$^{-1}$. All measurements are within the range previously reported by \citet{Ho1978} and \citet{Caselli2008}. The FWHM ranges from 0.38$\pm$0.07~km~s$^{-1}$ to 0.61$\pm$0.08~km~s$^{-1}$, which is systematically higher than the thermal broadening contributions (e.g., a mean value of 0.48$\pm$0.07~km~s$^{-1}$ is $\sim$40\% higher). The observed widths are statistically larger than the expected thermal width in five different offsets, including the central position. As mentioned earlier, we attribute this difference to non-thermal contributions. With the exception of one offset at $\sim$40$\arcsec$ North, the non-thermal contributions are higher in the center compared to the rest of the offsets. The line is generally up to $\sim$90\% broader than expected from thermal broadening alone. Finally, the highest value of $T_A^*$ (0.55$\pm$0.02~K) is found in the central position, while it generally decreases with increasing offset. The fact that at certain offsets, the $T_A^*$ varies by a factor $>$ 2 and beyond the errors is indicative of local inhomogeneities in the medium (in density and/or temperature).

Given that N$_{2}$H$^{+}$ 4-3 emission was detected towards only two positions, we do not show a plot associated with its analysis but summarize our results as follows. The $V_{LSR}$ is found to be $\sim$7.20$\pm$0.04~km~s$^{-1}$, i.e., within the uncertainties of the corresponding H$_{2}$D$^{+}$ value. As in the case of H$_{2}$D$^{+}$, thermal broadening does not fully account for the FWHM of N$_{2}$H$^{+}$ (0.4$\pm$0.1~km~s$^{-1}$), as the observed FHWM is broader by a factor of $\sim$4. In the case of N$_{2}$H$^{+}$, non-thermal widths are higher by at least a factor of $\sim$2.5 compared to those of H$_{2}$D$^{+}$. Therefore, N$_{2}$H$^{+}$ emission appears to stem from a medium of higher non-thermal motions compared to H$_{2}$D$^{+}$ or that these mechanisms somehow affect N$_{2}$H$^{+}$ more. Finally, both detections of N$_{2}$H$^{+}$ have very low $T_A^*$ values, with a maximum value of 0.12$\pm$0.03~K, a factor of 5 lower than the H$_{2}$D$^{+}$ peak intensities. 

{\bf{L1521f:}} This core (see Figure~\ref{fig.plot_L1521f}) is the only known protostellar core in the sample \citep{Bourke2006} and the one out of two for which N$_{2}$H$^{+}$ 4-3 is detected. The line analysis of H$_{2}$D$^{+}$ is presented in Figure~\ref{fig.plot_L1521f}. The $V_{LSR}$ ranges between 6.41$\pm$0.02~km~s$^{-1}$ and 6.54$\pm$0.03~km~s$^{-1}$. The FWHM values range from 0.34$\pm$0.05~km~s$^{-1}$ to 0.55$\pm$0.07~km~s$^{-1}$. The FWHM can be explained by thermal broadening at the central position and higher offsets ($>$30$\arcsec$). At intermediate offsets ($\sim$20$\arcsec$), however, the FWHM is up to a factor of 2 larger than that expected from thermal broadening. While magnetic and turbulent contributions are also likely to play a role towards this core, we recall that this core is at a later evolutionary state compared to the rest of the sample. Indeed, its known compact outflow \citep{Takahashi2013} may also contribute to the observed line broadening, although H$_{2}$D$^{+}$ is not known to trace outflow activity. $T_A^*$ fluctuates between 0.10$\pm$0.02~K and 0.30$\pm$0.05~K without showing a specific trend. We attribute the observed variations to local changes in density and/or temperature (e.g., a clumpy medium).

As in L1544, N$_{2}$H$^{+}$ emission was detected towards only two positions, and therefore we do not show an associated offset plot. The $V_{LSR}$ took values of 6.47$\pm$0.03~km~s$^{-1}$ and 6.56$\pm$0.04~km~s$^{-1}$, are again within the uncertainties of the corresponding H$_{2}$D$^{+}$ values. At the central position, the FWHM is 0.6$\pm$0.1~km~s$^{-1}$, therefore broader by a factor of 5 compared to the width expected from thermal broadening. The non-thermal contributions decrease by 65\% at the offset position. Therefore, the non-thermal effects are higher in the center of the core. This behavior is in contrast to what we saw in the H$_{2}$D$^{+}$ case, where the central position did not require non-thermal effects to explain the observed FWHM. Therefore, as in the case of L1544, the non-thermal contributions seem to have a more significant effect on N$_{2}$H$^{+}$ emission compared to H$_{2}$D$^{+}$ emission. In the case of L1521f, this behavior could indicate that N$_{2}$H$^{+}$ emission originates from outflow cavities, as opposed to H$_{2}$D$^{+}$ emission, or that the magnetic pressure and turbulence affect the two species differently. Finally, at the offset position, the measured $T_A^*$ of the N$_{2}$H$^{+}$ line (0.30$\pm$0.04~K) is very similar to that of the H$_{2}$D$^{+}$ line. 

  \section{RATRAN results - original dataset}
  \label{ratran_ori}

In Figures~\ref{RATRAN_L183}--\ref{RATRAN_L1521f_n2hp}, we present the original, before azimuthal averaging, maps of H$_{2}$D$^{+}$ 1$_{10}$-1$_{11}$ and N$_{2}$H$^{+}$ 4-3 towards the cores, overplotted with the modelled ones for Static, LP, SIS and QE-BES models from RATRAN.
  
\begin{figure*}[ht]
\begin{center}
\includegraphics[scale=0.5]{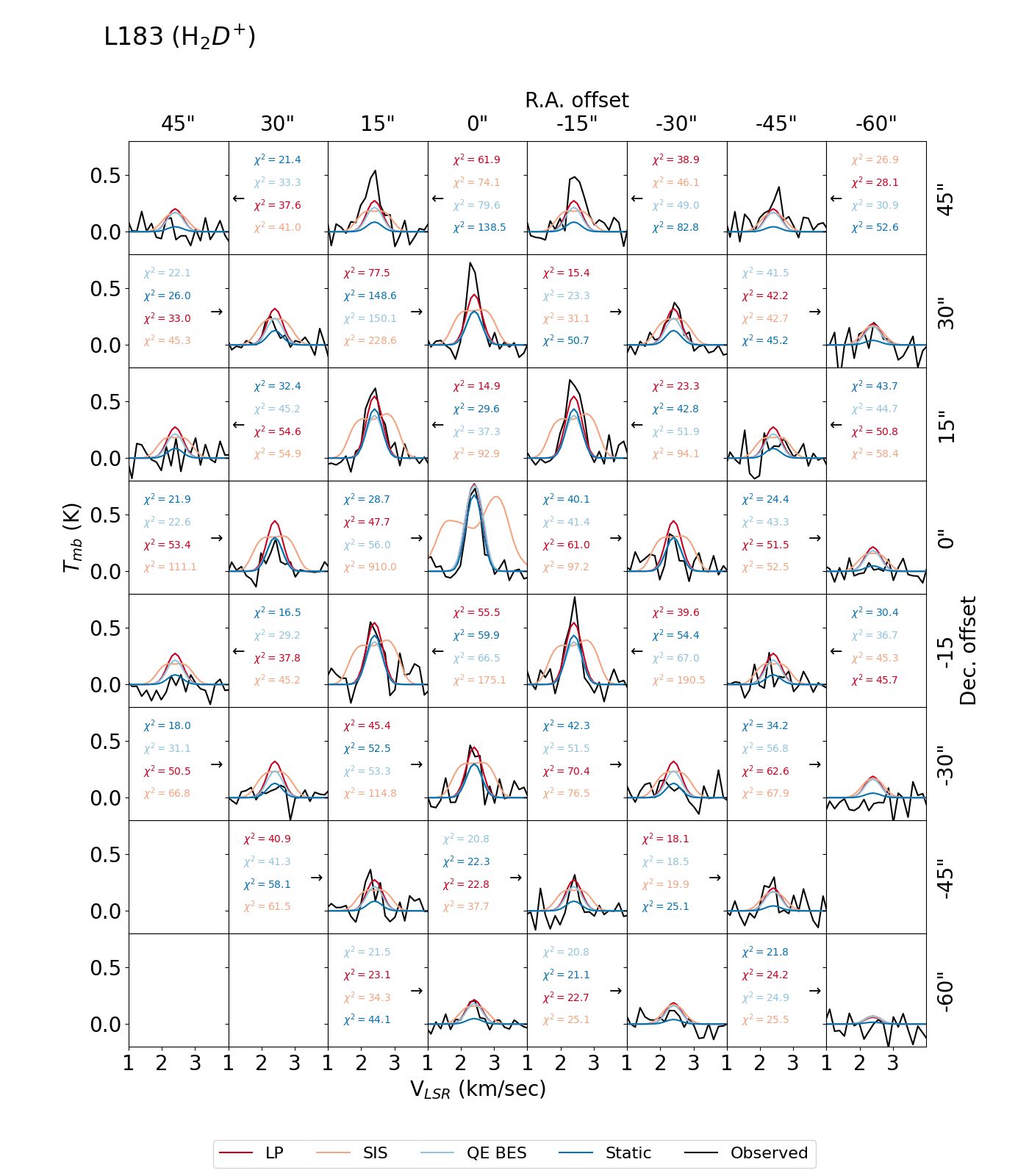}
\end{center}
\caption{Observed line profiles before azimuthal averaging for each position of H$_{2}$D$^{+}$ maps towards L183, overplotted with the modelled ones for Static, LP, SIS and QE-BES models from RATRAN.}
\label{RATRAN_L183_ori}
\end{figure*}


\begin{figure*}[ht]
\begin{center}
\includegraphics[scale=0.5]{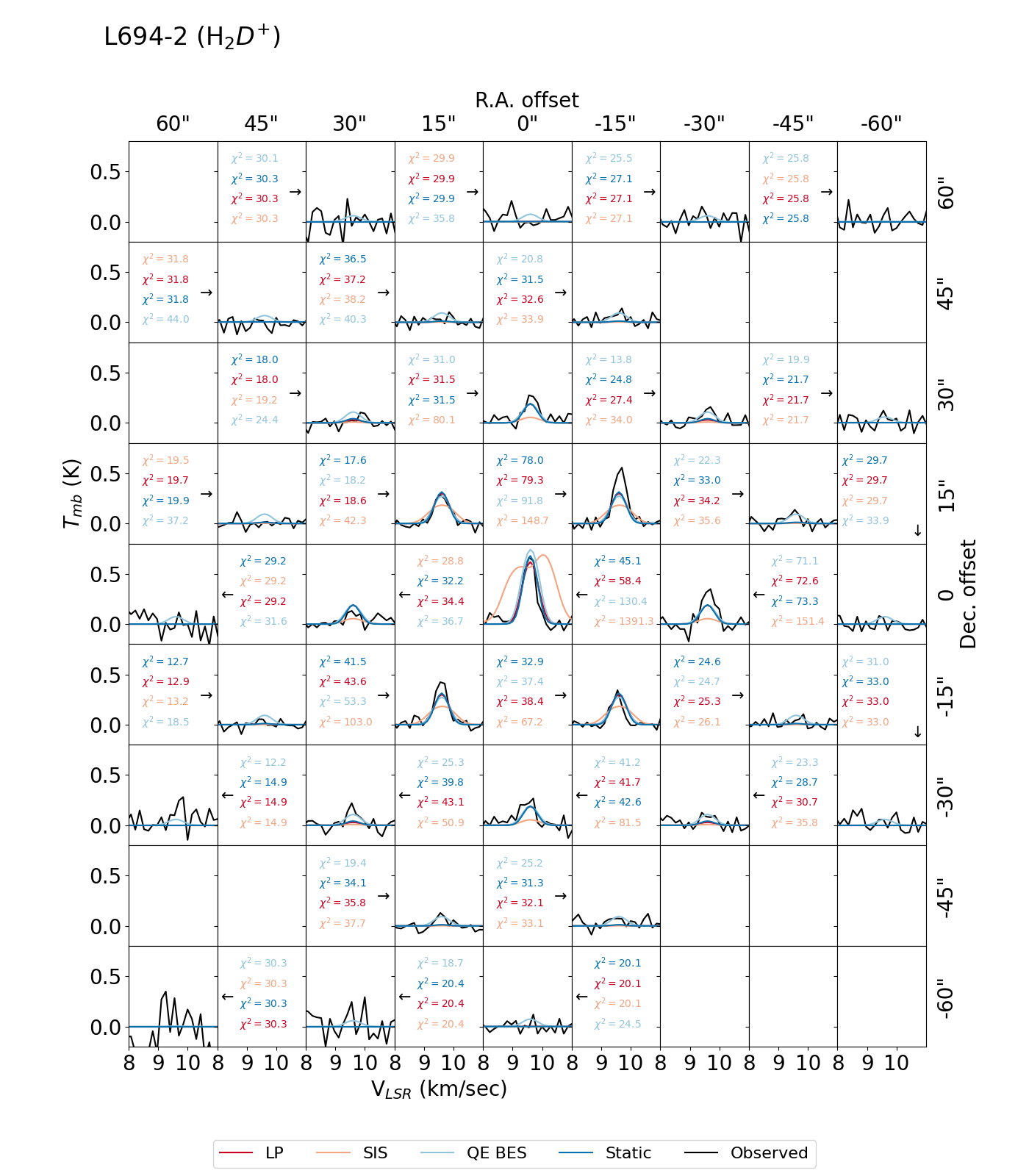}
\end{center}
\caption{As Fig.~\ref{RATRAN_L183_ori}, but for L694-2.}
\label{RATRAN_L694-2_ori}
\end{figure*}


\begin{figure*}[ht]
\begin{center}
\includegraphics[scale=0.5]{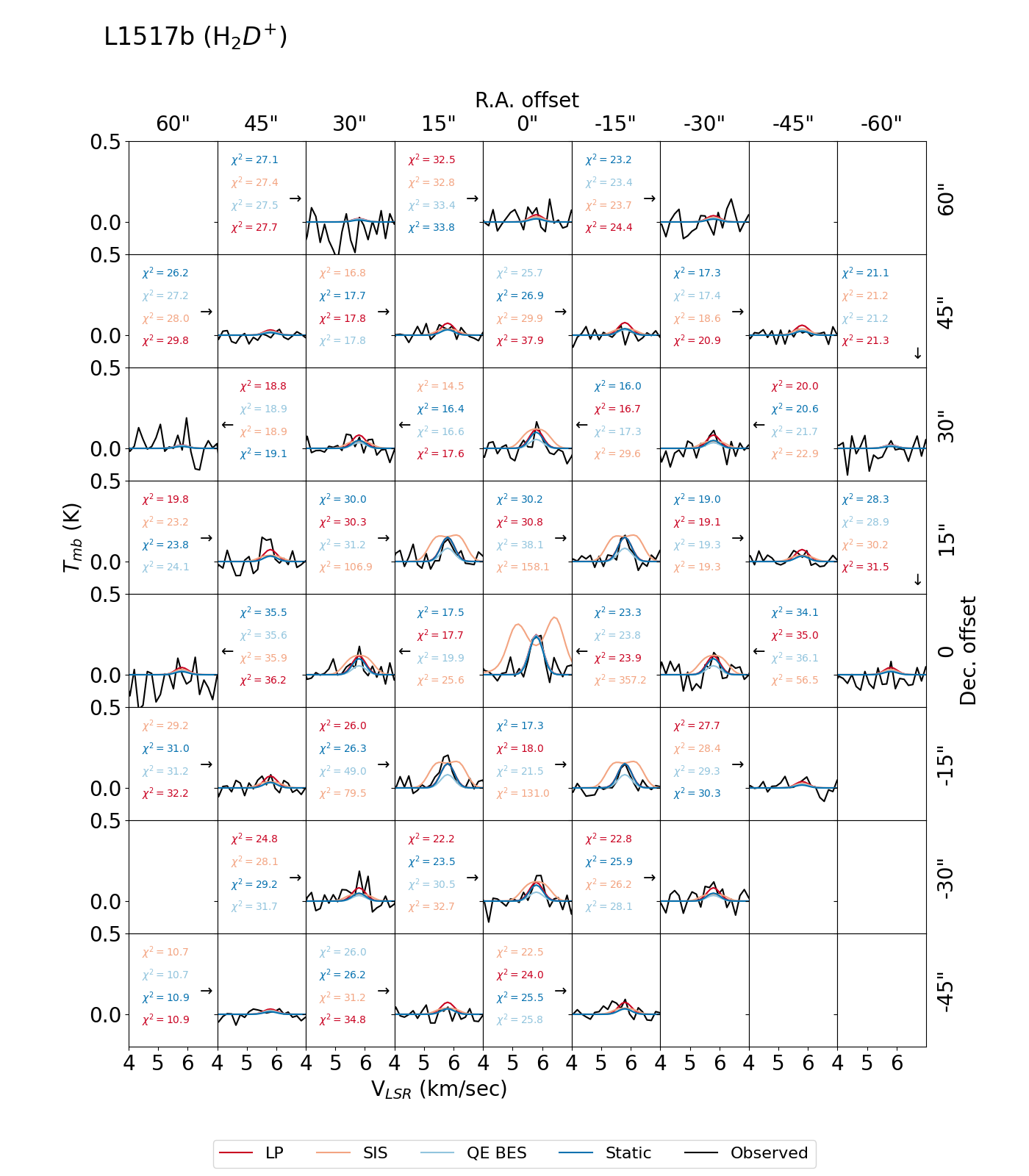}
\end{center}
\caption{As Fig.~\ref{RATRAN_L183_ori}, but for L1517B.}
\label{RATRAN_L1517B_ori}
\end{figure*}

\begin{figure*}[ht]
\begin{center}
\includegraphics[scale=0.5]{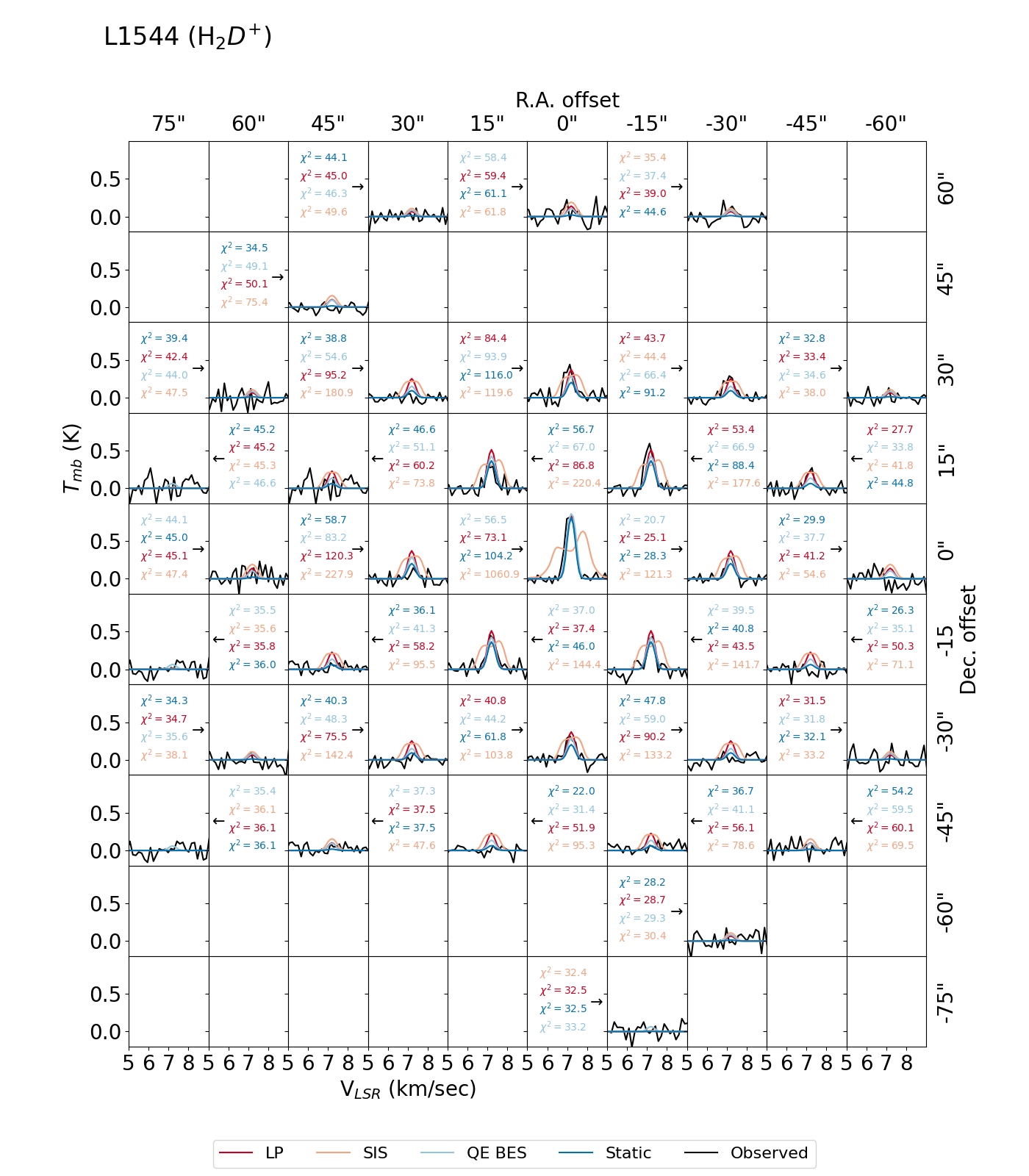}
\end{center}
\caption{As Fig.~\ref{RATRAN_L183_ori}, but for L1544.}
\label{RATRAN_L1544_ori}
\end{figure*}

\begin{figure*}[ht]
\begin{center}
\includegraphics[scale=0.5]{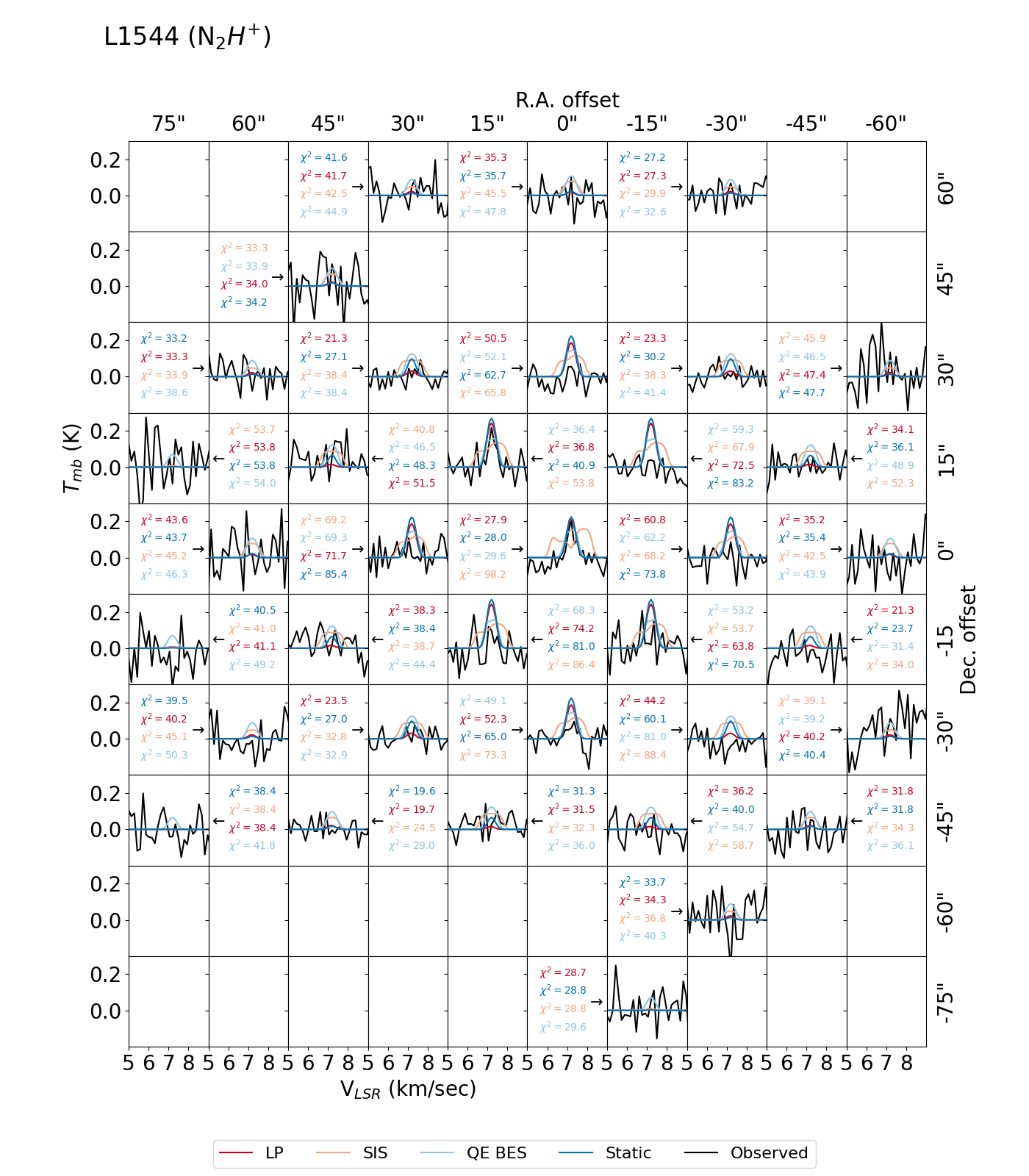}
\end{center}
\caption{As Fig.~\ref{RATRAN_L183_ori}, but for N$_{2}$H$^{+}$ towards L1544.}
\label{RATRAN_L1544_n2hp_ori}
\end{figure*}

\begin{figure*}[ht]
\begin{center}
\includegraphics[scale=0.5]{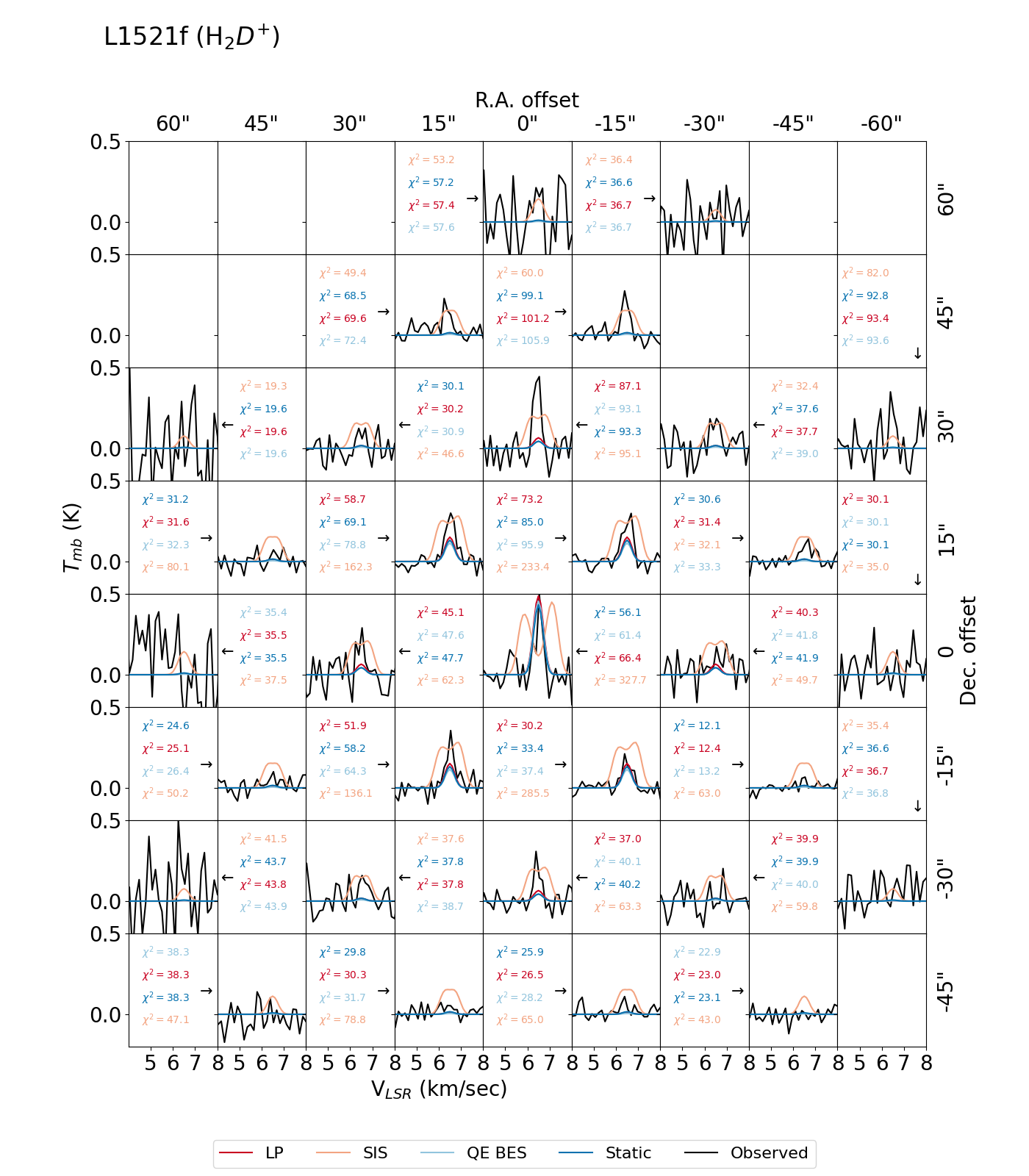}
\end{center}
\caption{As Fig.~\ref{RATRAN_L183_ori}, but for L1521f.}
\label{RATRAN_L1521f_ori}
\end{figure*}

\begin{figure*}[ht]
\begin{center}
\includegraphics[scale=0.5]{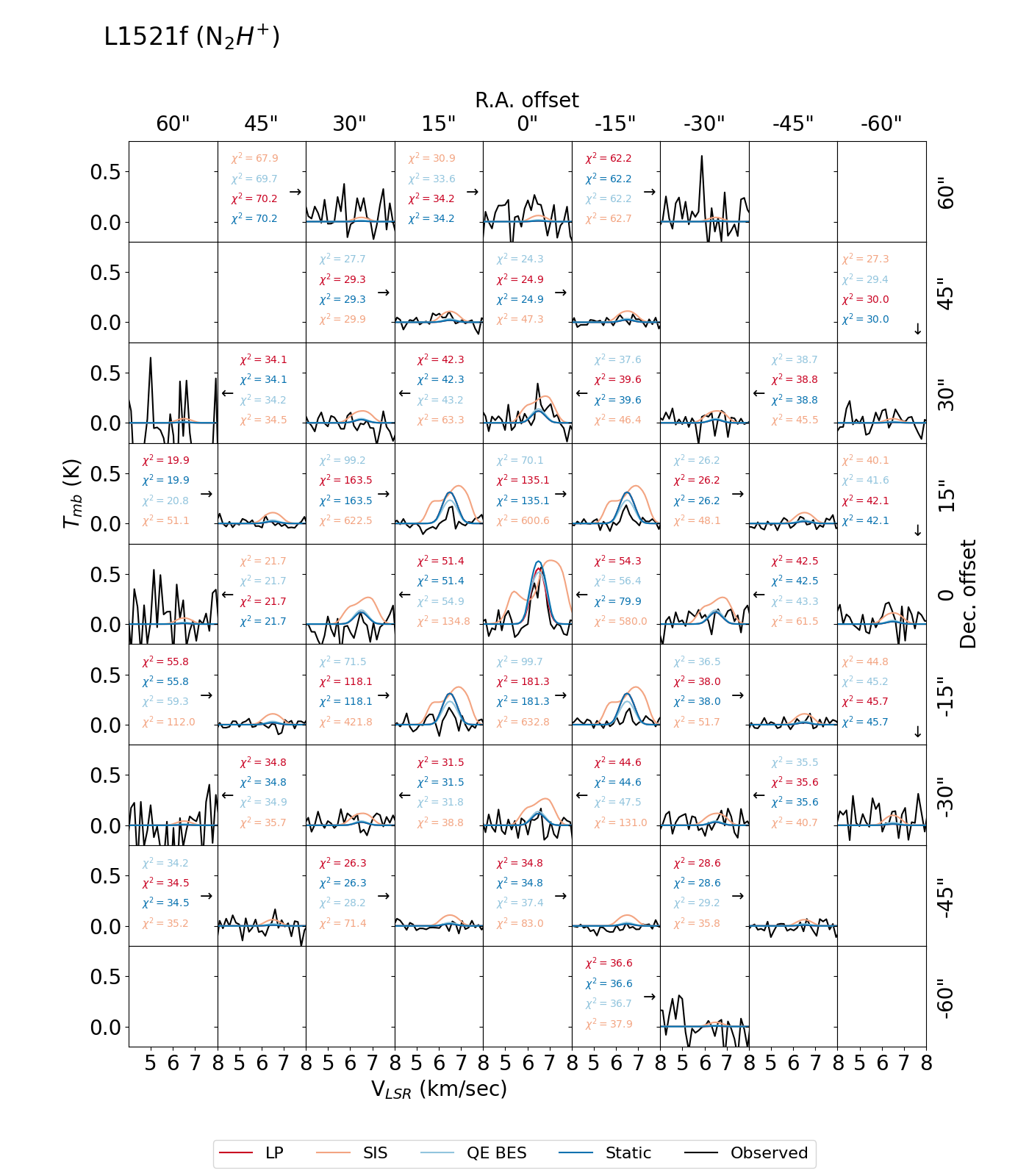}
\end{center}
\caption{As Fig.~\ref{RATRAN_L183_ori}, but for N$_{2}$H$^{+}$ towards L1521f.}
\label{RATRAN_L1521f_n2hp_ori}
\end{figure*}

\end{appendix}

\end{document}